\documentclass[a4paper,11pt]{article}
\usepackage{amsmath}
\usepackage{amssymb}
\usepackage{epsfig}
\DeclareMathOperator{\re}{Re}
\DeclareMathOperator{\im}{Im}
\DeclareMathOperator{\tr}{tr}
\DeclareMathOperator{\dete}{Det}

\newcommand{\dgr}{^{\rm o}}
\def\nostrocostrutto#1\over#2{\mathrel{\mathop{\kern 0pt \rlap% 
  {\raise.2ex\hbox{$#1$}}}
  \lower.9ex\hbox{\kern-.190em $#2$}}}
\def\gsim{\nostrocostrutto > \over \sim}   %greater or around...
\catcode`@=11
\newcount\@tempcntc
\def\@citex[#1]#2{\if@filesw\immediate\write\@auxout{\string\citation{#2}}\fi
  \@tempcnta\z@\@tempcntb\m@ne\def\@citea{}\@cite{\@for\@citeb:=#2\do
    {\@ifundefined
       {b@\@citeb}{\@citeo\@tempcntb\m@ne\@citea\def\@citea{,}{\bf ?}\@warning
       {Citation `\@citeb' on page \thepage \space undefined}}%
    {\setbox\z@\hbox{\global\@tempcntc0\csname b@\@citeb\endcsname\relax}%
     \ifnum\@tempcntc=\z@ \@citeo\@tempcntb\m@ne
       \@citea\def\@citea{,}\hbox{\csname b@\@citeb\endcsname}%
     \else
      \advance\@tempcntb\@ne
      \ifnum\@tempcntb=\@tempcntc
      \else\advance\@tempcntb\m@ne\@citeo
      \@tempcnta\@tempcntc\@tempcntb\@tempcntc\fi\fi}}\@citeo}{#1}}
\def\@citeo{\ifnum\@tempcnta>\@tempcntb\else\@citea\def\@citea{,}%
  \ifnum\@tempcnta=\@tempcntb\the\@tempcnta\else
   {\advance\@tempcnta\@ne\ifnum\@tempcnta=\@tempcntb \else \def\@citea{--}\fi
    \advance\@tempcnta\m@ne\the\@tempcnta\@citea\the\@tempcntb}\fi\fi}
\catcode`@=12
\begin{document}
%\input{gb1v4.tex}

%****************************************************************

% Titlepage

%****************************************************************

\thispagestyle{empty}
\begin{titlepage}

% BUTP Nr.
\vspace*{-1cm}
\hfill \parbox{3.5cm}{BUTP-98/27 \\ MPI-PhT/98-89\\
%hep-ph/9811518\\
%revised 16.2.1999
}   
%\vspace*{1.0cm}
\vfill

% Title
\begin{center}
  {\large {\bf
%      \hspace*{-0.2cm}The least massive glueballs and
%      \protect\vspace*{0.2cm} \\
%      \protect\hspace*{0.2cm}the scalar meson nonet} 
     Identification of the glueballs and the scalar meson nonet\\ of lowest
    mass}
    % the scalar $q\overline q$ mesons}
      \footnote{Work
      supported in part by the Schweizerischer Nationalfonds.}  }
%  \vspace*{3.0cm} \\
\vfill
% Authors
{\bf
    Peter Minkowski } \\
    Institute for Theoretical Physics \\
    University of Bern \\
    CH - 3012 Bern, Switzerland
   \vspace*{0.3cm} \\  
   and \vspace*{0.3cm} \\
{\bf
    Wolfgang Ochs } \\
    Max Planck Institut f\"ur Physik \\
    Werner Heisenberg Institut \\
    D - 80805 Munich, Germany\\
%   \vspace*{0.5cm} \\  

% Date
%XX January 1998

%\vspace*{4.0cm}
\vfill

% Abstract
\begin{abstract}
\noindent
%****************************************************************
We discuss the theoretical expectations and phenomenological evidence
for the lightest glueballs and the members of the meson nonet
with quantum numbers $J^{PC}=0^{++}$.  
We reconsider the recent evidence for candidate states with masses below
$\sim$1700 MeV, but include also the results from earlier 
phase-shift analyses. 
Arguments are presented to classify the scalars $f_0(980)$ and $f_0(1500)$ 
as members of the $0^{++}$ nonet, with a mixing rather similar
to that of the pseudoscalars $\eta'$ and $\eta$. The S-wave states 
called $f_0(400-1200)$ and $f_0(1370)$ are considered as 
different signals from a
single broad resonance, which we take to be the lowest-lying $0^{++}$ glueball. 
This state together with  
$\eta(1440)$ and  $f_J(1710)$ with spin $J=2$ form the basic triplet of binary 
gluonic bound states. We argue that these hypotheses are consistent with
what can be expected theoretically. 
%****************************************************************
\end{abstract}
\end{center}

\end{titlepage}

%****************************************************************

% End of Titlepage 

%****************************************************************

\newpage
%\tableofcontents
%\vspace*{4cm}
%\newpage
%****************************************************************

% Main Text

%****************************************************************

\pagestyle{plain}

%****************************************************************

%   Introduction

%****************************************************************

\section{Introduction}
\label{sec1}

An early prediction of QCD concerns the existence of a spectrum 
of glueballs, i.e. mesonic bound states of two or more constituent 
gluons, in addition to the spectrum of $q\overline q$ mesons with 
characteristic restrictions in the accessible quantum numbers \cite{HFPM}. 
Such glueball states have been
searched for extensively. 
The first 20 years of searches have seen some interesting candidates
\cite{Close},  especially in gluon rich processes 
such as radiative $J/\psi$ decays,
but no clear and convincing evidence for the two kinds of spectroscopy
have emerged, despite many efforts \cite{Heusch}. 

In the recent years these studies have entered a more optimistic phase.
On one hand, the theoretical predictions from lattice QCD 
were claimed to become more accurate. For the purely gluonic theory
these calculations put the lightest glueball 
into the mass region around 1600 MeV (for recent reviews, see 
\cite{Michael,Teper}). On the other hand, there are new 
experimental investigations with high statistical precision
aiming at a better understanding of the spectroscopy in the 1000-2000 MeV
region, especially by the Crystal Barrel Collaboration
\cite{Ams,Amseta,Ams4,Ams5,Ams2,abele}  in the analysis 
of $p\overline p$ annihilation at rest and by the WA102 Collaboration
\cite{Barb,Barb1} 
studying central production in high energy $pp$ collisions. 

The analysis of the recent results with inclusion of certain older experimental
data has apparently brought a new consensus supporting the lattice result
with a $J^{PC}=0^{++}$ glueball 
around 1600 MeV. In this channel more states are
reported than are
expected for the $q\overline q$ nonet 
\cite{Landua}. Prime candidates for the lightest glueball 
are the $f_0(1500)$ and the $f_J(1710)$ with spin taken as $J=0$. 
 As
the decay branching ratios of these states  do not follow 
closely the expectations for a glueball it is proposed that
these states and also the $f_0(1370)$ represent mixtures of the
glueball with the members of the $J^{PC}=0^{++}$ nonet. 
With this mixing scheme various experimental results can be
described \cite{amsclo,aps,clo}.

In this paper we begin with a discussion of the theoretical expectations
in Section 2. In particular it is pointed out that the first 
results from unquenched lattice calculations show large effects from
sea quarks with the tendency to decrease the glueball mass with decreasing
quark mass.
The spectral sum rules
require a gluonic contribution at low mass around 1 GeV. 
We discuss possible mass patterns for the
scalar $q\overline q$ states   % within the $\sigma$ model which arise from
in a model with general QCD potential and
explicit chiral symmetry breaking by quark masses. 

A closer look at the real world reveals a surprisingly complex experimental
situation and we implement the data with several 
{\em phenomenological hypotheses} :

\begin{description}
\item a) Despite the eventually strong mixing between quark and gluon states
it is possible to classify the ``more'' quark like states with
$L_{ q  \bar{q}} \ = \ 1$ into four nonets. 

\item b) The four isoscalar members of the respective nonets with largest mass
do not exceed (by more than $\sim$100 MeV) the mass of $f'_{ 2}  (  1525 )$ .

\item c) All members of the four nonets in question have been observed,
possibly with incorrect assignment of quantum numbers \cite{PDG} .

%\item d) The relevant spectroscopic quantity is mass square.
\end{description}

%Any one of the above hypotheses can be false. We strive
%for a clear situation at the risk of error as preferrable to a confused one
%admitting many conclusions equivalent to none.

In Section 3 we begin our phenomenological analysis and
discuss in some detail the evidence for resonance states in the mass
range below $\sim$1700 MeV. We pay particular attention to the earlier results
from elastic and inelastic $\pi\pi$ scattering which allow in principle the
determination of the amplitude phase. 
It is the evidence for the moving phase of the Breit-Wigner amplitude 
which is necessary to establish a resonant state. The restriction to the
study of resonance peaks may become misleading, especially if 
the state is broad and above ``background''. The present analysis confirms
the amplitude ``circles'' for the $f_0(1500)$ whereas we do not accept
the $f_0(1370)$ listed by the Particle Data Group (PDG) \cite{PDG}
as a genuine resonant state.   

In Section 4 we study the additional information provided by the
various couplings in production and decay in order to identify the members
of the $J^{PC}=0^{++}$ nonet. The satisfactory
solution includes the $f_0(980)$ and $f_0(1500)$. There is the broad object
seen in $\pi\pi$ scattering, often called ``background'', which extends from
about 400 MeV up to about 1700 MeV. This object we consider as a single broad
resonance\footnote{we refer to it as ``red dragon''} 
which we identify as the lightest glueball with quantum numbers
$J^{PC}=0^{++}$ as will be discussed 
in Section 5. %the $q\overline q$ multiplets   
Two further states with $J^{PC}=0^{-+}$ and $J^{PC}=2^{++}$ complete the basic triplet
of binary gluon states (Section 6). 

The conclusions are drawn in Section 7,
in particular, we compare our spectroscopic conclusions with the theoretical
expectations.

%****************************************************************

%\input{gb2v12.tex}
%\*** File 'gb2v12.tex'

\section{Theoretical expectations}
\label{sec2}
The purpose of this section is to clarify the possible mass patterns
of the lightest mesons %with the quantum numbers $J^{PC}=0^++$ of the vacuum
which are either bound states of quark-antiquark or of gluons.
We assume the dynamics to be reducible to chromodynamics with three
light flavors u, d and s. 

\subsection{Properties of low mass glueballs}
\label{glueballs}

\subsubsection{Scenarios for glueball and $q\overline q$
spectroscopy}

We consider first the spectroscopy in the chiral and antichiral limits
\begin{equation}
  \label{eq:1}
  \begin{array}{lll}
  \chi_{\ 3} \ : &\ \lim \ m_{\ u, d , s} &\ = \ 0
  \vspace*{0.3cm} \\
  \chi_{\ 2} \ : &\ \lim \ m_{\ u, d}& \ = \ 0
  \hspace*{0.3cm} ; \hspace*{0.3cm}
  m_{\ s} \ > \ 0 % \mbox{unchanged}
  \vspace*{0.3cm} \\
  \overline{\chi} \ : &\ \lim \ m_{\ u, d , s}& \ \rightarrow \ \infty
  \end{array}
\end{equation}
In the antichiral limit $\overline{\chi}$ the %quark-antiquark and 
gluon states become separately visible. The quantum numbers of these states,
scenarios for their masses as well as the decay properties
 have been discussed in Ref. \cite{HFPM}.  
Here we  consider the basic triplet of binary glueball states $gb_i$ 
which can be formed 
by two ``constituent gluons'' and correspond to the three invariants
which can be built from 
the bilinear expression of gluon fields $F_{a}^{\mu\nu}F^{a\rho\sigma}$
with $J^{PC}$ quantum numbers  
\begin{equation}
gb_0(0^{++}),\quad gb_1( 0^{-+})\quad \mbox{and} \quad 
     gb_2( 2^{++}) \label{triple}
\end{equation}
corresponding to the helicity wave functions of the two ``constituent
gluons''
$|11\rangle+|-1-1\rangle$, $|11\rangle -|-1-1\rangle$ and
$|1-1\rangle$ (or $|-11\rangle$). 
Theoretical calculations to be discussed below (bag model, sum rules,
lattice)
suggest the $0^{++}$ state to be the lightest one
\begin{equation}
%\overline{\chi}: \qquad
m_{gb_{0}}\ < \ m_{gb_{1}}\ ,
        \  m_{gb_{2}} \label{gbmin}
\end{equation}
\noindent
and these three states dominate 
the low energy dynamics. In the antichiral limit $\overline{\chi}$
the mass $m_{gb_{0, \infty}}$ of
the scalar glueball meson
defines the mass gap in the purely gluonic world;
 in this limit at least the lightest 
scalar and pseudoscalar  glueballs are stable.

In the chiral limits $\chi_2$ and $\chi_3$ the $q\overline q$ multiplets 
may partly overlap in mass with the glueball states. Of special
interest is the multiplet with the quantum numbers of the vacuum $0^{++}$
as its members have the same  quantum numbers as the glueball of lowest
mass. We  focus on the two 
alternatives for the glueball mass $m_{gb_0}$ and the  mass of the lightest
particle $m_{a_0}\sim 980$~MeV of the scalar $q\overline q$ nonet
(see Fig. \ref{gbfig1})
\begin{description}
\item 1) $m_{gb_0  } \lesssim \ m_{ a_{ 0}}$ corresponding to a ``light'' 
glueball
\item 2) $m_{gb_0 \ } \ \gg \ m_{ a_{ 0}}$ corresponding to a ``heavy''
 glueball.
%  \hspace*{0.3cm}
This condition is considered to be met  if $m_{gb_0  }$
exceeds $\sim 1500$ MeV.
\end{description}
The first alternative is an extension of scenario A, 
the second one of scenario(s) B(C) 
discussed in Ref. \cite{HFPM}.

In case 2) the basic triplet of binary glueballs is in the high mass
region. Then  their width is expected 
to be small according to perturbative
arguments (``gluonic Zweig rule'' \cite{HFPM}). Also in this case the glueball
states may be well separated in mass from the states in the $q\overline q$
nonet of lowest mass.

In case 1) which we favor  the width  
of $gb_{ 0}$ could be large.
First, the gluonic Zweig rule 
cannot be invoked any more as the coupling $\alpha_s$ 
at low energies may become large.
Secondly, the main decay mode is the pseudoscalar
channel $\pi\pi$ and at higher mass also 
$K\overline K$ and $\eta\eta$ and
there is a dynamical argument based on the
overlap of the wave functions between the external pseudoscalar states and the
intermediate gluon states:
The angular momentum between the constituent 2 gluons is dominated
by S waves ($L_{ gg} \ = \ 0$) and so are 
 the open 
pseudoscalar decay channels such as $\pi \pi$.
This alignment of S wave dominance in constituent quanta and 
two body decay channels distinguishes
$gb_{ 0}$ from the $0^{++}$ $q \overline{q}$ states which form an intermediate
P-wave. The same is also true for the lowest 
$q \overline{q}$ vector mesons where  the
intermediate S wave contrasts with the external P waves. We therefore expect
\begin{equation}
\Gamma_{gb_0}\ \gg  \ \Gamma_{q\overline q-hadron}. \label{gammaglu}
\end{equation}
Both arguments, the large coupling and the large overlap of internal and
external wave functions
 lead to the expectation of a broad $0^{++}$ glueball if it is light.

\subsubsection{Bag models}
A dynamical calculation of hadron masses has been achieved in models where
quarks and massless gluons are confined in spherical bags
of similar size.
If the $gg$ interaction is neglected
one expects \cite{JJ} the lightest glueball states with even parity 
to be degenerate
in mass and the same holds for the states with odd parity:
\begin{equation}
  \label{glbag}
  \begin{array}{ll}
\alpha_s=0:\quad & m_{gb}(0^{++})\ \sim\ m_{gb}(2^{++})\ \sim\ 0.87 \ \mbox{GeV}\\
                &  m_{gb}(0^{-+})\ \sim\ m_{gb}(2^{-+})\ \sim\ 1.3 \ \mbox{GeV} 
 \end{array}
\end{equation}
Inclusion of the $gg$
interaction leads to a hyperfine splitting and typically a mass ordering
\cite{Close}
\begin{equation}
\alpha_s\ne 0:\qquad\qquad\ \ \ \ 
             m_{gb_{0}}\ < \ m_{gb_{1}}\ <  \  m_{gb_{2}}.
   \qquad \label{gbmasses}
\end{equation}
The energy shifts in $O(\alpha_s)$ are calculated \cite{Barnes}
in terms of two parameters, the coupling $\alpha_s$ and the cavity radius
$a$. Reasonable values for these parameters
consistent with $q\overline q$ spectroscopy ($\alpha_s=0.748,
\ a^{-1}=0.218$ GeV) led to an identification of the $0^{-+}$
glueball with $\eta(1440)$ and of $2^{++}$ with $f_J(1720)$ in today's
nomenclature. The mass for the $0^{++}$ glueball was then {\em predicted} as
\begin{equation}
m_{gb_{0}}\ \sim 1\ \mbox{GeV}. \label{m0bag}
\end{equation}
and because of the self energy of the gluons in the bag this mass can hardly
become smaller. So in this unified treatment of $q\overline q$ and gluonium
spectroscopy the ``light glueball'' scenario 1) is preferrred.   

\subsubsection{QCD Spectral sum rules}

In a recent application of the sum rule approach \cite{svz} 
the basic quadratic gauge boson bilinear operators with quantum numbers 
$J^{ P C} \ = 0^{ ++}, \ 0^{ -+}$ and  $2^{ ++}$ have been  analysed by
 Narison \cite{Nar}
together with the lowest quark antiquark operators.
Constraints for the masses of a sequence of states are obtained by
saturating the spectral sum rules. It is interesting to note that
in the $0^{++}$ channel not all sum rules can be satisfied by a single 
glueball at a mass of around
1500 MeV -- as suggested by quenched lattice calculations.
Rather one is forced to include contributions from lower mass
 around 1~GeV with a large width. A consistent
solution is found with two states $\sigma_B(1000),\sigma'_B(1370)$
which both have  large gluonic couplings. A sum rule fit which includes
a light glueball
with mass around 500 MeV besides a heavier one at 1700 MeV 
has been presented by Bagan and Steele \cite{bagan}.  
We take these results as a further hint towards the need of a light
glueball in agreement with our findings. 

On the other hand, the 
spectrum of the next heavier gluon states differs
from our suggestions. Also our assignment of the scalar $q\overline q$ nonet 
is different from the one in \cite{Nar} and not along the OZI rule.

\subsubsection{Results from lattice gauge theory}
A serious tool to assess the spectral location
of glueball states -- in particular in the antichiral limit, where all quark
masses are sent to infinity -- comes from simulations of pure $SU(3)_{ c}$
Yang Mills theories on a lattice \cite{latt,latt2,Morning}. First results
from full QCD including sea quarks became available recently 
\cite{lattunq1,lattunq2}.

In the calculations without quarks one finds
the lowest lying scalar glueball 
$gb_{ 0}$  in the
mass range 1500 - 1700 MeV
which corresponds to our high mass scenario 2) discussed above.
In the scenario suggested by Weingarten \cite{latt2} 
the members of the scalar 
 $q\overline q$ nonet are taken to be the observed states 
listed in Table~\ref{eq:3b}. 
The quark composition is assumed along the OZI rule.
The actually observed particles with $0^{++}$ quantum numbers
$a_0(980)$ and $f_0(980)$ at lower energies 
are considered as ``irrelevant to glueball
spectroscopy'' and not taken as candidates for the scalar nonet. 
% and taken as $K\overline K$ molecules.
%
\begin{table}[ht]
\caption{Assignment of the (bare) scalar $\overline{q} q$ nonet
%{\bf \hspace*{2.85cm}
according to Weingarten \protect\cite{latt2}.}
\[ %  \begin{array}{l}
  \begin{array}{lccc}
\hline 
 \mbox{name} &  \overline{q} q & 
  \mbox{mass}^{ 2} \ \left \lbrack \mbox{GeV}^{ 2} \right \rbrack &
  \mbox{mass} \ \left \lbrack \mbox{GeV} \right \rbrack 
  \\ 
\hline %& & &  \\
  f_{ 0}        (1390)\qquad  
  & \frac{1}{\sqrt{2}} 
  \ \left ( 
  \ \overline{u} u \ + \ \overline{d} d 
  \ \right ) 
  & 1.932  & 1.390  \\
  a_{ 0}^{ 0}        (1450) \qquad 
  & \frac{1}{\sqrt{2}} 
  \ \left ( 
  \ \overline{u} u \ - \ \overline{d} d 
  \ \right ) 
  & 2.102  & 1.450 \\
  K_{ 0}^{ *}  (1430) \qquad
  & \overline s q, \ s\overline q
  & 2.042 & 1.429  \\
  f_{ 0}        (1500) \qquad
  &  \overline{s} s 
  & 2.265 & 1.505  
 % \\  & & &
  \\ \hline
%  \end{array}
  \end{array} \]

 \label{eq:3b}
\end{table}
%{\bf \hspace*{0.9cm} Table 1 : Assignment of the scalar $\bar{q} q$ nonet ;} \\
%{\bf \hspace*{2.85cm} according to Weingarten \cite{latt2}.}
%
%
%\vspace*{0.1cm}
%
%\noindent
%"Figure 1 shows the established $0^{ ++} \ , \ 1^{\ ++}$ and $2^{\ ++}$
%resonances and their strange $\ 0^{\ +} \, , \, 1^{\ +}$ and $2^{\ +}$ partners,
%{\em with the omission of $f_{\ 0} \ (980) \ , \ a_{\ 0} \ (980)$ and
%$f_{\ 1} \ (1420)$ , all irrelevant to glueball spectroscopy."}
%\hspace*{0.2cm} (our italics).
%
In a variant of this phenomenological scheme \cite{amsclo,Teper,latt2} 
one includes the $f_J(1720)$ with spin assignment $J=0$ and assumes 
the three observed $0^{++}$ states to be a superposition of the bare
glueball and the two bare isoscalar $q\overline q$ states.
In this mixing scheme one 
 can take into account the observed small $K\overline K$
branching ratio of the $f_0(1500)$.

First calculations in unquenched QCD
including two flavors of quark have been carried out by Bali et al.
\cite{lattunq1,lattunq2}. Results from a $16^3\times 32$ lattice
with an inverse lattice spacing
of $2  -  2.3 \ \mbox{GeV}$ show a definite dependence 
of the results on the quark mass
and correspondingly on the pion mass. If the pion mass is lowered from
about 1000  to  700 MeV the $0^{++}$ 
glueball mass decreases from 1400 to about 1200 MeV (using the 
data in \cite{lattunq2}). The quark masses are still quite large but 
in any case the 
glueball mass becomes smaller  than in case of the quenched
approximation without quarks
(see Table \ref{eq:43}). On the other hand, the calculations \cite{lattunq2}
also indicate a significant dependence on the volume. For a $24^3\times 40$
lattice the glueball mass goes up again to the larger number 
of the quenched calculation.

\begin{table}[ht]
\caption{Glueball masses with statistical and systematic errors in quenched 
lattice approximation whereby the first two determinations are based on
data in \protect\cite{latt}
(upper part) and with inclusion of sea quarks
for different spatial lattice sizes $L_S$ (lower part).}
\[
%  \begin{array}{l} 
  \begin{array}{lccc} 
\hline
  \mbox{author} & m_{ gb_{ 0}} \ \left \lbrack  \mbox{MeV} 
        \right \rbrack \quad
  & n_{ fl} \ &  \ m_{ \pi}
  \ \left \lbrack  \mbox{MeV}  \right \rbrack 
  \\ \hline %\vspace*{-0.5cm}
 % \\ % & & & \\
 \mbox{Teper}\cite{Teper} & 1610 \pm 70\pm 130 & 0 & \\
 \mbox{Weingarten} \cite{latt2} & 1707  \pm \ 64 &  0&  \\
 \mbox{Morningstar et al.} \cite{Morning} & 1630  \pm 60 \pm 80 &0& \\ 
\hline
 \mbox{Bali et al.} \cite{lattunq1,lattunq2}  & \sim 1200 \ (L_S=16) & 
          2    &     700 - 1000 \\
 \mbox{Bali et al.} \cite{lattunq2} & \sim 1700 \ (L_S=24) & 2  &  \\ \hline
%  \end{array}
  \end{array} \]

 \label{eq:43}
\end{table}
%

%These two results not only differ in the mass value, which may
%be the result of the nontrivial quark flavor environment in the two calculations,
%but most importantly, the calculation of Bali et al. \cite{latt} shows a 
%very pronounced dependence of $m_{ gb_{ 0}}$ on $m_{ \pi}$ which%
%corresponds to the different quark mass parameters compared :%
%an increase of $m_{ \pi}$ by 7 per cent - from the lowest val%ue
%chosen $m_{ \pi} \ \sim \ 600 \ \mbox{MeV}$ - causes an increase
%of $m_{ gb_{ 0}}$ by 14 per cent! If we use the relation
%
%\begin{equation}
%  \label{eq:44}
%  \begin{array}{l} 
%  m_{ gb_{ 0}} \ \propto \ m_{ \pi}^{2}
%  \end{array}
%\end{equation}
%
%\noindent
%which should be still valid for $m_{ \pi} \ \sim \ 600 \ \mbox{MeV}$ 
%to a good approximation and extrapolate the result of Bali et al. \cite{latt} 
%to zero quark mass we find $m_{ gb_{ 0}} \ \rightarrow \ 0$.

\noindent
We conclude from the mass values in Table \ref{eq:43} that the 
quenched calculation
supports the ``high mass region'' for the lightest glueball,
i.e. our alternative 2), while the results with sea quarks do not exlude the
opposite,
 i.e.  $gb_{ 0}$ placed into the ``low mass region'', in view of
the large values of the quark masses in the calculation 
and the observed decrease of the glueball
mass with the quark mass. 

It is also of great interest to compute 
the mass of the lightest scalar state $a_0(0^{++})$ in lattice QCD.
First results have been 
obtained recently in quenched approximation with non-perturbatively 
$O(a)$ improved Wilson fermions \cite{Gockel}
%[M. G\"ockeler et al., hep-lat/9707021]
for two values of the coupling $\beta$,
see Table \ref{scalar}.

\begin{table}[ht]

\begin{center}
\begin{tabular}{lccc}
\hline
% & & & \\
%\rule[-2ex]{0ex}{5.5ex}
$ \beta$  &    ${a^2}/{K}$  & $ R={aM_{a0}}/{a\sqrt{K}}$  &  $ R\sqrt{K}$ 
[GeV]\\ \hline
%& & & \\
%\rule[-2ex]{0ex}{5.5ex}
6.0   &   0.048  &   3.72  $\pm$  0.15    &     1.59  $\pm$  0.06 \\
%& & & \\
%\rule[-2ex]{0ex}{5.5ex}
6.2   &   0.026  &   2.86  $\pm$  0.13    &     1.22  $\pm$  0.05 \\
\hline
\end{tabular}
\end{center}
\caption{The ratio $R$ of the
$a_0$-mass and the string tension % $\protect\sqrt{K}$
as function of the square of the lattice
spacing $a$ in units of the string tension. 
The last column shows the ratio $R$ multiplied with the physical value of
the string tension % $\protect\sqrt{K}=$ 
0.427 GeV \protect\cite{Gockel}.}
\label{scalar}
\end{table}
The ratio $R\sqrt{K}$ in the last column of Table \ref{scalar}
extrapolates to the
physical mass in the continuum limit
 $a^2 \to 0$. As can be seen these 
mass values decrease in the approach of this limit, the lowest mass value
being $M_{a_0} \sim 1.2$~GeV.  A reliable extrapolation from two data points
cannot be expected. If one extrapolates nevertheless one finds
$M_{a_0} \sim 0.8$ GeV. 
These results seem to be consistent with
the mass 0.98 GeV for the lightest scalar meson, but the heavier mass 1.45 GeV
as suggested for this state by Weingarten
cannot be  excluded  on the basis of only two
measurements.\footnote{We thank D. Pleiter and S. Aoki for helpful
discussions of these results}
With improved accuracy such calculations could provide an interesting hint
towards the classification of the $a_0(980)$ state as the lowest mass scalar
meson.

The results reported here indicate that our hypothesis of a light glueball
with mass around 1 GeV accompanied by a scalar nonet with particles
around 1 GeV
is not necessarily in contradiction with lattice QCD results.
We also wish to point out that the parametrization of the $0^{++}$ spectrum
in terms of one resonance only may not be appropriate; this was found in
case of QCD sum rules and may be true in particular
if the lightest state is very broad.

%We restate here the main purpose of this paper : to establish alternative 1) ,
%i.e. to find $m_{ gb_{ 0}}$ in the ``low mass region''.
%%\vspace*{0.1cm}

\subsection{The scalar nonet % and pseudoscalar nonets 
and effective $\Sigma$ variables} 
\label{Sigma}

An important precondition for the assignment of glueball states 
is the understanding of the low mass $q\overline q$ spectroscopy.
As the lightest glueball is expected with $J^{PC}=0^{++}$ quantum numbers
we focus here on the expectations for the lightest scalar $q\overline q$
nonet. The lightest particles with these quantum numbers
are $a_0(980)$ and $f_0(980)$,
approximately degenerate in mass. Some authors consider one or both of these
states as $K\overline K$ molecules \cite{molecule} and take $a_0(1450)$
and  $f_0(1370)$ (or a broad $f_0(1000)$) as members of the scalar nonet.
The next (uncontroversial) candidate for the nonet is $K^*_0(1430)$;
the last member of the nonet is a heavier isoscalar state $f_{0>}$,
possibly $f_0(1500)$ or $f_J(1720)$,
 which can mix with the lighter $f_0(980)$. 

An attractive theoretical approach to the scalar and pseudoscalar mesons is
based on the ``linear sigma models'' 
which realize the spontaneous chiral symmetry breakdown (for reviews, see
\cite{ls1}). The requirement of renormalizability provides a considerable
restriction in the functional form of the effective potential compared to
what would be generally allowed. 
In a recent application  
T\"{o}rnqvist \cite{Torn} considered a  renormalizable Lagrangian 
for the scalar and pseudoscalar sector. 
In his solution for the scalar nonet the 
OZI rule holds exactly for the bare states with a broad isoscalar
non-strange ``$\sigma$''
resonance below 1~GeV and the $f_0(980)$ as the lowest
$s\overline s$ state. The resulting mass spectrum, however, 
is considerably modified by unitarisation effects.

In an  alternative approach \cite{dmitra,klempt,burakovsky} 
one starts from a 3-flavor Nambu-Jona-Lasinio
model but includes an  effective action for the sigma fields
with an instanton induced axial $U(1)$ symmetry-breaking
determinant term (proportional to $I_3$ in Eq. (\ref{eq:6}) below), 
along the suggestion by t'Hooft \cite{thooft}, which keeps the Lagrangian
renormalizable. This corresponds again to a linear sigma
model but now the scalars are close to the singlet and octet states, and 
they do not
split according to the OZI rule; the sign of the mass splitting in the
scalar and pseudoscalar sectors is reversed. This suggests $f_0(1500)$ 
to be near the octet state whereas different options are pursued
for the lighter isoscalar $f_0$ and the isovector $a_0$ by the authors
\cite{dmitra,klempt,burakovsky}.

In our approach we do not follow the $K\overline K$
molecule hypothesis for the $f_0(980)$ and the $a_0(980)$  (see also
the remarks in the next section)
but take them as genuine members of the $q\overline q$ scalar nonet.
In the rest of this section we discuss
what can be derived about the mass of $f_{0>}$ and the mixing pattern 
in the scalar nonet from the most general effective QCD potential 
for the  $\Sigma$-variables pertaining to the scalar and pseudoscalar
mesons, whereby we do not restrict the analysis to renormalizable
interaction terms. In this way we explore the consequences of chiral
symmetry in the different limits in (\ref{eq:1}) in a general QCD framework.

Thereafter we turn to the phenomenological analysis of data where we 
try to  minimize the theoretical preconditions as suggested 
by the present section.
 
\subsubsection{$\Sigma$ variables and chiral invariants}

We assume that the glueball states do not
affect in an essential way the remaining effective degrees of freedom 
at low energy. 
Then all degrees of freedom can be integrated out. 
The variables are those of a linear sigma model 
\begin{equation}
  \label{eq:3}
  \begin{array}{c}
   \Sigma_{\ st} \ = \ ( \ \sigma_{\ st} \ - \ i \ p_{\ st} \ )
\vspace*{0.3cm} \\
 \sum_{\ c} \ \overline{q}_{\ s}^{\ c} \ q_{\ t}^{\ c} \ \leftrightarrow
  \ \sigma_{\ st}
  \hspace*{0.3cm} ; \hspace*{0.3cm}
  \sum_{\ c} \ \overline{q}_{\ s}^{\ c} \ i \ \gamma_{\ 5} \ q_{\ t}^{\ c} 
  \ \leftrightarrow \ p_{\ st}
%  \hspace*{0.3cm} ; \hspace*{0.3cm}
%  s,t \ = \ u,d,s
  \end{array}
\end{equation}
where the indices $s,t$ refer to the flavors $u,d,s$.
We do not require interactions to be renormalizable, 
rather we study the general effective action of QCD restricted 
to the sigma variables \cite{PM}. 
The resulting mass spectra and mixings are then less restricted
than in the renormalizable Lagrangian models: 
for example, OZI splitting is possible but not particularly favored. 
%Our subsequent phenomenological analysis in fact does not
%support the popular 
%spectral assignment of isoscalar resonances along the OZI rule
%but rather one with small singlet-octet splitting.

In Eq. (\ref{eq:3}) we chose
the normalization of the complex (nonhermitian) field
variables $\Sigma_{\ st}$ such that in the chiral limit
$\chi_{\ 3} \ (  \lim \ m_{\ u, d , s} \ = \ 0$ ) the vacuum expected value
corresponds to the (real) unit matrix :
\begin{equation}
  \label{eq:4}
  \begin{array}{l}
\chi_{\ 3}
  \hspace*{0.3cm} : \hspace*{0.3cm}
\ \left \langle \ \Omega \ \right | \ \Sigma_{\ st} \ ( \ x \ ) 
\ \left | \ \Omega \ \right \rangle \ \rightarrow \ \delta_{\ st}.
  \end{array}
\end{equation}
\noindent
We propose to discuss the {\em general} form of the effective
potential, more precisely its real part -- restricted only to the first order
approximation with respect to the strange quark mass term in the two flavor
chiral limit $\chi_{\ 2}$ ($ \lim \ m_{\ u, d} \ = \ 0$)
\begin{equation}
  \label{eq:5}
  \begin{array}{l}
\chi_{\ 2} 
  \hspace*{0.3cm} : \hspace*{0.3cm} 
  V \ ( \ \Sigma \ ) \ \rightarrow \ V_{\ 0} \ - \ \mu_{\ s} \ \re \ \Sigma_{\ 33}
  \hspace*{0.3cm} ; \hspace*{0.3cm} \mu_{\ s} \ \propto \ m_{\ s}.
  \end{array}
\end{equation}
\noindent
The quark mass parameter $\mu_{\ s}$ in Eq. (\ref{eq:5}) is to be expressed
in appropriate units ($\mbox{mass}^{4}$). $V_{\ 0}$ refers to the
chiral limit $\chi_{\ 3}$;
it depends in an a priori arbitrary way on four base variables
for which we can choose
\begin{equation}
  \label{eq:6}
  \begin{array}{l}
  I_{\ 1} \ = \ \tr \ \Sigma  \ \Sigma^{\ \dagger} 
  \ - \ \tr \ {\bf 1}
  \hspace*{0.3cm} ; \hspace*{0.3cm}
  I_{\ 2} \ = \ \tr \ ( \ \Sigma  \ \Sigma^{\ \dagger} \ )^{\ 2}
  \ - \ \tr \ {\bf 1}
  \vspace*{0.3cm} \\
  I_{\ 3} \ = \ \re \ \dete \ \Sigma \ -  1
  \hspace*{0.3cm} ; \hspace*{0.3cm}
  I_{\ 4} \ = \ \im \ \dete \ \Sigma 
  \end{array}
\end{equation}
\noindent
If we introduce  
 the shifted variables
\begin{equation}
\Sigma \ = \ {\bf 1} \ + \ Z
  \hspace*{0.3cm} ; \hspace*{0.3cm}
  Z \ = \ s \ - \ i \ p
\label{shift}
\end{equation}
we can express the four invariants defined in Eq. (\ref{eq:6}) as
\begin{equation}
  \label{eq:6a}
  \begin{array}{l}
  I_{\ 1} \ = \ 2 \ \tr \ s \ + \ \ \ \tr \ s^{\ 2} \ + \ \ \ \tr \ p^{\ 2}
  \vspace*{0.3cm} \\
  I_{\ 2} \ = \ 4 \ \tr \ s \ + \ 6 \ \tr \ s^{\ 2} \ + 2 \ \tr \ p^{\ 2}
              \ + \ 4 \ \tr \ s^{\ 3} \ + \ 4 \ \tr \ s \ p^{\ 2}
  \vspace*{0.3cm} \\
  \hspace*{7.5cm}      \ + \ \tr \ ( \ Z \ Z^{\ \dagger} \ )^{\ 2}
  \vspace*{0.3cm} \\
  I_{\ 3} \ = \ \ \ \tr \ s \ + \ \frac{1}{2} 
  \ \left (
  \begin{array}{l}
  ( \ \tr \ s \ )^{\ 2} \ - \ \tr \ s^{\ 2} 
  \ - \ ( \ \tr \ p \ )^{\ 2} \ + \ \tr \ p^{\ 2} 
  \end{array}
  \ \right )
  \vspace*{0.3cm} \\
  \hspace*{2.2cm}      \ + \ \re
  \ \frac{1}{6} 
  \ \left (
  \ ( \ \tr \ Z \ )^{\ 3} \ - \ 3 \ \tr \ Z \ \tr \ Z^{\ 2}
  \ + \ 2 \ \tr \ Z^{\ 3}
  \ \right )
  \vspace*{0.3cm} \\
  I_{\ 4} \ = - \ \tr \ p \ - 
  \ \left (
  \ \tr \ s \ \tr \ p \ - \ \tr \ s \ p 
  \ \right )
  \vspace*{0.3cm} \\
  \hspace*{2.2cm}      \ + \ \im
  \ \frac{1}{6} 
  \ \left (
  \ ( \ \tr \ Z \ )^{\ 3} \ - \ 3 \ \tr \ Z \ \tr \ Z^{\ 2}
  \ + \ 2 \ \tr \ Z^{\ 3}
  \ \right )
  \end{array}
\end{equation}
%
%================================================================
\noindent
There is no loss of generality - {\em concentrating on scalar mass terms only} -
to restrict $\Sigma$ to the hermitian matrix $s$ whereby the four variables
in Eq. (\ref{eq:6}) reduce to three :
\begin{equation}
  \label{eq:7}
  \begin{array}{l}
  I_{\ 1} \rightarrow \ 2 \ \tr \ s \ + \ \ \ \tr \ s^{\ 2} 
  \vspace*{0.3cm} \\
  I_{\ 2} \ \rightarrow \ 4 \ \tr \ s \ + \ 6 \ \tr \ s^{\ 2} 
              \ + \ 4 \ \tr \ s^{\ 3}
           \ + \ \tr \ s^{\ 4}
  \vspace*{0.3cm} \\
  I_{\ 3} \ \rightarrow \ \ \ \tr \ s \ + \ \frac{1}{2} 
  \ \left ( 
  \  ( \ \tr \ s \ )^{\ 2} \ - \ \tr \ s^{\ 2} 
  \ \right )
  \vspace*{0.3cm} \\
  \hspace*{2.2cm}      \ +
  \ \frac{1}{6} 
  \ \left (
  \ ( \ \tr \ s \ )^{\ 3} \ - \ 3 \ \tr \ s \ \tr \ s^{\ 2}
  \ + \ 2 \ \tr \ s^{\ 3}
  \ \right ).
  \end{array}
\end{equation}

\subsubsection{Scalar mass terms to order $\mu_{\ s}$}

To the precision required we need the expansion of $V_{\ 0} \ ( \ s \ )$
to third order in the matrix variable $s$. 
To third order in $s$ the three base variables
in Eq. (\ref{eq:7}) can be replaced by the simple power basis
$\ \tr \ s$ , $\ \tr \ s^{\ 2}$ , $\ \tr \ s^{\ 3}$.
\noindent
As a consequence $V_{\ 0} \ ( \ s \ ) \ $ is of the form
\begin{equation}
  \label{eq:9}
  \begin{array}{l}
  V_{\ 0} \ =
  \begin{array}[t]{l}
  \frac{1}{2}
  \ \left ( 
  \ A \ \tr \ s^{\ 2} \ + \ B \ ( \ \tr \ s \ )^{\ 2}
  \ \right )
  \ + 
  \vspace*{0.3cm} \\
  \frac{1}{3}
  \ C \ \tr \ s^{\ 3} \ +
  \ \frac{1}{2}
  \ D \ ( \ \tr \ s \ ) \ \tr \ s^{\ 2} \ +
  \ \frac{1}{3}
  \ E \ ( \ \tr \ s \ )^{\ 3} \ + \ O \ ( \ s^{\ 4} \ ).
  \end{array}
  \end{array}
\end{equation}
\noindent
We shall neglect the terms of order $s^{ 4}$ in the following.
To first order in the strange quark mass the vacuum expected values
are shifted from their values in Eq. (\ref{eq:4}) according to Eq. (\ref{eq:5})
\begin{equation}
  \label{eq:10}
  \begin{array}{l}
  \left \langle \ \Omega \ \right | \ \Sigma
  \ \left | \ \Omega \ \right \rangle
  \ = 
  \ {\bf 1} \ + 
  \ \left \langle \ s \ \right \rangle
  \hspace*{0.3cm} ; \hspace*{0.3cm}
  s \ = \ \left \langle \ s \ \right \rangle \ + \ x
  \vspace*{0.3cm} \\
  A \ \left \langle \ s \ \right \rangle
  \ + 
  \ B \ \tr \ \left \langle \ s \ \right \rangle \ {\bf 1}
  \ = 
  \ \mu_{\ s} \ P_{\ 3}
  %\vspace*{0.4cm} \\
  \hspace*{0.3cm} ; \hspace*{0.3cm}
  P_{\ 3} \ =
  \ \left (
  \ \begin{array}{lll}
  0 & 0 & 0 \\
  0 & 0 & 0 \\
  0 & 0 & 1
  \end{array}
  \ \right )
  \vspace*{0.4cm} \\
  \left \langle \ s \ \right \rangle \ =
      \ \displaystyle{
      \frac{\mu_{ s}}{A}
  \ \left (
  \ P_{ 3} \ -
      \frac{B}{A+3B}
  \ {\bf 1}
  \ \right ) }
  %\hspace*{0.3cm} ; \hspace*{0.1cm}
  \end{array}
\end{equation}
\noindent
Thus the quadratic parts with respect to $x$ of $V$ to first order
in the strange quark mass are of the form
\begin{equation}
  \label{eq:11}
  \begin{array}{l}
  V^{\ (2)} \ =
  \ \begin{array}[t]{l}
  \frac{1}{2}
  \ \left ( 
  \ A \ \tr \ x^{\ 2} \ + \ B \ ( \ \tr \ x \ )^{\ 2}
  \ \right ) \ +
  \vspace*{0.3cm} \\
  C \ \tr \ \left \langle \ s \ \right \rangle \ x^{\ 2} \ +
  \ D \ ( \ \tr \ x \ ) \ \tr \ \left \langle \ s \ \right \rangle \ x \ +
  \vspace*{0.3cm} \\
  \frac{1}{2}
  \ D \ ( \ \tr \ \left \langle \ s \ \right \rangle \ ) 
  \ \tr \ x^{\ 2} \ +
  \ E \ ( \ \tr \ \left \langle \ s \ \right \rangle \ )
  \ ( \ \tr \ x \ )^{\ 2}.
  \end{array}
  \end{array}
\end{equation}
\noindent
The first two terms composing $V^{\ (2)}$ in Eq. (\ref{eq:11}) describe
singlet and octet masses (squares) 
$m_{\ (1)}$ and $m_{\ (8)}$ in the u-d-s chiral limit $\chi_{3}$ ,
whereas the remaining terms contain the further mass splittings to
first order in the strange quark mass.
Introducing the quantities
\begin{equation}
  \label{eq:12}
  \begin{array}{rlrl}
  m_{\ (1)}^{\ 2} & = \ A \ + \ 3 \ B
  \hspace*{0.3cm} ; &
  m_{\ (8)}^{\ 2} & = \ A \vspace*{0.2cm} \\
  R & = \ {\displaystyle 
       \frac{ m_{\ (8)}^{\ 2}}
        {m_{\ (1)}^{\ 2}}\ ; \ \qquad }  &
        ( \ c \ , \ d \ , \ e \ ) & = {\displaystyle 
      \ \frac{\mu_{s}}{A}
\   ( \ C \ , \ D \ , \ E \ ) }\\
  \end{array}
\end{equation}
\noindent
the expression for $V^{\ (2)}$ in Eq. (\ref{eq:11}) becomes
\begin{equation}
  \label{eq:13}
  \begin{array}{l}
  V^{\ (2)} \ =
  \ \begin{array}[t]{l}
  \left \lbrack
  \ A \ + \ \frac{2}{3} \ c \ ( \ R \ - \ 1 \ )
  \ + \ d \ R
  \ \right \rbrack
  \ \frac{1}{2} \ \tr \ x^{\ 2} \ + 
  \vspace*{0.3cm} \\
  \left \lbrack
  \ B \ + \ \frac{2}{3} \ d \ ( \ R \ - \ 1 \ ) \ +
  \ 2 \ e \ R
  \ \right \rbrack
  \ \frac{1}{2} \ ( \ \tr \ x \ )^{\ 2} \ +
  \vspace*{0.3cm} \\
  c \ \tr \ P_{\ 3} \ x^{\ 2} \ +
  \ d \ ( \ \tr \ x \ ) \ \tr \ P_{\ 3} \ x. 
  \end{array}
  \end{array}
\end{equation}
\noindent
{\em Remark on the (semi)classical interpretation of} $V^{\ (2)}$\\
\vspace*{0.1cm}
\noindent
We assume here and in the following that the (semi)classical interpretation
of $V^{\ (2)} \ ( \ x \ )$ as a quadratic function of the shifted
field variables $x$ actually describes the real part of the mass (square) term
pertaining to scalar mesons and can be extended to pseudoscalar mesons,
while we neglect the specific $m_{\ s}$ dependence of the kinetic energy term,
which within the same (semi)classical interpretation is 
 in general
simplified to remain unperturbed, i.e. of the form
\begin{equation}
  \label{eq:25}
  \begin{array}{l}
  {\cal{L}}_{\ kin} \ = \ \frac{1}{4} \  f_{\ \pi}^{\ 2}
  \ \tr
  \ \left (
  \ \partial^{\ \varrho} \ \Sigma^{\ \dagger} \ \partial_{\ \varrho} \ \Sigma
  \ \right ).
  \end{array}
\end{equation}
\noindent
In Eq. (\ref{eq:25}) $f_{\ \pi} \ \sim \ 93 \ \mbox{MeV}$ denotes the
pseudoscalar decay constant in the three flavor chiral limit.

The simplified form of the kinetic energy term in Eq. (\ref{eq:25}) can always
be achieved after a nonlinear transformation of the $\Sigma$ variables.
The corresponding chiral (Noether) currents are then only proportional to
the respective quark bilinear currents modulo explicitely $m_{\ s}$
dependent factors as visible in the ratio of physical
pion to kaon decay constants, far away from the flavor symmetric limit 1.

\subsubsection{Mass square patterns for the scalar nonet}

It follows from the structure of the mass terms in Eq. (\ref{eq:13})
that the (nearly perfect) degeneracy of the $f_{0} \ (980)$ and
$a_{0} \ (980)$ isosinglet and isotriplet levels can only be realized 
independently of $m_{\ s}$ if the constant $B$ prevailing in the $\chi_{\ 3}$
limit vanishes. We adopt thus $B \ = \ 0$ in the following, which
implies that the entire scalar nonet is degenerate in mass in the
chiral limit $\chi_{\ 3}$.
Thus the expression for $V^{\ (2)}$ in Eq. (\ref{eq:13}) becomes
\begin{equation}
  \label{eq:14}
  \begin{array}{l}
  V^{\ (2)} \ =
  \ \begin{array}[t]{l}
  \left \lbrack
  \ A \ + \ d 
  \ \right \rbrack
  \ \frac{1}{2} \ \tr \ x^{\ 2} \ +
  %\vspace*{0.3cm} \\
  \ e \ ( \ \tr \ x \ )^{\ 2} \ +
  \vspace*{0.3cm} \\
  c \ \tr \ P_{\ 3} \ x^{\ 2} \ +
  \ d \ ( \ \tr \ x \ ) \ \tr \ P_{\ 3} \ x.
  \end{array}
  \end{array}
\end{equation}
\noindent
The first term on the right hand side of Eq. (\ref{eq:14})
yields a common mass square to the entire nonet. Hence, if we
consider all mass squares relative to $m^{\ 2} \ ( \ a_{0} \ )$
all contributions are contained in the last three terms
composing $V^{\ (2)}$ , which we denote by 
$\Delta \ m^{\ 2} \ = \ m^{\ 2} \ - \ m^{\ 2} \ ( \ a_{0} \ )$
\begin{equation}
  \label{eq:15}
  \begin{array}{l}
  \Delta \ m^{\ 2} \ =
  c \ \tr \ P_{\ 3} \ x^{\ 2} \ +
  \ d \ ( \ \tr \ x \ ) \ \tr \ P_{\ 3} \ x \ +
  \ e \ ( \ \tr \ x \ )^{\ 2}
  %\hspace*{0.3cm} ; \hspace*{0.3cm}
  \vspace*{0.3cm} \\
  \tr \ P_{\ 3} \ x^{\ 2} \ =
  \ \overline{K} \ K
  \ + \ \frac{2}{3} \ S_{\ (8)}^{\ 2} \ - 
  \ 2 \ \frac{\sqrt{2}}{3}
  \ S_{\ (8)} \ S_{\ (1)} 
  \ + \ \frac{1}{3} \ S_{\ (1)}^{\ 2} 
  \vspace*{0.3cm} \\
  ( \ \tr \ x \ ) \ \tr \ P_{\ 3} \ x \ =
  - \ 2 \ \frac{1}{\sqrt{2}} \ S_{\ (8)} \ S_{\ (1)} 
  \ + \ S_{\ (1)}^{\ 2} 
  \vspace*{0.3cm} \\
  ( \ \tr \ x \ )^{\ 2} \ =
  \ 3 \ S_{\ (8)}^{\ 2}.
  \end{array}
\end{equation}
\noindent
In Eq. (\ref{eq:15}) $S_{\ (1)}$ , $S_{\ (8)}$ denote the (hermitian)
singlet, octet component fields within the scalar nonet respectively.
Furthermore
\begin{equation}
  \label{eq:15a}
  \begin{array}{l}
x_{\ 3 3} \ = \ \frac{1}{\sqrt{3}} \ S_{\ (1)} 
\ - \ \frac{2}{\sqrt{6}} \ S_{\ (8)}.
  \end{array}
\end{equation}

\noindent
From the structure of $\Delta \ m^{\ 2}$ in Eq. (\ref{eq:15})
we obtain the mass of the $K \ \overline{K}$ system 
\begin{equation}
  \label{eq:16a}
  \begin{array}{l}
   \Delta \ m^{\ 2} \ ( \ K \ ) \ = \ c
  \end{array}
\end{equation}
\noindent
as well as the
mass and mixing pattern involving the two isosinglets
$S_{\ (1)}$ and $S_{\ (8)}$. We introduce the octet-singlet 
mixing matrix $\Delta \ m^{\ 2}_{\ 8-1} \ \equiv \ \Delta \ M^{\ 2}$,
which generates the quadratic form in $S_{\ (8)} \ , \ S_{\ (1)}$
in Eq. (\ref{eq:15}) 
\begin{equation}
  \label{eq:16}
  \begin{array}{l}
  %\hspace*{0.3cm} ; \hspace*{0.3cm}
  %\vspace*{0.3cm} \\
   \Delta \ M^{\ 2} \ =
  \ \left ( 
   \ \begin{array}{lr}                 
   \frac{4}{3} \ c &
   - \ \sqrt{\ 2 \ } 
   \ \left ( \ \frac{2}{3} \ c \ + \ d \ \right )
  \vspace*{0.3cm} \\
   - \ \sqrt{\ 2 \ } 
   \ \left ( \ \frac{2}{3} \ c \ + \ d \ \right ) &
   \frac{2}{3} \ c \ + \ 2 \ d \ + \ 6 \ e
      \end{array}
   \ \right )
  \end{array}
\end{equation}
\noindent
which we can transform into 
\begin{equation}
  \label{eq:17}
  \begin{array}{l}
  \Delta \ M^{\ 2} \, = \, \frac{4}{3} \ c
  \ \left ( 
   \ \begin{array}{lr}                 
   1 &
   - \ \frac{1}{\sqrt{2}} \ ( \ 1 \ + \ \delta \ )
  \vspace*{0.3cm} \\
   - \ \frac{1}{\sqrt{2}} \ ( \ 1 \ + \ \delta \ ) \ &
   \ \frac{1}{2} \ \left ( 
   \ ( \ 1 \ + \ \delta \ )^{\ 2}
   \ + \ \varepsilon \ - \ \delta^{\ 2}
   \ \right )
      \end{array}
   \ \right )
  \vspace*{0.3cm} \\
  \delta \ = \ % {\displaystyle
    \frac{3}{2} \frac{d}{c} % }%end display
             \ ; \quad
  \varepsilon \ = \ %{\displaystyle 
  \frac{9\ e}{ c} % }%end display
 \ ; \quad
  \dete \ \Delta \ M^{\ 2} \ = \ \frac{8}{9} \ c^{\ 2} 
  \ ( \ \varepsilon \ - \ \delta^{\ 2} \ ).
  \end{array}
\end{equation}

\noindent
The mass square differences of the lighter and heavier
isoscalar $f_{0 \ <}$ , $f_{0 \ >}$ are obtained as eigenvalues
of $\Delta \ M^{\ 2}$.
We note that the observed (approximate) degeneracy 
of $f_{0} \ (980)$ and $a_{0} \ (980)$, i.e.
$\Delta \ m \ ( \ f_{0 \ <} \ ) \ \sim \ 0$,
corresponds through first order in $m_{\ s}$ to the vanishing
determinant of $\Delta \ M^{\ 2}$
\begin{equation}
  \label{eq:18}
  \begin{array}{l}
  \dete \ \Delta \ M^{\ 2} \ = \ 0
  \hspace*{0.3cm} \leftrightarrow \hspace*{0.3cm}
  \varepsilon \ = \ \delta^{\ 2}
  \vspace*{0.3cm} \\
  \Delta \ M^{\ 2} \ = \ \frac{4}{3} \ c
  \ \left ( 
   \ \begin{array}{ll}                 
   1 \ & \ k  
  \vspace*{0.3cm} \\
   k \ & \ k^{\ 2}
  \end{array}
   \ \right )
  \hspace*{0.3cm} ; \hspace*{0.3cm}
  k \ = \ - \ \frac{1}{\sqrt{2}} \ ( \ 1 \ + \ \delta \ ).
  \end{array}
\end{equation}
\noindent
Introducing the mixing angle $\Theta$ by
\begin{equation}
  \label{eq:19a}
  \begin{array}{rl}
  f_{0 \ >} & = 
  \ \cos \ \Theta \ S_{ (8)} \ +
  \ \sin \ \Theta \ S_{ (1)} \\
%  \hspace*{0.1cm} ; \hspace*{0.1cm}
  f_{0 \ <} & = \
   - \ \sin \ \Theta \ S_{ (8)} \ +
  \ \cos \ \Theta \ S_{ (1)}
  \end{array}
\end{equation}
\noindent
we find the mass square and mixing pattern due to $\Delta \ M^{\ 2}$ 
in Eq. (\ref{eq:18}), with $k=\tan \Theta$, to be given by
\begin{equation}
  \label{eq:19}
  \begin{array}{l}
  \Delta \ m^{\ 2} \ ( \ f_{0 \ >} \ ) \ = {\displaystyle 
  \ \frac{4}{3} \ c
       \ \frac{1}{
  \cos^{ 2} \ \Theta  } \ ;} % end displaystyle
  \hspace*{0.3cm} 
  \Delta \ m^{\ 2} \ ( \ f_{0 \ <} \ ) \ = 
  \ 0.
  \end{array}
\end{equation}
\noindent
Note that in the present approximation there is the inequality 
\begin{equation}
\Delta \ m^{\ 2} \ ( \ f_{0 \ >} \ ) \ > \ \ \frac{4}{3}
  \ \Delta \ m^{\ 2} \ ( \ K \ ).
\label{ineqnonet}
\end{equation}
Finally, we consider
two limiting patterns for the mass square of scalar mesons :

\begin{description}
\item I) No or small singlet octet mixing.
%\vspace*{0.1cm}

\noindent {\it a) No mixing}: \\
This assignment corresponds to 
\begin{equation}
  \label{eq:21a}
  \begin{array}{l}
  k \ = \ 0 
  \hspace*{0.3cm} ; \hspace*{0.3cm}
  \delta \ = \ - \ 1
  \hspace*{0.3cm} ; \hspace*{0.3cm}
  d \ = \ - \ \frac{2}{3} \ c.
  \end{array}
\end{equation}
In the following discussion we
use as unit of mass square the $K^*_0 - a_0$ splitting constant $c$
in (\ref{eq:16a})   
%
%\hspace*{3.0cm} $c \ = \ m^{\ 2} \ ( \ K^*_0 \ ) \ - \ m^{\ 2} \ ( \ a_{0} \ )$
%
and denote the common nonet mass in the $\chi_{\ 3}$ limit by
$m_{\ (9)}$.
Relative to $m_{\ (9)}^{\ 2}$ the four degenerate
states $f_{0 \ <} \ , \ a_{0}$ are lower in mass square by
$\frac{2}{3}$ units, the $K^*_0  , \ \overline{K^*_0}$ states are
higher by $\frac{1}{3}$ unit, whereas $f_{0 \ >}$ is raised
by $\frac{2}{3}$ units. To first order in $m_{ s}$
the Gell-Mann-Okubo mass square formula is valid within the octet 
\begin{equation}
  \label{eq:21b}
%        \begin{array}{l}
  3 \ \Delta \ m^{\ 2} \ ( \ f_{0 \ >} \ ) 
  \ = \
  4 \ \Delta \ m^{\ 2} \ ( \ K^*_0 \ ) 
\end{equation}
and yields a prediction for the mass of the heavier isoscalar
\begin{equation}
  m \ ( \ f_{0 \ >} \ ) \ \sim \ 1550 \ \mbox{MeV}.
\label{f0hms}
\end{equation}
This mass pattern is also displayed in 
Fig.  \ref{gbfig2} (Ia) together with the one for the pseudoscalars
for comparison. According to (\ref{ineqnonet}) the mass value (\ref{f0hms})
is the lower limit for $m(f_{0>})$ under the condition $m(a_0)=m(f_{0<})$.

\noindent %\item Ib) 
{\it b) Small mixing as in the pseudoscalar nonet}\\
\noindent
This mixing pattern is suggested by our phenomenological
analysis 
in the following sections and corresponds to
\begin{equation}
  \label{eq:21}
  \begin{array}{l}
  k \ = \ \frac{1}{\sqrt{8}}
  \hspace*{0.3cm} ; \hspace*{0.3cm}
  \Theta \ = \ \arcsin \ \frac{1}{3}
  \ \sim \ 19.5\dgr
  \vspace*{0.3cm} \\
  \Delta \ m^{\ 2} \ ( \ f_{0 \ >} \ ) \ =
  \ \frac{3}{2} \ \Delta \ m^{\ 2} \ ( \ K^*_0 \ )
  \hspace*{0.3cm} \rightarrow \hspace*{0.3cm}
  m \ ( \ f_{0 \ >} \ ) \ \sim \ 1600 \ \mbox{MeV}.
  \end{array}
\end{equation}
\noindent
Relative to $m_{\ (9)}^{\ 2}$ the four degenerate
states $f_{0 \ <} \ , \ a_{0}$ are now lower in mass square by
one unit, the $K^*_0  , \ \overline{K^*_0}$ states are
at the same level, whereas $f_{0 \ >}$ is raised
by $\frac{1}{2}$ units (see Fig. \ref{gbfig2}).

\item II) Strict validity of the OZI rule.

Flavor mixing according to the OZI-rule corresponds to
$\delta \ = \ 0$ and thus to
\begin{equation}
  \label{eq:20}
  \begin{array}{l}
  k \ = \ - \ \frac{1}{\sqrt{2}}
  \hspace*{0.3cm} ; \hspace*{0.3cm}
  \Theta \ = \ - \ \arcsin \ \frac{1}{\sqrt{3}}
  \ \sim \ - \ 35.3\dgr
  \vspace*{0.3cm} \\
  \Delta \ m^{\ 2} \ ( \ f_{0 \ >} \ ) \ =
  \ 2 \ \Delta \ m^{\ 2} \ ( \ K^*_0 \ )
  \hspace*{0.3cm} \rightarrow \hspace*{0.3cm}
  m \ ( \ f_{0 \ >} \ ) \ \sim \ 1770 \ \mbox{MeV}.
  \end{array}
\end{equation}
In this case
the four degenerate
states $f_{0 \ <} \ , \ a_{0}$ remain at the same level 
as $m_{\ (9)}^{\ 2}$, 
the $K^*_0 \ , \ \overline{K^*_0}$ states are
higher by one unit, whereas $f_{0 \ >}$ is raised
by two units (see Fig. \ref{gbfig2}).
\end{description}

\noindent
We conclude that the degeneracy in mass of $f_0  (980)$ and $\ a_0  (980)$
indeed implies a full degenerate nonet in the $\chi_{\ 3}$ chiral limit.
It is important to note, however,  that
the contributions of order $m_{ s}$ can respect the
$f_0-a_0$ mass degeneracy, without splitting necessarily the nonet
according to the OZI-rule, i.e. according to flavor, as often assumed.
Furthermore we point out,
that an eventual similarity
of singlet octet mixing  for scalars {\em and} pseudoscalars as outlined in
Ib) is by no means excluded. 
Approximate singlet-octet 
%or $\eta$ $\eta '$ 
mixing is known to prevail for the latter -- with
a mixing angle near 19.5$\dgr$ as in (\ref{eq:21})  \cite{Bij}.

Only case I) is compatible with our phenomenological analysis         
in the subsequent sections and we assign
\begin{equation}
f_{0 \ >} \ \rightarrow \ f_0(1500).
\label{f0gt}
\end{equation}
The observed mass is slightly lower than the masses theoretically calculated
in the lowest order of the strange quark mass $m_s$.
We emphasize here, that the splitting between
$a_{0}$ and $K^*_0$ is considerable and thus there will be non-negligible
corrections of higher order in $m_{ s}$,
in particular to the $K^*_0, \ \overline{K^*_0}$ square masses.
These corrections can easily account for the violation
of the inequality  (\ref{ineqnonet}) and the larger $f_{0 \ >}$ masses
predicted.

The alternative choice II would treat $f_0(980)$ as purely nonstrange state
which is not attractive phenomenologically as will be discussed below.
The pure $s\overline s$ state  $f_{0 \ >}$ with mass as in (\ref{eq:20}) 
could then be associated with the $J=0$ component of $f_J(1710)$ 
(see, Sect. \ref{basic_triplet}) but not much is known about the flavor
properties of this state. In any case, the mass ordering of the three spin
triplet states 
\begin{equation}
  \label{eq:37}
  \begin{array}{l}
  f_{\ J \rightarrow 0} \ (1710) %  \ \leftrightarrow
                                    % \ f_{\ 0} \ (1500)
  \hspace*{0.3cm} ; \hspace*{0.3cm}
  f_{\ 1} \ (1510)
  \hspace*{0.3cm} ; \hspace*{0.3cm}
  f_{\ 2}^{\ '} \ (1525)
  \end{array}
\end{equation}
\noindent
%their spectroscopic properties then become quite different. In particular,
%the natural ordering in mass 
would be upset by $\sim \ 200 \ \mbox{MeV}$ within this scheme.

%We now turn to the more detailed phenomenological discussion of the
%spectroscopic results.
%
%--------------------------------------------------------------
%
%\input{gb4v3.tex}
% section spectroscopy   gb4v3.tex
%date 8.5.98 -> 30.10.98
%
\section{Spectroscopy of light isoscalar $J^{PC}=0^{++}$ states}
\label{spectr}

Next we turn to the more detailed phenomenological discussion,
first concerning the  lowest mass $q\overline q$ nonet
and the  lightest glueball.
Much effort has been devoted 
to clarify the experimental situation. To this end a variety of
reactions has been studied in considerable detail
\begin{equation}
\begin{array}{l}
\label{reactions}
\begin{array}{rl}
1.&  \pi^+\pi^- \to \pi^+\pi^- ,\  \pi^0\pi^0 \\
2.&  \pi^+\pi^- \to K^+K^- ,\ K^0 \overline{K^0} \\
3.&  \pi^+\pi^- \to \eta \eta,\ \eta \eta'  \\
4.&  p\overline p \to 3 \pi^0,\ 5 \pi^0 ,\ \pi^0 \pi^0 \eta,\ \eta \eta \pi^0,
      \eta \eta'\pi^0 \\
5.&   J/\psi \to \phi\pi\pi,\ \phi K \overline K,\ \omega\pi\pi,\ \omega K \overline K \\
6.&  J/\psi \to \gamma \pi \pi,\ \gamma K \overline K,\ \gamma \eta \eta,\ 
    \gamma \eta \eta'\\
7.&  pp \to pp\ + \ X_{{\rm central}} \\
8.&   \psi' \to J/\psi \pi\pi,\ Y' \to Y\pi\pi, \ Y''  \to Y\pi\pi \\
9.&   \gamma \gamma \to \pi \pi, \ K \overline K \\
\end{array}
\end{array}
\end{equation}
Our knowledge about the
first three reactions 
comes from the analysis of  peripheral $\pi N$ collisions
in application of
 the one-pion-exchange  model; these reactions
represent the oldest source of information on the scalar resonances.
The fourth one, $p\overline p$ annihilation at threshold, 
has been studied in recent years with high statistics at the LEAR facility
at CERN and has improved our understanding of the spectroscopy above 1 GeV
in particular; data from
 higher primary energies have been obtained at FERMILAB.
The states recoiling against the $\phi$ and the $\omega$ in reaction 5
should have a large
strange or nonstrange $q\overline q$ component respectively.
The reactions 6,7  and 8
are expected to provide a gluon rich environment 
favorable for glueball production (for a review, see Ref. \cite{Close}),
whereas in the last one (9) the glueball production 
is suppressed if the mixing with $q\overline q$ states is small. 

In the search for resonances one usually looks first for peaks in the
mass spectrum. If several states are overlapping, or in the presence of
coherent ``background'', 
the peak position may be shifted or the resonance may 
even  appear as a dip in the mass spectrum. The crucial
characteristics of a   resonance is therefore the energy
dependence of the
corresponding complex partial wave amplitude which moves along  a full 
loop inside the ``Argand diagram'': besides the mass peak the phase
variation has to be demonstrated.   

Such results are obtained from energy independent
phase shift analyses which try to determine  
the individual partial waves for a sequence of energy values. Usually
such  analyses are plagued by
ambiguities. To start with, one can obtain a description of the scattering
data in an energy dependent fit from an ansatz 
with a superposition of resonances. Such global fits to the mass
spectra of mesonic systems in a broad
range up to about 1700 MeV and including an increasing number of different
reactions in (\ref{reactions}) 
have been performed by several groups,
starting with the CERN-Munich Collaboration \cite{CERN-Munich1,CERN-Munich2},
 then by Au, Morgan and Pennington (AMP) \cite{amp,mp}, 
Lindenbaum und Longacre (LL) \cite{linden}
 and more recently by Bugg, Sarantsev and Zou (BSZ) \cite{bsz} and by Anisovich,
Prokoshkin and Sarantsev (APS)
\cite{aps}. 

A survey of results from these representative global fits 
are given in  Table \ref{tabres}.
All these fits include the narrow $f_0(980)$, probably the only
uncontroversial and well located $f_0$ state.  
Furthermore, they all show
one rather  broad state of more than  500 MeV width, called now 
$f_0(400-1200)$ by the PDG \cite{PDG}; this state is considered as resonance
$f_0(1000)$ in \cite{mp}, otherwise it is just refered to as ``background''.
In addition, states of higher mass are required by the fits but with masses
which fluctuate from one fit to another. 
The PDG in their recent summary table includes the
$f_0(1370)$ and the $f_0(1500)$ which also represents
 the states quoted earlier, 
the $f_0(1300)$ and $f_0(1590)$.

\begin{table}[t]
%  \begin{array}{l}
  \begin{tabular}{cccc}
 \hline
% & & & \\
authors & broad state & other states & 
reactions   \\ \hline 
%   & & & \\
%
CM\ \cite{CERN-Munich1} &  \ 1049\ -\ $i$\ 250\ MeV  &
$f_0(980),\ f_0(1537) $
& 1a\\
 %  & & & \\
%
AMP\ \cite{amp} & \ 910\ -\ $i$\ 350  \ MeV &$\ f_0(988),\ [f_0(991)],
\ f_0(1430)$ & 1a, \ 2, \ 5 \\
  & & & 7a,b\ 8,\ 9 \\
  % & & & \\
%
LL\ \cite{linden} & \  1300\ -\ $i$\ 400 \   
&$f_0(980),\ f_0(1400), \ f_0(1720)$ & 1,\ 2,\ 3,\ 6 \\
   & & & \\
BSZ\ \cite{bsz} & \ A:~571\ -\ $i$\ 420 \ MeV \  &$f_0(980),\ f_0(1370), \
f_0(1500)$ & 1,\ 2,\ 4\\
               & \ B:\ 1270\ -\ $i$\ 530 \ MeV  & & \\ 
  % & & & \\
%
APS\ \cite{aps} &  \ 1530\ -\ $i$\ 560 \ MeV  
& $f_0(980),\ f_0(1370),\ f_0(1500), $& 1,\ 2,\ 3,\ 4\\  
& & $f_0(1780)$ & \\ \hline
\end{tabular}
%  \end{array}
\caption{Isoscalar states included in various global 
energy dependent fits to
reactions (\protect\ref{reactions}) with channels a-d. 
For the broad state the poles of the scattering
amplitude at $m -  i~\Gamma/2$ is given. Only one of the two states near
the $f_0(980)$ found by AMP are kept in \protect\cite{mp}.}
\label{tabres}
\end{table}

In  Fig. \ref{gbfig3} we show some recent results on
the mass dependence
%In our approach the broad state is to be associated with the lowest $0^+$
%glueball.
%\begin{equation}
%f_0(400-1200) \quad\to \quad gb(1000) \label{gluedef}
%\end{equation} 
of the  isoscalar (I=0)  S-wave  $\pi\pi$ cross section as obtained by
BNL-E852 \cite{gunter} and GAMS Collaborations \cite{aps}.
This mass spectrum with three peaks (the ``red dragon'') 
will be interpreted by us as a very broad state centered around
1 GeV (glueball) which interferes with
the resonances $f_0(980)$ and $f_0(1500)$ whereby the dips 
near the respective resonance positions are generated.

In the following we will reexamine the evidence for resonances claimed 
in the different mass intervals, especially in the peak regions in Fig.
\ref{gbfig3} by studying the   
phase shift analyses in different
processes and in particular the phase variation near the 
respective resonance masses.  

%**************************************************************

%figure 2

%****************************************************************

%\clearpage
%\thispagestyle{empty}
%\hoffset 1.5cm
%\voffset 1.0cm
%\begin{figure}[htb]
%\begin{center}
%\vskip 1mm
%\hskip -20mm
%\leavevmode
%%\epsfxsize=60mm
%%\epsfbox[40 50 530 590]{fig1.eps}
%
%\resizebox{!}{14.5cm}{%
%\includegraphics{prokoko.eps}
%}
%
%\resizebox{!}{8.9cm}{%
%\epsfbox{higgsf1.eps}}

%\epsfbox{higgsf1.eps} 

% where you want to insert a vbox for a figure

%\vskip -90mm
%\voffset -1.0cm
%\end{center}
%{\center{
% \hspace*{0.2cm}
%\begin{minipage}{10.5cm}
%\center{\caption{\hspace*{-0.3cm} \mbox{} 
% \label{fig:swaves}
%}} 
%\begin{flushleft}
%{\bf
%\caption{ 
%The $I \ = \ 0$ S wave $\pi \pi \ \rightarrow \ \pi \pi$ cross section.
%}
%}
%\end{flushleft}
%\end{minipage} }}

%\label{fig:ZZ}
%\end{figure}
%\clearpage
%\hoffset 0,0cm
%\voffset 0.0cm
%****************************************************************

%Text after figure 2

%****************************************************************
   
\subsection{The  low energy $\pi\pi$ interaction 
($m_{\pi\pi}\lesssim 1000$ MeV) and the
claim for a narrow  $\sigma(770)$ resonance}
\label{spectr1}

At low energies only the $\pi\pi$ channel is open.
According to the common view 
which emerged  in the mid of the 70's the isoscalar S-wave 
has negligable coupling to inelastic channels below the $K\overline K$
threshold and the phase shift $\delta_{\ell}^{I}$ with $\ell=0,\ I=0$ 
 moves much more
slowly  through the $\rho$ meson region than the P-wave.
This strong $\pi\pi$ interaction is interpreted either as 
``background'' or as a very broad state as discussed above. Only 
 near the $K\overline K$ threshold the phase varies rapidly 
because of the presence of the 
$f_0(980)$ resonance. There is an old claim for the existence of
a narrow resonance $\sigma(770)$ under the $\rho$ meson which has been put
forward again more recently arguing with results from polarized target.

Most results on the $\pi\pi$ S-wave obtained more than 20 years ago have been 
derived from the reactions
\begin{equation}
\mbox{(a)} \quad \pi N \to \pi\pi N \qquad  
\mbox{(b)} \quad \pi N \to \pi\pi \Delta
\label{pinucleon}
\end{equation}
in the $\pi^+\pi^-$ charge mode 
with unpolarized target in application of variants of the 
 one-pion exchange (OPE) model (for a general review, see \cite{mms}, for the low
energy $\pi\pi$ interactions, see \cite{ochs}, for example).
In this charge mode there is a twofold ambiguity
(``up-down'') for each energy interval
 which corresponds to either a narrow or a very broad resonance
under the $\rho$ meson. 
From the study of the $\pi^+\pi^-$  \cite{flatte,CERN-Munich1,CERN-Munich2}
and  $\pi^0\pi^0$ data \cite{apel} the narrow resonance solution has
finally been excluded \cite{flatte,em73,CERN-Munich1,CERN-Munich2,mp}.
 
The measurement of reaction (\ref{pinucleon}a) with polarized target
by the CERN-Cracow-Munich Collaboration \cite{CKM1,CKM2} has made possible a
more detailed investigation of the production mechanisms but 
the analysis also leads to
a new class of ambiguities in phase shift analysis.

In a recent reanalysis of these data Svec \cite{svec1} finds 
in the modulus of one of the transversity amplitudes  a  narrow peak
near 750 MeV, whereas in case of the other one a broad mass spectrum
is observed. In Breit-Wigner fits
to these mass spectra an extra state $\sigma(750)$ of width $\Gamma=150$
MeV -- or in the preferred fits even two $\sigma$ states --  are included 
 besides the $f_0(980)$ resonance.
In these considerations no attempt has been made to  
fit the amplitude phases nor to respect the partial wave unitarity which is
important in particular near the inelastic threshold.

These constraints are taken into account in the subsequent 
analysis of the polarized target data 
by Kami$\acute{\rm n}$ski, Le$\acute{\rm s}$niac and Rybicki 
(KLR) \cite{klr}. In the
region below 1000 MeV they found four different solutions
duplicating the old up-down ambiguity:
\begin{equation}
 \mbox{(a) up-steep \qquad   (b) up-flat \qquad 
  c) down-steep \qquad  (d) down-flat.}
\label{solution}
\end{equation}
Furthermore, a separation into pseudoscalar ($\pi$) and pseudovector ($a_1$)
exchange amplitudes has been carried out. 

The solution (\ref{solution}c) is excluded immediately as it leads to a strong
violation of partial wave unitarity. 
The solution (\ref{solution}d) is essentially consistent
with the previous result from the unpolarised target 
data \cite{CERN-Munich1} up to $m_{\pi\pi} \sim 1400$ MeV
and the phase shift deviates by not more than $30\dgr$ in the
region above this mass.

The solution (\ref{solution}a) which is consistent with a narrow
$\sigma(750)$ as in \cite{svec1} also shows a systematic violation of unitarity 
and is therefore
considered as ``somewhat queer'' by KLR but not excluded. However, both up
solutions suffer from similar problems already 
discussed in connection with the old analyses:

%\begin{description}
\noindent
{\it (a) Comparison with the $\pi^0\pi^0$ final state}\\
In this case the P-wave is forbidden and therefore the up-down
ambiguity does not show up.
The recent very precise 
data on the reactions 4 
in (\ref{reactions}) by the Crystal Barrel Collaboration
\cite{Ams,abele} can be interpreted in terms of $\pi\pi$ amplitudes using an 
isobar model for the annihilation process. The  striking effects from 
the $f_0(980)$ state are clearly
visible but the mass spectra  around 750 MeV are
rather structureless and there is no sign of a narrow resonance.
In particular, the existence of a resonance with width around 250 MeV
has been excluded in \cite{abele}.

\noindent
{\it (b) Rapid variation of phase near $K\overline K$ threshold}\\
The GAMS-Collaboration \cite{GAMS} 
has presented results on the S-wave magnitude of 
reaction 1b in (\ref{reactions}) obtained from process (\ref{pinucleon}a).  
Their results show a sudden decrease 
of the S-wave magnitude above a mass of 850
MeV with a narrow dip at 970 MeV. A dip of comparable type is also obtained
for the KLR down-solution (Fig. 2a of \cite{klr}); the position of the dip
is slightly moved upwards, presumably because of different isospin 
I=2 contributions.
On the other hand, the up-solution reaches the minimum cross section already
at  the lower mass around 900 MeV in qualitative difference to the GAMS data.
The GAMS collaboration 
so far has not yet published the original experimental results in terms of
spherical harmonic moments. Once available the 4 different solutions from the
polarized target experiment could be compared directly to the moments from
the $\pi^0\pi^0$ final state which should determine the unique
solution. The GAMS data are consistent again with fits 
which properly take into account
unitarity at the threshold such as in Refs. \cite{bsz,aps}.
A similar behaviour in the mass region below $\sim 1000$ MeV is shown by the
BNL-E852 data \cite{gunter} (see also Fig.~\ref{gbfig3}). 
%\end{description}
%
These arguments favor the down-flat-solution which agrees with 
the results obtained previously. All other choices would lead to serious
inconsistencies with general principles or with other experimental results.
Nevertheless, it would be desirable to obtain the complete 
results and a common description 
of the reactions (\ref{pinucleon}) in the $\pi^+\pi^-$ and
$\pi^0\pi^0$ charge modes.

\subsection{How reliable are $\pi\pi$ scattering results from unpolarised
target experiments?}
\label{OPE}

It may be surprising at first sight  that the results from
polarized and unpolarized target are so similar, as found by KLR \cite{klr}.
In fact, it is occasionally claimed (especially in \cite{svec1}) 
that the analyses from unpolarized target experiments are obsolete
because of the importance of $a_1$-exchange besides $\pi$-exchange.
Most results obtained so far on elastic and inelastic $\pi\pi $ interactions
1-3 in (\ref{reactions}), which are important in the subsequent discussion,
 have been obtained from unpolarized target experiments.
Therefore, it is appropriate
at this point to contemplate the consequences to be drawn from the
experiments with polarized target.

Motivated by 
the OPE model with absorptive corrections,
the commonly applied procedure to extract the production amplitudes
from the unpolarized target experiment, has been based on the following two
assumptions \cite{ochs2,fm,em0} concerning 
the nucleon helicity flip and non-flip 
amplitudes 
$ f^{\pm}_{\ell,\mu}$ and $ n^{\pm}_{\ell,\mu}$ with natural ($+$) and
unnatural ($-$) parity exchange for production of a mesonic system with
spin $\ell$ and helicity $\mu$:
\begin{description}
\item (i) Spin-flip dominance: 
the non-flip amplitudes $n^{\pm}_{\ell,\mu}$ vanish, at least
the $(-)$ amplitudes, which are {\em not} generated by absorbed OPE at high
energies (this allows for $a_2(2^{++})$-exchange but not for
$a_1(1^{++})$-exchange).
\item (ii) Phase coherence: The phases of the production amplitudes at fixed
mass $m_{\pi\pi}$ and momentum transfer $t$ between the
incoming and outgoing  nucleons 
depend only on $\pi\pi$
 spin $\ell$ and not on the helicities. 
\end{description}
A further simplification can be obtained if $t$-integrated moments are used
in the ``$t$-channel frame'' \cite{owag}.
These assumptions yield an overdetermined system of equations. It can be
solved for the amplitudes up to some discrete ambiguities 
whereby the constraints are found to be well satisfied \cite{CERN-Munich1};
results from  ``Chew-Low extrapolation'' in $t$  and from 
$t$ averaged moments yield comparable
results \cite{em1}.

The polarized target experiment has clearly demonstrated the existence of
 the $a_1$ exchange process \cite{CKM1,CKM2} which contributes to the
non-flip amplitudes invalidating assumption (i).
However, for the amplitude analyses  carried out in an unpolarized target 
experiment -- such as in \cite{CERN-Munich1,em1} -- 
weaker assumptions than the ones above 
are sufficient  to obtain the same results
\cite{ochs2}, namely
\begin{description}
\item (i$'$) 
nucleon helicity flip and non flip amplitudes $f$ and $n$ are proportional 
 \begin{equation}
n^{+}_{\ell,\mu}=\alpha^{(+)} f^{+}_{\ell,\mu}, \qquad
n^{-}_{\ell,\mu}=\alpha^{(-)} f^{-}_{\ell,\mu}
 \label{flip}
\end{equation}
for natural ($+$) and unnatural ($-$)
exchange separately for any dipion spin
and helicity $\ell,\mu$.
\item (ii$'$) as in (ii) but there may be an overall phase difference between
$(+)$ and $(-)$ amplitudes.
\end{description}
Then also the transversity up and down amplitudes $g$ and $h$ 
which are determined in the polarized target experiment
\begin{equation}
g^{\pm}_{\ell,\mu}\ =\ %\ \sqrt{\frac{1}{2}} (n^{\pm}_{\ell,\mu} \mp
                       \frac{1}{\sqrt{2}} (n^{\pm}_{\ell,\mu} \mp
                     f^{\pm}_{\ell,\mu}), \qquad
h^{\pm}_{\ell,\mu}\ =\ %\sqrt{\frac{1}{2}} (n^{\pm}_{\ell,\mu} \pm
                       \frac{1}{\sqrt{2}} (n^{\pm}_{\ell,\mu} \pm
                     f^{\pm}_{\ell,\mu}) \label{transvers}
\end{equation}
should be proportional
\begin{equation}
\frac{|g^{\pm}_{\ell,\mu}|}{|h^{\pm}_{\ell,\mu}|} = 
\frac{|\alpha+i|}{|\alpha-i|}.  \label{gdh}
\end{equation}
This relation is approximately fulfilled and the ratio
is found $\frac{|g|}{|h|} \approx 0.6$ for
S,P,D and F waves over the full mass range explored in the small $|t|<0.15$
GeV region (Fig. 6 in \cite{CKM2});
however, the fluctuation of the data is quite large and local
deviations cannot be excluded. 
In a restricted analysis using only S and P
waves below 900 MeV some trend of this amplitude ratio with mass was found
\cite{CKM1} but the D-waves can certainly not be neglected here.

It is pointed out by KLR \cite{klr} that in the narrow 
regions where the S-wave
magnitude is small, i.e. around 1000 MeV and 1500 MeV the $a_1$
contribution may become as large as the $\pi$ exchange contribution, whereas
otherwise it amounts to only about 20\% of it. 

With one pion exchange only (modified by absorption)
the ratio (\ref{gdh}) 
would have to be unity, so the modification (\ref{flip})
amounts only to a change of the overall adjustment of
normalization in the energy dependent fits. 

It is very satisfactory that the down-solution for the S-wave 
which we preferred above is also consistent 
with the energy independence of the amplitude ratio $\frac{|g|}{|h|} \approx
0.6$ (Fig. 2b in \cite{klr}) whereas the disfavored up solution
with the narrow $\sigma$ would lead to an increase of this ratio by 
up to a factor
of 2 just in the mass interval of ambiguity 800-1000 MeV. Such exceptional
behaviour of amplitudes is not plausible. 

As to the simplifying assumption (ii) on the phase coherence of amplitudes
the data from polarized target are confirming it in their general trend
 but there are overall shifts of amplitude phases 
of up to about
20$\dgr$, only some  relative phases involving the D-wave amplitudes
 indicate larger differences 
(Figs. 8,9,10 in (\cite{CKM2})).   

In summary, the original assumption (i) has been demonstrated by the
polarized target experiment to be clearly violated; the modified assumption
(i$'$) is still approximately correct within the given accuracy, whereas 
some moderate violations of phase coherence (ii$'$) have been seen.
This explains why the phase shift results from the polarized target
experiment -- looking at the preferred solution -- are not very different
from the previous findings, in particular, there is no evidence for entirely
new states, such as a $\sigma(750)$.

The proportionality (\ref{flip}) is expected, in particular, if the amplitudes
$a_1\pi \to \pi\pi$ and $\pi\pi \to \pi\pi$ are proportional and
appear as factors in the
production amplitudes. In general, such a relation may be violated as
different resonant states could have different couplings to the $\pi\pi$ and
$a_1\pi$ channels, also there could be different signs. However, as long as 
the $a_1$ exchange is small, such as  in the small $t$ region, 
the violation of the assumption can play only a role at this reduced level.
On the other hand, one has to be careful in applications of the above
assumptions in kinematic regions where OPE is not dominant
(for example, large $t$).

\subsection{Interference of the $f_0(980)$ with 
the ``background'' and the molecule hypothesis}
\label{spectr2}
 
In elastic $\pi\pi$ scattering the narrow  $f_0(980)$
interferes with the large ``background'', now also called $f_0(400-1200)$ 
and appears as a dip in the S-wave
cross section \cite{flatte,CERN-Munich1}. 
There are other processes where to the contrary the $f_0(980)$ 
appears as a peak. This phenomenon has been observed first in
pion pair production with large momentum transfer $|t|\gtrsim 0.3$ GeV
\cite{binnie} and more recently by GAMS \cite{GAMS}. Fits to the peak yield
values for the total width of $\Gamma = 48\pm 10$ MeV. 
A direct clear
evidence for the phase variation according to
 a Breit-Wigner resonance can be inferred from the
interference pattern of the rapidly varying resonance amplitude with the
tail of the $f_2(1270)$ in reaction (3a) at large $t$ as measured by the GAMS
Collaboration \cite{GAMS}. 

The interference of this  narrow resonance with the background 
varies from one reaction to the other. In this way one can see that this
``background'' has its own identity. 
The  reactions in (\ref{reactions}) with a $\pi\pi$ system in the final
state
can be classified roughly into 3 groups according to the 
different appearence of the $f_0(980)$ in the mass spectrum:
\begin{description}
\item (a) dip in reaction 1, indication 
of dip in reaction 4a \cite{Ams,Armstrong1};
\item (b) peak in reaction 1 in large t production, and in 
5a \cite{Gidal,falvard,Lockman}, 5c \cite{Augustin5pi}, 9a
\cite{Cball,Markii,JADE};
\item (c) an interference of the $f_0(980)$ Breit-Wigner amplitude with a
background amplitude of positive real part is suggested 
in 4b \cite{Amseta} and in a similar way in 7 \cite{gamspp,akesson}.
\end{description}
The different interference patterns are naturally attributed
to the different couplings of the $f_0(980)$ and 
of the ``background'' to the initial
channel. 

The dip is observed in the elastic $\pi\pi$ channel. In this case the
background amplitude is near the unitarity limit and the additional
resonance has to interfere destructively. The reaction 4a shows a small dip
around 950 MeV followed by a peak near 1000 MeV and fits into group (a) or
(b). All other processes are inelastic. 

In particular, the transmutation of a
dip into a peak in $\pi N\to \pi\pi N$ with increasing momentum transfer
can be explained by the assumption of an  increasing importance of  
$a_1$ exchange
over $\pi$ exchange with
\begin{equation}
|A(\pi a_1 \to f_0(980) \to  \pi\pi)|\quad \gg \quad
       |A(\pi a_1 \to f_0(400-1200) \to  \pi\pi)|.
     \label{a1exchange}            
\end{equation}                     
In this case the peak occurs, either because the background interferes
constructively, or because it becomes too small. 
 
There is some 
support for this interpretation from the KLR results  \cite{klr}
discussed above on the
polarized target data at small $|t|<0.2$ GeV.
For the favored
``down-flat'' S-wave solution the modulus of the $a_1$-exchange amplitude
 shows a peak (significance about 2$\sigma$) just in the 
mass interval 980-1000 MeV  whereas the pion exchange amplitude shows a dip
in the region 980-1060 MeV (see Fig. 7a in \cite{klr}).

Similar conclusions concerning different exchange mechanism has been drawn in
the recent paper by Achasov and Shestakov \cite{achsh} where detailed fits
including $a_1$-exchange are presented. 

A remarkable similarity is seen in the interference pattern of the two
reactions in group (c) where a small peak near 950 MeV is followed by a
large drop near 1000 MeV. 
In reaction 4b the initial $p\overline p$ state must be in a $\eta,\eta'$ type
state, so the $\pi\pi$ state couples to two isoscalars,
similarly in reaction 7 if the initial state is formed by two
isoscalar pomerons. This is in marked difference to the pattern seen in
$p\overline p \to 3\pi^0$ where four isovectors couple together. This shows that
the $f_0(980)$ and the ``background'' must have different flavor
composition although they have the same quantum numbers. 

Finally, we comment on the hypothesis \cite{molecule}, 
$f_0(980)$ could correspond to a $K\overline K$ molecule (or other 4q
system), which is adopted in
various contemporary classification schemes (see Sect. 2).
S-matrix parametrizations have been used to argue both ways, against \cite{mp}
or in favor \cite{locher} of such a hypothesis.
If the $a_0(980)$ and $f_0(980)$ are such bound states one 
has to worry that the successful quark model 
spectroscopy is not  
overwhelmed by a large variety of additional hadronic
bound states.
On the phenomenological side, 
if the $f_0(980)$ is a loosely bound system, then, 
in a violent collision with large
momentum transfer one would expect an increased probability for a break-up.
The GAMS data, however, demonstrate the opposite, the persistence of
$f_0(980)$ with respect to  the background.  Furthermore, a recent
investigation by the OPAL Collaboration \cite{OPALf0} has shown the production
properties of the $f_0(980)$ to be very similar to those of the $q\overline q$
states  $f_2(1270)$ and $\phi(1020)$ nearby in mass 
in a variety of measurements.
Therefore, we do not feel motivated to give up $f_0(980)$ 
as  genuine $q\overline q$ state but we suggest  
a flavor composition different from the one of the ``background''.

\subsection{The mass region between 1000 and 1600 MeV}
\label{spectr3}

This includes the mass range from the  $f_0(980)$ up to the  $f_0(1500)$.
Near both resonance positions there are 
dips in the elastic $\pi\pi$ S-wave cross section (see Fig. \ref{gbfig3}).
For this region the PDG lists -- besides the $f_0(400-1200)$ -- 
the $f_0(1370)$ state; there may actually be two states, one 
seen as a large effect in the elastic $\pi\pi$ scattering, 
the other one being strongly inelastic.
We will reconsider now the evidence from phase shift analyses for
states in this mass interval.

\subsubsection{Elastic and charge exchange $\pi\pi$ interaction}
Phase shift analyses have been performed using the CM data from unpolarized
target \cite{CERN-Munich1,CERN-Munich3,em1} and by the Omega spectrometer
group \cite{omega}. One finds here a number of ambiguous solutions which are 
discussed in terms of Barrelet zeros \cite{barrelet}. Namely, for a finite
number of partial waves the amplitude can be written as a polynomial in
$z=\cos \vartheta$. Then the measurement of the 
cross section differential in the scattering
angle $\vartheta$ determines the 
real parts and the moduli of the imaginary parts of
the amplitude zeros $z_i$. The different solutions can then be classified
according to the signs of $\im z_i$.

In \cite{CERN-Munich3} four solutions for elasticities $\eta_\ell^I$ and
phase shifts $\delta_\ell^I$
are presented distinguishing the signs
of $\im\ z_i$ at 1500 MeV for $i=1,2,3$ 
as  ($---$), ($-+-$), ($+--$) and ($++-$)
and assuming a sign change of $\im z_1$ at 1100 MeV. They
correspond to the solutions A, C, $\overline {\rm B}$ and  $\overline {\rm D}$ in 
\cite{em1}. Yet more solutions are given in \cite{em1,omega} corresponding
to different branches near 1100, 1500 and above 1800 MeV. The comparison
with $\pi^0\pi^0$ data \cite{shimada,omega} left the solutions C and D
as unfavored.
Solution A is also consistent with the energy dependent result of CM
\cite{CERN-Munich1} up to 1500 MeV where $\delta_0^0\approx 156\dgr$, whereas
solution $\overline {\rm B}$  reaches $\delta_0^0\approx 165\dgr$.
Some descendents $\alpha,\beta$ and $\beta'$ are obtained 
from solutions A and B if constraints from dispersion relations are taken
into account \cite{martinpenn}. 

The data from polarized target \cite{CKM2,klr}  essentially lead to a unique
solution in this mass range as the imaginary parts of zeros came out 
rather small
\begin{eqnarray}
\im\ z_1 \ \sim \  0 \quad & \textrm{for}& 
       \quad m_{\pi\pi}\ >\ 1100\ \textrm{MeV} \nonumber\\ 
\im\ z_2\ \sim \ 0 \quad & \textrm{for}& 
       \quad m_{\pi\pi}\ > \ 1400\ \textrm{MeV}  
\label{zeros}
\end{eqnarray}
so that the various solutions are not significantly different any more.
The results for the phase shifts $\delta_0^0$ in \cite{klr} are again 
similar to solution A in \cite{em1} or to the 
energy dependent phase shift solutions  in CM
\cite{CERN-Munich1} up to $m_{\pi\pi}\sim 1400$ MeV; some additional
variation is indicated above this energy in both $\delta_0^0$ 
and $\eta_0^0$. 

Furthermore we note two aspects of the polarized target results
\begin{description}
\item (a) The S-wave is near the unitarity limit in 
$1150\lesssim m_{\pi\pi} \lesssim 1450$ MeV and  drops to
zero at $m_{\pi\pi} \sim 1500$ MeV (Fig. 2 in \cite{CKM2}).
\item (b) The phase difference of S and D wave amplitudes changes sign 
in both $g$ and $h$ transversity amplitudes with
\begin{eqnarray}
\varphi_S-\varphi_D\ >\ 0 \quad &\textrm{for}&\quad 
           m_{\pi\pi}\ <\ 1250\ \textrm{MeV},
   \nonumber\\
\varphi_S-\varphi_D\ <\ 0 \quad &\textrm{for}&\quad 
           m_{\pi\pi}\ >\ 1350\ \textrm{MeV}.
\label{phchange}
\end{eqnarray}     
\end{description}
The phase differences (Fig. 9,10 in  \cite{CKM2}) 
are best met by the previous solution $\beta'$, the result
(b) excludes solution B. The drop of intensity (a) is not so well reproduced
by the previous phase shift analyses of unpolarized target experiments.

Next we turn to the charge exchange reaction $\pi^+\pi^-\to\pi^0\pi^0$
which should help to select the unique solution for the S-wave.
There are three relevant experiments: (a) IHEP-NICE \cite{apel2}, 
(b) GAMS \cite{GAMS} and (c) BNL -- E852 (preliminary results
\cite{gunter}). 

In (a) the amplitude $S_0$ is obtained after subtraction of the
 I=2 contribution. One finds two different solutions for $S_0$, 
one inside the unitarity circle with $0.5\lesssim\eta_0^0\lesssim1$, 
and another one with larger modulus which
exceeds by far the unitarity limit. The physical solution has
\begin{equation}
\varphi_S-\varphi_D\ <\ 0 \ \quad \textrm{for} \ 
                       \quad m_{\pi\pi}\ >\ 1100\ \textrm{MeV}.
\label{phchange1}
\end{equation}
The two solutions branch around $m_{\pi\pi}=1100$ MeV where 
$\cos(\varphi_S-\varphi_D)\approx 1$. 
The sign change is similar to the result (\ref{phchange}) above in the
charged $\pi\pi$ mode.   

The first results from (c) at small momentum transfer $t$ 
 with higher statistics give a similar picture: The
amplitude with the smaller modulus is the physical solution. The modulus
shows again a sharp drop at 1000 and 1500 MeV.
There is a
strong inelasticity immediately above 1000 MeV and a small phase shift
rise by about $30\dgr$ if $m_{\pi\pi}$ increases from 1000 to 1300 MeV, in
qualitative agreement with the CM result \cite{CERN-Munich1}.

In the GAMS experiment (b) with high statistics
one finds again the two solutions with large and
small amplitude modulus peaking at $m_{\pi\pi}\sim 1200$ MeV and the 
dips at 1000 and
1500 MeV. There is no attempt, however, to
determine the phase shifts and to consider the role of unitarity. In fact, the
solution with large modulus is declared as the physical one. 
The phase difference
$\varphi_S-\varphi_D$ stays positive in the full mass range:
% contrary to 
%the results (\ref{phchange}) and  (\ref{phchange1}) obtained from the phase
%shift analyses. This phase difference obtained by GAMS 
it is falling for  $m_{\pi\pi}<1100$ MeV as in (\ref{phchange}) but it is
rising again in the
range from 1100 to 1400 MeV -- contrary to all solutions A-D in \cite{em1} and 
the one in \cite{CKM2}. Such behaviour 
would imply the S wave phase $\varphi_S$ to rise more
rapidly than the resonant D wave phase $\varphi_D$ which is highly
unplausible. Possibly 
a transition from one
solution to the other one around $m_{\pi\pi}\sim 1200$ MeV -- consistent with
(\ref{phchange}) --
is allowed within the errors which would resolve this 
conflict.\footnote{Taking the
data on $|S|^2$, $|D|^2$ and $|S||D|\cos (\varphi_S-\varphi_D)$ 
presented in \cite{anis} at face value 
we obtain unphysical values $\cos(\varphi_S-\varphi_D)>1$ at
$m_{\pi\pi}\sim 1200$ MeV. }
%Unfortunately, neither the angular distribution
%moments nor the details of the analysis method are presented so that 
%we cannot study further these discrepencies.} 
One should also consider the possibility of small phase shift differences
due to $a_1$ exchange as suggested by the polarized target experiment
and discussed in Sect. \ref{OPE}. 
Such systematic effects could be studied further from the angular
distribution moments which have not been presented so far.

In summary, the various phase shift 
analyses of $\pi\pi$ scattering suggest the
isoscalar S wave under the $f_2(1270)$ to be slowly
moving with an inelasticity  between 0.5 and 1. This does not
correspond to a resonance of usual width. 
We can estimate the width $\Gamma$ 
of a hypothetical resonance 
near $m_{\pi\pi}\sim 1300$ MeV from the energy slope of the phase shifts
$\delta_0^0$. Using the data from  CM \cite{CERN-Munich1} 
or KLR \cite{klr} we find
\begin{equation}
\frac{d\delta_0^0}{dm_{\pi\pi}}=\frac{2}{\Gamma}\frac{1+\eta^0_0}{2 \eta^0_0}
\approx 3.7\ \textrm{GeV}^{-1} \label{gamma}
\end{equation}
in the mass range $1100\lesssim m_{\pi\pi}\lesssim 1400$ MeV 
which yields a lower limit on the width
\begin{equation}
\Gamma[f_0(1300)]\ > \ 540\  \textrm{MeV}. 
\label{gam1300} 
\end{equation}
where the inequality corresponds  to $\eta^0_0<1$.
This estimate shows that the slow movement of the phase 
does not allow the interpretation in terms of a usual narrow resonance.
On the other hand, there is evidence for a rapid drop in the S-wave
intensity near 1500 MeV indicating a Breit-Wigner type narrow resonance
around this mass. 
An analysis which would treat both the
elastic and charge exchange $\pi\pi$ scattering data 
together in this mass range has not yet been carried out but
is highly desirable.

\subsubsection{The reaction $\pi\pi\to K\overline K$}
It is argued by BSZ \cite{bsz} that the influence of the additional 
scalar state $f_0(1370)$ is marginal for the $\pi^+\pi^-$ (CM) 
data whereas it becomes
essential in the $K\overline K$ final states.
The study of this final state is more difficult than $\pi\pi$  
as resonances with both G parities can  occur in a production experiment where
besides one pion exchange also exchanges with $G=+1$ particles may
contribute.  

Phase shift analyses of the $K\overline K$ system have been carried out
(a) at Argonne \cite{argonne}, (b) by the Omega spectrometer experiment
 at CERN
\cite{costa} and (c) at BNL \cite{bnl}; (d) the moduli of the partial wave 
amplitudes in the $K^+K^-$ final state have been obtained also from 
a CERN-Krakow-Munich experiment with polarized target \cite{CKM3}.

In the Argonne experiment (a) a comprehensive analysis of three reactions
\begin{equation} 
\pi N\to K \overline K N' 
\end{equation}
in different charge states has allowed the separation
of partial waves in both isospin states I=0 and I=1. In the mass region below 
1500 MeV S, P and D waves have been included and 8 different
partial wave solutions  have been obtained
in the beginning. Requirements of charge symmety,
reasonable t-dependence for the pion exchange reactions and approximate 
 behaviour
of the P waves as tails of the $\rho$ and $\omega$ resonances have finally
selected a unique solution (called Ib). The absolute phase of the amplitude is
determined relative to the Breit-Wigner phase of the D wave resonances 
$f_2$(1270) and $a_2$(1320). It is remarkable that the P waves 
of the preferred solution are compatible both in
magnitude and phase with what is expected for the  tails of the vector mesons.

The preferred solution shows two features
\begin{description}
\item (i)
The modulus of the $S_0$ amplitude has a 
narrow peak near threshold -- possibly caused by the $f_0(980)$ --  and shows 
an enhancement near 1300 MeV before
it drops to a small value near $m_{K\overline K}\sim 1600$ MeV (see Fig. 
\ref{gbfig3}b).
\item (ii) The phase stays nearly constant up to 
$m_{K\overline K}\sim 1300$ MeV,
thereafter it  advances by $\Delta \delta_0^0\sim 100\dgr$ 
when approaching $m_{K\overline
K}\sim 1500$ MeV and then drops. 
\end{description}
This phase variation is  similar to the one in elastic
$\pi\pi$ scattering  ($\Delta \delta_0^0\sim 100\dgr$ in KLR \cite{klr} and
$\Delta \delta_0^0\sim 70\dgr$ in CM \cite{CERN-Munich1}) and 
lead to the conclusion \cite{argonne} that the
phase variation in $K\overline K\to K\overline K$ is small and that
the phase variation in $\pi\pi\ \to \ K\overline K$ 
is related to a resonance $\epsilon(1425)$  which couples
mainly to $\pi\pi$.

The other two experiments (b) and (c) extend their analysis towards higher
masses which leads to more ambiguities. In the mass region considered here
below 1600 MeV there is close similarity with the Argonne results in the
two features (i) and (ii) above: in (b) one of the two ambiguous solution
agrees whereas the second corresponds to a narrow resonance 
in the S-wave with $m_{\pi\pi}=1270$ MeV
and $\Gamma=120$ MeV whereas in (c) the favored solution agrees except for
an additional phase variation below 1200 
MeV by 30$\dgr$. In addition the
continuation to higher masses with a suggested resonance at 1770 MeV is
presented. 
In the region below 1200 MeV  (cos$(\phi_S-\phi_D)\approx -1$ at 1200 
MeV) another
solution should be possible with $\phi_S-\phi_D \ \to \
\pi-(\phi_S-\phi_D)$; this choice would yield a slightly decreasing phase
$\phi_S$ similar to the one in (a); the  D-wave phase $\phi_D$ 
presented in (c) in
this region is decreasing which would contradict the expected threshold
behaviour.
In (a) %\cite{argonne} 
the phase near threshold has been fixed by
comparison with the P-wave taken as tail of the $\rho$ meson.
All these experiments are consistent with a solution with slow phase
variation around 1300 MeV.

Only experiment (d) shows a different result which resembles the 
alternative solution in (b)
without peak in the S-wave near 1300 MeV emphasized in (i) 
and has been rejected in \cite{argonne} and also in
\cite{bnl} where no ambiguity exists contrary to
the $K^+K^-$ channel. No discussion of ambiguities in terms of
Barrelet zeros nor results
of a phase shift analysis has been presented. For the moment we assume that
the analysis in (d) did not find all ambiguous solutions and 
so does not invalidate the 
preferred  solution quoted above.  

We summarize the results from experiments (a)-(c) on the  parameters of the
resonance $\epsilon(1420)$
as obtained from the fits to the phases in Table \ref{KKbar}.
\begin{table}

%\begin{center}
%\label{scalar}
\begin{tabular}{lcccc}
\hline
% & & & \\
\rule[0ex]{0ex}{3.5ex}
 group  &   mass range  & mass  & width & $\Gamma_{K\overline K}/\Gamma_{\pi\pi}$\\
\rule[-2ex]{0ex}{3.5ex}       
 & of fit [MeV] & [MeV]  &  [MeV]& \\
 \hline
%& & & \\
\rule[-2ex]{0ex}{5.5ex}
Argonne \cite{argonne}   & 1000  - 1500   &   1425 $\pm$ 15   & 
          160 $\pm$ 30   & 0.1 - 0.2\\
%& & & \\
\rule[-2ex]{0ex}{5.5ex}
OMEGA \cite{costa}  & $<$1550     &  $\sim$ 1400   &   $\sim$ 170  &  \\
\rule[-2ex]{0ex}{5.5ex}
BNL \cite{bnl}  & 1000 - 2000   & 1463 $\pm$ 9     &  $118^{+138}_{-16}$ 
   &   \\ \hline
\rule[-2ex]{0ex}{5.5ex}
PDG \cite{PDG}: $f_0(1500)$   &  & 1500 $\pm$ 10     &  112 $\pm$ 10 
   & 0.19 $\pm$ 0.07   \\
\hline
\end{tabular}
%\end{center}
\caption{Different determinations of resonance parameters in $K\overline K$
final state near 1500 MeV.}
\label{KKbar}
\end{table}
The comparison of the $\epsilon$ resonance parameters with the 
PDG numbers in the last
line suggests the identification
\begin{equation}
\epsilon(1420) \to f_0(1500) \label{epsf0}
\end{equation}
because of comparable width and small $K\overline K$
coupling 
although there is a small shift in mass; the latter is smallest in a energy
dependent fit over the full mass range. 

%\begin{figure}
%\vspace{2cm}
%\caption{Amplitudes for isoscalar S-waves (a) $\pi\pi\  \to\  K\overline K$ 
%representing the data by Etkin et al. \protect\cite{bnl} (b) 
%$\pi\pi\  \to\  \eta\eta$, estimated from the data of Binon et al.
%\protect\cite{IIL1} on the modulus $|S_0|$
%and relative phase $\phi_D-\phi_S$ with an
%estimate of $\phi_D$. The numbers along the curves are the $\pi\pi$ cms
%energies in MeV, the dashed curves indicate the background amplitudes.} 
%\label{S0KK}
%\end{figure}
For further 
illustration we show in Fig. \ref{gbfig4}a the S-wave in an Argand diagram
we obtained as smooth interpolation of the results 
by Etkin et al. \cite{bnl}. After an initial decrease
of the amplitude from its maximum near threshold 
a resonance circle develops in the region
1200-1600 MeV  with small phase velocity at the edges and largest velocity
in the interval 1400-1500 MeV suggesting a resonance pole with negative
residue. The small dip  at 1200 MeV appears, as the resonant amplitude moves
first in negative direction of $\re{S_0}$ (to the ``left'') 
above the background which moves slowly to the ``right''.
This interference phenomenon yields the peak near 1300 MeV in
Fig. \ref{gbfig3}b, but  
there is no evidence for an extra loop corresponding to 
the additional state $f_0(1370)$.
The Argand diagram presented by Cohen et al. \cite{argonne} for masses below
1600 MeV shows a qualitatively similar behaviour. 

%On the other hand,
%in the region around 1300 MeV with the small bump there is no 
%or little phase variation. 
%We would then interpret  this bump  again as part of a broader structure  
%which we associate with the glueball state.

\subsubsection{The reactions $\pi\pi\to \eta\eta,\ \eta\eta'$}

These reactions have been studied by the IHEP-IISN-LAPP Collaboration
\cite{IIL1,IIL2} again 
in $\pi p$ collisions 
applying the OPE model with absorptive corrections.
The partial wave decomposition of the $\eta\eta$ channel
yields an S wave with an enhancement which
peaks near 1200 MeV and a second peak with parameters
\begin{equation}
m_G=(1592\pm 25)\ \textrm{MeV},\qquad \Gamma_G=(210\pm 40)\ \textrm{MeV}.\
\label{G0mass}
\end{equation}
This state G(1590) has been  considered as a glueball
candidate by the authors 
 as in this mass range there are no resonance signals from $K\overline K$
nor from $\pi^0\pi^0$. The  phase difference 
$\varphi_D-\varphi_S$ varies with mass in a similar way as the one in the   
$K\overline K$ final state: it rises from 210$\dgr$ at 1200 MeV up to the maximum
of 300$\dgr$ at around 1500 MeV and then drops
(see, for example \cite{bnl}). Therefore a similar
interpretation is suggested. The $f_2(1270)$ interferes with the S-wave
which is composed of one Breit-Wigner resonance around 1500 - 1600 MeV
above a background with slowly varying phase. 

There is nevertheless a major difference between both channels
if the S-wave magnitude is considered. At its peak
value  in $\eta\eta$ near 1600 MeV there is a minimum
in $K\overline K$ and the opposite behaviour around 1300 - 1400 MeV, namely a
peak in $K\overline K$ and a dip in $\eta\eta$
(see Fig. \ref{gbfig3}b,c). Both phenomena can be explained
by  a change in the relative sign between 
the background  and the 
resonance amplitude. 

In Fig. \ref{gbfig4}b we show the behaviour of the S-wave amplitude 
which we obtained using the
data by Binon et al. \cite{IIL1}.
Assuming a
Breit-Wigner phase for the $f_2(1270)$ one finds that the S-wave phase
around 1300 MeV is slowly varying again. 
At higher energies a contribution of about 20\% from $f_2(1520)$ 
is expected in the 
D-wave as in the $\pi\pi\ \to \ K\overline K$ channel.
 Now the resonance circle in Fig. \ref{gbfig4}b 
corresponds to a pole in the amplitude with positive residue and this
explains the rather different mass spectra 
emphasized above in the $K\overline K$ and
$\eta\eta$ channels.

%If for $m<m_{res}$ the background interferes destructively
%in the real part of the Breit-Wigner amplitude
%we get the $\eta\eta$ case with the upwards shift of the resonance peak,
%and vice versa for $K\overline K$ reaction. 

The situation may be illustrated by the following simple model  
for the $S$ and $D$-wave amplitudes 
%\[
\begin{equation}
\label{Sf0f2}
\begin{array}{lll}
\pi\pi\to K\overline K \quad\ & S= B_K(m) - x_K^{(0)} f_0(m)e^{i\phi_K};
 & \quad\ D=x_K^{(2)} f_2(m)\\% \nonumber \\
\pi\pi\to \eta\eta:\quad\ & S= B_\eta(m)+x_\eta^{(0)} f_0(m)e^{i\phi_\eta};
  &  \quad\ D=x_\eta^{(2)} f_2(m) 
\end{array} %\]
\end{equation}
where $B_i(m)$ denotes the slowly varying 
background amplitudes with $\re{B_i}<0$ and
 $f_\ell(m)$ the  Breit-Wigner amplitudes for spin $\ell$ 
with $f_\ell(m_{res})=i$
and elasticities $x_i^{(\ell)}$; $\phi_i$ is an additional small 
phase due to background.

Then, despite the rather large mass difference between
 $\epsilon(1420)$ and $G(1590)$ of 170 MeV, we can
consider both states as representing the same resonance interfering with
the broad background and therefore suggest
\begin{equation}
G(1590)\to f_0(1500). \label{Gf0}
\end{equation}
The $\eta\eta$ experiment has been repeated at higher beam energy which
allows the study of higher mass states. In the lower mass region the
previous results are essentially recovered \cite{IILL}.
 
The $\eta\eta'$ mass spectrum \cite{IIL2} 
shows a threshold 
 enhancement around 1650 MeV which the authors interpret as another
signal from G(1590). 

\subsubsection{$p\overline{p}$ annihilation and the $f_0(1370)$
and $f_0(1500)$ states}

The Crystal Barrel (CB) Collaboration at the LEAR facility at CERN
has measured the $p\overline{p}$ annihilation reaction at rest. In this case 
the initial $p\overline{p}$ 
system is in either one of the three $J^{PC}$ states
\begin{equation}
^{ 1}S_{ 0} \ ( 0^{ - +} ), \quad
     ^{ 3}P_{ 1} \ ( 1^{ + +} ), \quad
^{ 3}P_{ 2} \ ( 2^{ + +} )\quad  \textrm{with}\quad 
  I^{ G} \ = \ 1^{ -} \ \textrm{or} \ 0^{ +}.
 \label{ppbarqz}
\end{equation}
Another experiment has been carried out at the Fermilab antiproton
accumulator (E-760) at the $cms$ energies $\sqrt{s}$ of 3.0 and 3.5 GeV.
The following final states have been investigated
(with $I^{ G}$ in brackets assuming isospin conservation)
\begin{gather}
(a)\ \pi^0 \pi^0 \pi^0\ (1^-)\qquad (b)\ \pi^0 \pi^0 \eta\ (0^+)\qquad 
(c)\ \eta \eta \pi^0\ (1^-) \nonumber\\
 (d)\ \eta \eta' \pi^0\ (1^-)\qquad 
(e)\ \eta \eta \eta\ (0^+)
\label{ppbarfs}
\end{gather}

Reaction (a) has been studied by CB with the very high
statistics of 712 000 events \cite{Ams}.
The Dalitz plot shows clearly the narrow band of low density (region A)
corresponding to the $f_0(980)$ as also seen in the elastic $\pi\pi$
cross section. In the projection to the $\pi^0\pi^0$ mass the broad
structureless bump peaking near 750 MeV and the sharper peak due to the
$f_2(1270)$ are seen. The new feature is the peak at around 1500 MeV
which corresponds to a band of about constant density in the Dalitz plot;
therefore its quantum numbers are determined to be $J^{PC}=0^{++}$ and 
the state is now called $f_0(1500)$.
A peak of similar position and width is seen at the higher energies
of E-760 \cite{Armstrong1} together with an increased $f_2(1270)$ 
signal and a decreased 
low mass bump. It is therefore likely that the same state $f_0(1500)$
is observed here although the authors consider it as a  $f_2(1520)$
state. This conflict could be solved by means of a phase shift analysis. A
weak signal of the same state is seen in reaction (b) at the higher
energies.

The state $f_0(1500)$ is also identified 
by CB in reaction (c) where it appears 
as clear peak in the $\eta\eta$ mass spectrum and band in the Dalitz plot
\cite{Ams4,Ams3}. A significant signal is again seen at the higher 
$p\overline p$ energies \cite{Armstrong2}.
Furthermore, Amsler et al. relate the threshold enhancement seen in the
reaction (d) to the $f_0(1500)$ \cite{Ams5}.
It is suggestive to identify this state with the resonance discussed before
in this section
where also the phase variation 
has been observed directly 
and to consider the $f_0(1500)$ as a genuine Breit-Wigner resonance.

In the region between the $f_0(980)$ and $f_0(1500)$ Amsler et al. claim the
existence of a further scalar state, the $f_0(1370)$. In reaction (a) it is
required by the fit as background under the $f_2(1270)$ very much as in
$\pi\pi$ scattering. In reaction (c) it actually appears as clear bump
with rather broad width of $\Gamma=380$ MeV. On the other hand this bump has
disappeared entirely at the higher $cms$ energies \cite{Armstrong2}, 
whereas the $f_0(1500)$ stays. This
indicates a different intrinsic structure of both states; the disappearence
of the $f_0(1370)$ at higher energies 
is reminiscent to the disappearence observed by GAMS of the peak at 700
MeV at production with large $t$ in comparison with the $f_0(980)$
(see Sect. \ref{spectr2}).

In the $p\overline p$ reaction a direct phase shift analysis as in the $\pi\pi$
scattering processes 
is not possible. The amplitudes for the different initial states
(\ref{ppbarqz}) are constructed with the angular distributions specified in 
\cite{Zemach} and an ansatz for the $\pi\pi$ amplitudes within 
the framework of an isobar
model. Unitarity constraints cannot be strictly enforced here as in 
case of two body
$\pi\pi$ scattering processes. 
The evidence for the $f_0(1370)$ is  based on the fits of this model to the
Dalitz plot density. 
%From the interference of the different isobar channels  
%some phase sensitivity is given. However, it  seems to be less selective
%as in two body processes. 

%For illustration we refer to two analyses of the CB data on reactions (a)-(c)
%by Anisovich et al. \cite{as1,anis}.
%In the first analysis \cite{as1}, also including
%CERN-Munich high mass moments, phase shifts and inelasticities 
%of the S-wave are obtained in close agreement with the solution ($-+-$) above
%1400 MeV in \cite{CERN-Munich3} which corresponds to solution C in
%\cite{em1}. This solution, however, can been excluded
%because of the rather slowly varying S-wave
%magnitude which does not show the drop near 1500 MeV observed by CKM
%\cite{CKM2} and GAMS \cite{GAMS}.
%In the later work \cite{anis} the GAMS data
%have been included in addition; 
%this leads to a different S-wave amplitude which becomes rather small 
% in this mass region at an acceptable fit quality for the $p\overline p$ data.
%So the $\pi^0\pi^0$ spectra in the
%$p\overline p$ reactions are not as selective in this case as
%in the two body $\pi\pi$ process.

In the fit by BSZ \cite{bsz} the CB data have been described
with inclusion of the
$f_0(1370)$. Their fit predicts a rapid decrease of the phase 
in the channel
$\pi\pi\to K\overline K$ near 1200 MeV; this variation is consistent with the
BNL data \cite{bnl} within their large errors but not
with the slowly varying  phases determined by Cohen et al. \cite{argonne}
with smaller errors.
Furthermore a small dip is expected for the S-wave magnitude $|S|$
near the top which is neither observed in the GAMS \cite{GAMS} nor in the
BNL \cite{gunter} experiments on the $\pi^0\pi^0$ final state.

For the moment, we do not accept the $f_0(1370)$ effect as a genuine
Breit-Wigner resonance. It appears to us that the Dalitz plot analysis of the
$p\overline p$ data -- although some phase sensitivity is given -- is less
selective than the phase shift analysis of two-body processes.
%is not sensitive enough to the phase variations.
The $\pi\pi$, $K\overline K$ and $\eta\eta$ data discussed in the previous
subsections (Figs. \ref{gbfig3},\ref{gbfig4}) speak against a resonance
interpretation of the peak at 1370 MeV. 
%
%---------------------------------------------------------------------
%\input{gb5v4.tex}
%gb5v4.tex

\section{The  $J^{PC}=0^{++}$ nonet of lowest mass}
\label{sectnonet}

After the reexamination of the evidence for scalar states in the region up to
about 1600 MeV we are left with the $f_0(980)$ and $f_0(1500)$ resonances
where we have clear evidence for
a peak and for the phase variation
associated with a Breit-Wigner resonance. 
The identification of states in this mass region is so difficult because of
their interference with the large background. 
As explained in the previous section, we do not
consider the $f_0(1370)$ signal as evidence for a Breit-Wigner resonance 
in between the two $f_0$ states, but rather as the reflection of a yet
broader object or the ``background''.

As members of the scalar nonet we consider then the two $f_0$ states
besides the well known $a_0(980)$ and
the $K_0^*(1430)$. We will now have to clarify
whether this assignment provides a consistent picture for the 
various observations
at a given singlet-octet mixing angle to be determined. 
Such observations, together with further consequences of our hypotheses, 
 will be discussed next.

\subsection{Properties of $f_{0}  (980)$ and 
$f_{0}  (1500)$ from $J/\psi \ \rightarrow \ V \ f$  decays}

There are three (primary) mechanisms -- all without 
full analytic understanding --
which contribute to purely hadronic 
decays of $J/\psi$ into noncharmed final states:
%\vspace*{0.1cm}

\begin{enumerate} 
\item $c \ \overline{c}$ annihilation into three gluons; 
%\vspace*{0.1cm}

\item $c \ \overline{c}$ mixing with noncharmed virtual vector mesons  
($\omega_{V}  , \ \varphi_{V}  , \ 3g_{V}$),\\ 
%\vspace*{0.1cm}
here $3g_{V}$ denotes a three gauge boson ``state'' 
with $J^{\ P C_{n}} \ = \ 1^{ - -}$;
%\vspace*{0.1cm}

\item $c \ \overline{c}$ annihilation into a virtual photon.
\end{enumerate}
%\vspace*{0.1cm}

\noindent
In each channel above the hadronization into a given 
exclusive, hadronic final state has to be included.

We will {\it assume} that the mechanism 1. above is the dominant one 
for the decay modes listed in Table \ref{eq:52}. 
\begin{table}[ht]
\[ % \begin{array}{l} 
  \begin{array}{rlcl} 
   \hline
   \mbox{decay}&\mbox{modes} & \mbox{symbol}&
               \mbox{branching ratios} \ \times \ 10^{\ 4}
\\ \hline
 J/\psi \ \rightarrow &  \varphi \ \eta^{\ \prime} \ (958) & 
       X_\varphi^{(-)}& 3.3 \ \ \pm \ 0.4 \\
   & \varphi \ f_{ 0} \ (980) & X_\varphi^{(+)} &  3.2 \ \  \pm \ 0.9 \\
   &  \omega \ \eta^{\ \prime} \ (958)  & X_\omega^{(-)} & 1.67 \ \pm \ 0.25 \\
   & \omega \ f_{ 0} \ (980) & X_\omega^{(+)} & 
                                 1.41 \ \pm \ 0.27 \ \pm \ 0.47 
\\     \hline
 % \end{array}
  \end{array} \]
\caption{Branching ratios $X_V^{(\pm)}$ 
for decays of $J/\psi $ into vectormesons $V$ and scalar (+) or pseudoscalar
($-$) particles according to
the PDG \protect\cite{PDG} }
\label{eq:52}
\end{table}
\noindent
This Table involves only the pseudoscalar-scalar associated pair
($ \eta^{\ \prime} \ (958) \ , \ f_{0} \ (980)$), i.e. only $f_{0} \ (980)$ from the
scalar nonet, whereas
no information is extracted at present for the associated decay modes   
into $\varphi \ f_{ 0} \ (1500)$ and $\omega \ f_{0} \ (1500)$
%
%\begin{equation}
%  \label{eq:53}
%  \begin{array}{l} 
%  \begin{array}{l|c|c} 
%   & \mbox{decay modes} & \mbox{branching fraction} \ \times \ 10^{\ 4}
%  \\ \hline \vspace*{-0.4cm}
%  \\ & & \\
%   & \varphi \ f_{0} \ (1500)         &  ?  \\
%   J/\Psi \ \rightarrow & & \\
%   & \omega \ f_{\ 0} \ (1500)          &  ? 
%  \end{array}
%  \end{array}
%\end{equation}
%
%\noindent
by the PDG \cite{PDG}.
%\vspace*{0.1cm}

A word of caution on the list in Table \ref{eq:52} is in order. The data
for $\omega \ f_{0} \ (980)$ are based upon a single experiment 
(DM2 \cite{Augustin5pi}) and therein
essentially on a single data point.
%tail of the $f_2(1270)$. 
The result is supported however by 
the Mark II experiment \cite{Gidal} in which the recoil spectrum against
the $f_{0} \ (980)$ is measured. 
The ratio of the $\omega$ and $\varphi$ peaks are
consistent with the ratio following from Table \ref{eq:52}
 although there is some uncertainty
about the  background under the  $\varphi$ meson. 

With this remark in mind we have 
a closer look at  Table \ref{eq:52}. 
The entries for the decays into the scalar and pseudoscalar particles
show an indicative {\it pattern} \footnote{We are indebted
to C. Greub for pointing out that the relevant decay modes of $J/\psi$ comprise
both $\varphi$ {\it and} $\omega$.}:  
\begin{equation}
  \label{eq:54}
 X^{\ (+)}_{\ V} \ \approx \ X^{\ (-)}_{\ V} \ \rightarrow \ X_{\ V}
  \hspace*{0.3cm} ; \hspace*{0.6cm}
  X_{\ \varphi} \ \approx \ 2 \ X_{\ \omega},
\vspace*{0.5cm} 
%
%  \left .
%  \begin{array}{l} 
% X^{\ (-)}_{\ V}\ = 
%  \  Br \ ( \ J/\Psi \ \rightarrow \ V \ \eta^{\ \prime} \ (958) \ ) 
%\vspace*{0.3cm} \\
%X^{\ (+)}_{\ V}\ = \   Br \ ( \ J/\Psi \ \rightarrow \ V \ f_{0} \ (980) \ )
%  \end{array}
%  \ \right \rbrace
%  \ \mbox{for} \ V \ = \ \omega \ , \ \varphi
%\\
%  \end{array}
\end{equation}
\noindent
i.e. the branching fractions into the 
$\eta^{\ \prime} \ (958)$ and the $ f_{0} \ (980)$ are very
similar which then suggests a similar quark composition.\footnote{We
neglect here the phase space effects of $\lesssim 15\%$ whereby it is
assumed that for the momenta $p\sim 1$ GeV the threshold behaviour of the
P-wave is reduced to that of the S-wave by formfactors.}
We thus decompose both states $ \eta^{\ \prime} \ (958)$ and $f_{0} \ (980)$ 
according to their respective strange (s) and nonstrange (ns) 
$q  \overline{q}$ composition,
neglecting their small mass difference, 
\begin{equation}
  \label{eq:55}
  \begin{array}{l} 
  \begin{array}{lll} 
  \eta^{\ \prime} \ (958) \ & \sim &
  \ c^{\ (-)}_{\ ns} \ u \overline{u} \ +  
  \ c^{\ (-)}_{\ ns} \ d \overline{d} \ +  
  \ c^{\ (-)}_{\ s} \ s \overline{s}  
\vspace*{0.3cm} \\
  f_{ 0} \ (980) \ & \sim &
  \ c^{\ (+)}_{\ ns} \ u \overline{u} \ +  
  \ c^{\ (+)}_{\ ns} \ d \overline{d} \ +  
  \ c^{\ (+)}_{\ s} \ s \overline{s}  
  \end{array}
  \end{array}
\end{equation}
with normalization 
$  2 \ | \ c_{\ ns}^{\ \pm} \ |^{\ 2}
  \ + \ | \ c_{\ s}^{\ \pm} \ |^{\ 2} \ = \ 1.$
We retain {\it only} 
the approximate relations in  Eq. (\ref{eq:54}) and,
according to the mechanism 1. 
in the above list, we infer for the $f_{0} \ (980)$
\begin{equation}
  \label{eq:56}
%  \begin{array}{l} 
  \begin{array}{lll} 
  X_{\ \omega} \ \simeq  \ 2 \ | \ c_{\ ns} \ |^{\ 2} 
  \hspace*{0.3cm} & ; & \hspace*{0.3cm}
  X_{\ \varphi} \ \simeq \ | \ c_{\ s} \ |^{\ 2}
\vspace*{0.3cm} \\
c_{\ ns}\ = \  c^{\ (+)}_{\ ns} \ =  \ c^{\ (-)}_{\ ns} 
  \hspace*{0.3cm} & ; & \hspace*{0.3cm}
c_{\ s}\ = \  c^{\ (+)}_{\ s} \ = \ c^{\ (-)}_{\ s} .
  \end{array}
%  \end{array}
\end{equation}
The second equation in (\ref{eq:54}) is satisfied if we choose
$c_s=2c_{ns}$. Then we find for the vector 
$\vec{c} \ = ( \ c_{\ ns} \ , \ c_{\ ns} \ , \ c_{s} \ )$%
\footnote{The ``mnemonic'' approximate form of $\vec{c}$ %in Eq. (\ref{eq:56})
is due to H. Fritzsch.} 
in case of
$f_{ 0} \ (980)$ 
%\vspace*{0.5cm} \\
%  X_{\ \varphi} \ = \ 2 \ X_{\ \omega} \ \rightarrow
%  \ c_{\ s} \ = \ 2 
%  \ c_{\ ns}
%\vspace*{0.3cm} \\
%\vec{c} \ = ( \ c_{\ ns} \ , \ c_{\ ns} \ , \ c_{s} \ )
\begin{equation}
f_{ 0} \ (980):\qquad \vec{c} 
       \ = \ \frac{1}{\sqrt{\ 6}} \ ( \ 1 \ , \ 1 \ , \ 2 \ )
\label{f0low}
\end{equation}
and in case of $f_{ 0} \ (1500)$ accordingly the orthogonal composition
\begin{equation}
  \label{eq:60}
%  \begin{array}{l} 
  f_{ 0} \ (1500): \qquad  \vec{c}^{\ '} = 
  \ \frac{1}{\sqrt{\ 3}} \ ( \ 1 \ , \ 1 \ , \ -1 \ ).
%  \end{array}
\end{equation}
\noindent
These derivations % from Eqs. (\ref{eq:52}) - (\ref{eq:56}) 
reveal 
-- within the approximations adopted -- the pair 
$ \eta^{\ \prime} \ (958)$ and $f_{0} \ (980)$ as a {\it genuine}
parity doublet.
Thus $\eta$ $\eta'$ and $f_{0}\ (980)$ $f_{ 0} \ (1500)$ are related
and governed approximately by the same singlet-octet mixing angle
\begin{equation}
  \label{eq:58}
  \begin{array}{l} 
\Theta \ \approx \ \arcsin \ 1 / 3 \ = 19.47\dgr 
  \end{array}
\end{equation}
%
%$\vartheta_{\ 0-8} \ \approx \ \arcsin \ 1 / 3 \ = 19.47^{\ o}$ 
with respect to the  vectors 
$\vec{e}_0= \frac{1}{\sqrt{\ 3}} \ ( \ 1 \ , \ 1 \ , \ 1 \ )$ and
$\vec{e}_8= \frac{1}{\sqrt{\ 6}} \ ( \ 1 \ , \ 1 \ , \ -2 \ )$
\begin{equation}
  \label{eq:57}
%  \begin{array}{l} 
\vec{c} \ = \ \vec{e}_0 \ \cos \ \Theta 
\ - \ \vec{e}_8 \sin \ \Theta.
%  \end{array}
\end{equation}
\noindent
There is one difference though in the mass patterns of
 the two octets in that the $I=0$ state
closer to the octet is the lighter one in the pseudoscalar case ($\eta$) but
the heavier one in the scalar case ($f_{0} \ (1500)$); then we adopt the
following correspondence in the quark compositions (\ref{f0low}) and
(\ref{eq:60})
\begin{equation}
\eta\quad \leftrightarrow \quad f_{0} (1500) \qquad \mbox{and}
      \qquad \eta'\quad \leftrightarrow \quad f_{0} (980). 
\label{f0eta}
\end{equation}

With this flavor composition of the $ f_{0}(1500)$ we also predict
 the ratio of decay widths 
\begin{equation} 
R\ =\  \frac{ B(J/\psi \ \rightarrow \ \varphi \ f_{0} (1500))}{
   B(J/\psi \ \rightarrow \ \omega \ f_{0} (1500))} %( Eq. (\ref{eq:53}) )
\ = \ \frac{1}{2}   % \ (\varphi) \ : \ 2 \ (\omega)$ .
\label{r1500phi}
\end{equation}
which is inverse to the corresponding ratio for $f_0(980)$.
A measurement of this ratio would be an interesting test of our hypotheses.

\subsection{Mass pattern and Gell-Mann-Okubo formula}

We come back to the two extremal possibilities for mixing discussed 
in Sect. \ref{Sigma} within the context of the $\sigma$ model:
\begin{description}
\item I) quenched singlet octet mixing,
\item II) strict validity of the OZI rule.
\end{description}
We now conclude, that the  $q  \overline{q}$ scalar nonet
is nearer to case (I).  
Furthermore, we suggest a definite
deviation from I)
parametrized by the {\it approximate} mixing angle (\ref{eq:58}), the same as
found for the pseudoscalar nonet.

This conclusion identifies the scalar
nonet as second one -- besides its pseudoscalar partner -- showing a large
violation of the OZI-rule.
%\footnote{We
%found that the other three $L_{q\overline q}=1$ nonets with
%$J^{PC}\ =\ 1^{++},\ 2^{++}$ and $ 1^{+-}$ can be filled with the OZI rule
%being satisfied within $\sim 0.1$ GeV.}  

Next we compare these results with the
expected mass pattern
as discussed in Sect. \ref{Sigma}.
In case of quenched singlet octet mixing (case I)
%patterns for the mixing of the isoscalar states in the scalar nonet:
%In the first case %which is found in the pseudoscalar sector -- 
one predicts 
from the  Gell-Mann-Okubo mass formula for the members of an octet
the heavier scalar $I=0$ member  to appear 
in the mass range 1550-1600 MeV, if one takes 
the $a_0$ and $K_0^*$ masses as input. 
The deviation from the observed mass of the $f_0(1500)$ is 
7-14\% in the masses squared
which we consider as tolerable; the deviation is attributed to
effects of $O(m_s^2)$.
Then the $f_0(980)$ is close to the
singlet member of the nonet.

On the other hand, for a splitting according to the OZI rule the isoscalar
$s\bar s$ state would be expected at the  mass of $\sim$1770 MeV. 
In this case the $f_0(980)$ would be a purely non-strange state which is
hardly consistent with the large decay rate $J/\psi\ \rightarrow \ \varphi
f_0(980)$ in Table \ref{eq:52}.

\subsection{Decays into $\pi\pi$, $K\overline K$, $\eta\eta$ and
$\eta\eta'$}

These 2-body decays are 
again sensitive to the flavor composition of the  $J^{PC}=0^{++}$
particles.
For further analysis we consider the decay of a $q\overline q$ 
state with arbitrary flavor composition where we define the mixing angle
$\phi$ now with respect to the strange non-strange directions
\begin{equation}
q\overline q\ =\ n\overline n\ \cos \phi\ + \ s\overline s\ \sin \phi;
\qquad n\overline n\ =\ (u\overline u\ + \ d\overline d )/\sqrt{2}.
   \label{resmix}
\end{equation}
The decay amplitudes are calculated with flavor symmetry but the 
relative amplitude $S$
to adjoin an $s \overline{s}$  pair relative to a
nonstrange
$u  \overline{u}$ or $d  \overline{d}$ one is assumed
to deviate from symmetry.  We assume S to be real with
$0 \ \leq \ S \ \leq \ 1$ and $S \ \sim \ \frac{1}{2}$, but it may depend
also on the mass of the decaying system with a restoration of symmetry at
high energies. For a mixed state as in (\ref{resmix}) 
 this ansatz leads to
the decay amplitudes in
Table \ref{flavor} which agree with the results in \cite{as1}. Here we take
the decomposition of $\eta$ and $\eta'$ as in (\ref{f0low}) and (\ref{eq:60})
with (\ref{f0eta}).%
\footnote{For an analysis with $S=1$, see also \cite{amsclo}.}
We also give the prediction for the two $f_0$ states with the same mixing as
in the pseudoscalar sector as discussed above 
and for the glueball taken as colour singlet
state.\footnote{Contrary to Ref. \cite{as1} we assume that the creation of
quarks from the initial gluonic system is flavor symmetric and that the
$s$-quark suppression occurs only in the secondary decay by creation of a soft
$q\overline q$ pair.} We now examine how the predictions from our 
hypotheses on the flavor decomposition in Table \ref{flavor} compare with
experiment.\\

%\[        
%\begin{array}{|>{$}c<{$}|}
%probe \sin \phi \\ \hline
%\end{array}
%/]%

\begin{table}
%\begin{center}
%\vspace{4mm}
%\[
%\begin{array}{|c|c|c|c|c|}
% \begin{tabular}{|>{$}c<{$}|>{$}c<{$}|>{$}c<{$}|>{$}c<{$}|>{$}c<{$}|}
%\setlength{\extrarowhight}{4pt}
\begin{tabular}{ccccc}
 \hline     
%\text{channel} & q\overline q \text{decay}(\phi) & f_0 (980) & f_0 (1500)
%  & \text{glueball}\\
channel   & $q\overline q$ decay ($\phi$) & $f_0 (980)$ & $f_0 (1500)$
& glueball\\
\hline      
$\pi^0\pi^0$ & $1\to \cos\ \phi /\sqrt{2}$ &$ 1 \to 1/\sqrt{6}$ & 
$ 1 \to 1/\sqrt{3}$ &$ 1 \to 1/\sqrt{3} $\\
$\pi^+\pi^-$ & $1$ &$ 1 $ & $ 1$ & $ 1  $\\
$K^+K^-$  & 
$(\protect\sqrt{2}\tan \phi+S)/2$ &$ (2+S)/2 $ & $ (-1+S)/2$ & $ (1+S)/2  $\\
$K^0\protect\overline{K^0}$ &
$(\protect\sqrt{2}\tan \phi+S)/2$ &$ (2+S)/2 $ & $ (-1+S)/2$ & $ (1+S)/2$\\
$\eta\eta$ & 
$(2+\sqrt{2}S\protect\tan \phi)/3$ &$ 2(1+S)/3 $ & $
(2-S)/3$ & $ (2+S)/3 $\\
$\eta\eta'$ &
$(\sqrt{2}-2 S\protect\tan \phi)/3$ &$ \sqrt{2}(1-2S)/3 $ & $
\sqrt{2}(1+S)/3$ & $ \sqrt{2}(1-S)/3$\\
$\eta'\eta'$ & 
$(1+2\sqrt{2}S\protect\tan \phi)/3$ &$ (1+4S)/3 $ & $
(1-2S)/3$ & $ (1+2S)/3 $\\
\hline
%\end{array} /]
\end{tabular} 
\caption{Amplitudes for the decays into two pseudoscalar mesons
of states with flavor mixing as in 
(\protect\ref{resmix}),
using for the $ f_0 (980)$, the  $ f_0 (1500)$
and the glueball the mixing angles 
$\protect\sin\phi=\protect\sqrt{2/3}$, $\protect\sin\phi=-
\protect\sqrt{1/3}$
and $\protect\sin\phi=\protect\sqrt{1/3}$ respectively.  
The $\eta  -  \eta'$ mixing is according to
 Eqs. (\protect\ref{f0low}) and (\protect\ref{eq:60}), 
S denotes the relative 
$ s\protect\overline{s}$ amplitude. Normalization is to $\pi^+\pi^-$,
the first row also indicates after ($\to$) the relative
weights of $\pi\pi$ decays. 
For identical particles the width has to be multiplied 
by $1/2$.}
\label{flavor}
\end{table}

\noindent {\it Couplings of the $f_0(980)$}\\
\noindent
This state has a mass near the $K\overline K$ threshold. So the directly
measurable quantities are the reduced widths 
$ \Gamma_{red}$ into $\pi\pi$ and  
$K\overline K$ for which we predict according to  Table \ref{flavor}
\begin{equation}
  \label{f0decay}
  \begin{array}{l} 
  R_0
   \ = \  {\displaystyle 
   \frac{ \Gamma_{red} (f_{0}(980) \ \to \ K\overline K)}
         { \Gamma_{red} (f_{0}(980) \ \to \ \pi\pi)}   
\ = \ \frac{(2+S)^2}{3}  } %end displaystyle
  \end{array}
\end{equation}

The experimental determination from elastic and inelastic $\pi\pi$
scattering is difficult because of the unitarity constraints near the
$K\overline K$ threshold. This can be avoided in a measurement of reaction
(\ref{pinucleon}a) at larger momentum transfer $t$ where the $f_0(980)$ appears
as a rather symmetric peak without much coherent background 
as already emphazised. Binnie et al.
\cite{binnie} % and more recently by GAMS \cite{GAMS} 
used their data on the $\pi^+\pi^-$, $\pi^0\pi^0$ and $K^+K^-$ channels
with $|t|\gtrsim 0.3$ GeV to measure the ratio
(\ref{f0decay}) directly by fitting their distributions to the Breit-Wigner
formula
\begin{equation}
\sigma_{\pi,K}\propto \left|
\frac{\Gamma_{\pi,K}^{1/2}}{m_0-m-i(\Gamma_\pi+\Gamma_K)/2}\right|^2,
\label{bwres2}
\end{equation}
where $\sigma_{\pi,K}$ denote the cross sections in the $\pi\pi$ and 
$K\overline K$ channels, respectively. Furthermore,
$\Gamma_\pi=\gamma_{\pi}p_\pi$ and
$\Gamma_K=g_{K}p_{K^+}+g_{K}p_{K^0}$,
where $p_h$ are the momenta of $h$ in the $hh\ cms$. The reduced widths
are given by $\Gamma_{red,\pi}=\gamma_\pi$ and 
$\Gamma_{red,K}=2 g_K$. We enter the result by Binnie et al. into Table
\ref{tabR0}. We also show the result of the fits by 
Morgan and Pennington (MP) \cite{mp}
to $\pi\pi$ and $K\overline K$ final states from various reactions 
taking into account the coherent background 
and unitarity constraints.
As we are close to the $K\overline K$ threshold the $S$ parameter may be not
well defined, this leaves a range of predictions also presented in Table
\ref{tabR0}.
We see that the determination by Binnie et al. is comparable to the
theoretical expectation whereas the one by MP 
(given without error) comes out a bit smaller.
\begin{table}[ht]
\[
  \begin{array}{lllccc}
\hline
   & \ \mbox{exp. results} & &S  =  0 & S  =  0.5 & S =  1.0
  \\ \hline % \vspace*{-0.6cm}  %-0.4
  R_{ 0} &  \ 1.9 \pm \ 0.5 &\textrm{Binnie}\ 
\protect\cite{binnie} & \ 1.3 & \ 2.1 & \ 3.0
  \\  %& & & & &\vspace*{-0.4cm} \\
 & \  \simeq 0.85 & \textrm{MP} \ \protect\cite{mp} &  &  &
\\ \hline   
 \end{array} \]
\caption{The ratio $R_0$ defined in (\ref{f0decay}) as determined from
experiment and predicted for different strange quark amplitudes $S$.}
\label{tabR0}
\end{table}

We want to add that the determination of this ratio $R_0$ needs data on the 
$K\overline K$ process. The sensitivity to $g_K$ in the
denominator of (\ref{bwres2}) is very weak. Therefore from fits to the
$\pi\pi$ spectra alone conflicting results are obtained.\\

\noindent{\it Couplings of the $f_0(1500)$}\\
\noindent
With respect to the branching fractions of the $f_{0}(1500)$ into
two pseudoscalars we scrutinize the phase space corrected reduced
rate ratios  deduced by C. Amsler \cite{amsams}
\begin{equation}
  \label{eq:62b}
  R_{1} \ = \ {\displaystyle
     \frac{ \Gamma_{\ red} \ ( \ \eta \ \eta \ )}
          { \Gamma_{\ red} \ ( \ \pi \ \pi \ )},
\quad
  R_{2} \ = \ 
    \frac{   \Gamma_{\ red} \ ( \ \eta \ \eta^{\ '} \ )}
           { \Gamma_{\ red} \ ( \ \pi \ \pi \ )}, 
\quad
  R_{3} \ = \ 
           \frac{  \Gamma_{\ red} \ ( \ K \ \overline{K} \ )}
           { \Gamma_{\ red} \ ( \ \pi \ \pi \ )}
   } % end displaystyle
\end{equation}
\noindent where all charge modes of $\pi\pi$ and $K\overline K$ are counted.
The experimental determinations \cite{amsams} are presented in Table
\ref{R123}.
Using our amplitudes for the decays of the $f_0(1500)$ in Table \ref{flavor}
we  predict for these ratios
%For the identical singlet octet mixing angle $\vartheta_{\ 0-8}$
%in Eqs. (\ref{eq:56}) - (\ref{eq:58}) for scalar {\bf and} pseudoscalar nonets
%the $q \ \overline{q}$ counting rule leads to the following
%Ansatz for the ratios in Eq. (\ref{eq:62b})
%
\begin{equation}
  \label{eq:62c}
  \begin{array}{l} 
  R_{1}  = \ \frac{4}{27} \ ( \ 1 \ - \ \frac{1}{2} \ S \ )^{2}  
\hspace*{0.1cm} , \hspace*{0.1cm} 
  R_{2}  = \ \frac{4}{27} \ ( \ 1 \ + \ S \ )^{2}  
\hspace*{0.1cm} , \hspace*{0.1cm} 
  R_{3}  = \ \frac{1}{3} \ ( \ 1 \ - \ S \ )^{2}  
  \end{array}
\end{equation}
%
%\noindent
%The quantity S in Eq. (\ref{eq:62c}) denotes the relative amplitude
%to adjoin an $s \ \overline{s}$ (quark antiquark) pair relative to a nonstrange
%$u \ \overline{u}$ or $d \ \overline{d}$ one in the decay of 
%$ f_{0} \ (1500) \ \rightarrow \ 2 \ ps$ . We assume S to be real with
%$0 \ \leq \ S \ \leq \ 1$ and $S \ \sim \ \frac{1}{2}$ .
%\footnote{ A similar Ansatz has been discussed for general mixing angle by
%Amsler and Close \cite{amsclo} .}
%\vspace*{0.1cm} 
%
\noindent
A $\chi^{2}$ fit for S using the ``data'' in Table \ref{R123} yields
\begin{equation}
  \label{eq:62d}
%  \begin{array}{l} 
  %S \ = \ 0.463^{\ +0.106}_{\ -0.089}
  S \ = \ 0.352^{\ +0.131}_{\ -0.109}
%\hspace*{0.3cm} ; \hspace*{0.3cm} 
%\chi^{\ 2} \ / \ ( \ \# \ d.f. \ = \ 2 \ ) \ = \ 0.386
%\chi^{\ 2} \ / \ ( \ \# \ d.f. \ = \ 2 \ ) \ = \ 0.887
%\vspace*{0.5cm} \\
\end{equation}
with a satisfactory $\chi^{2} =  0.887$ (for n.d.f. = 2). 
The range of values for S obtained in the fit as well as
the reasonable agreement with the derived rates in Ref. \cite{amsams}
is compatible % | given all approximations besides the errors |
with the (approximately) identical mixing of {\it both}
scalar and pseudoscalar nonets according to our assignments.
\begin{table}[ht]
\[  \begin{array}{cccc} 
\hline
  \mbox{ratio} & \mbox{data} & S \ = \ 0.352 & S \ = \ 0.5
  \\ \hline %\vspace*{-0.4cm}
%R_{\ 1} &  \ 0.195 \pm \ 0.075 & \ 0.131 & \ 0.125
  R_{1} &  \ 0.195 \pm \ 0.075 & \ 0.101 & \ 0.083
  \\ & & & \vspace*{-0.4cm} \\
%R_{2} &  \ 0.320 \pm \ 0.114 & \ 0.337 & \ 0.347
  R_{2} &  \ 0.320 \pm \ 0.114 & \ 0.271 & \ 0.333
  \\ & & & \vspace*{-0.4cm} \\
%R_{\ 3} &  \ 0.138 \pm \ 0.038 & \ 0.144 & \ 0.125
  R_{3} &  \ 0.138 \pm \ 0.038 & \ 0.140 & \ 0.083
\\ \hline
  \end{array}\]
\caption{Reduced rate ratios $R_i$ as defined in (\protect\ref{eq:62b}):
experimental determinations by Amsler \protect\cite{amsams} 
and predictions for different strange quark amplitudes $S$ according to
 (\protect\ref{eq:62c}).}
\label{R123}
\end{table}

In particular we should note the large rate $R_2$ for 
 the $\eta \eta^{'}$ decay mode, which
strengthens the octet assignment of the $f_0(1500)$ (for a flavor singlet
-- the glueball in Table \ref{flavor} -- 
this decay mode would disappear in the flavor symmetry limit $S=1$).
On the other hand, the 
ratio $R_3$ is rather small which is in contradiction to a 
pure $s\overline s$
assignment looked for in some classification schemes. The smallness of $R_3$
is now naturally explained by the negative
interference between the nonstrange and strange components of the $f_0(1500)$
in Table \ref{flavor} which would actually be fully destructive 
in the $SU(3)_{fl}$ symmetry limit.\\

\subsection{Decays into two photons}

In this paragraph we focus first on the $f_0(980)$, 
which according to our hypotheses is a $q \overline{q}$ resonance (mainly).
We distinguish the $q\overline{q} $ compositions of $f_0$
again according to the
two cases I) -- dominantly singlet -- and II) -- nonstrange -- 
discussed before
and include the third alternative -- $f_0 \ \sim \ \overline{s} s$ --
denoted T) , historically in the foreground, as proposed
again recently by T\"{o}rnqvist \cite{Torn}. For the decay into two photons
we compare with the neutral component of the isotriplet $a_0(980)$. 

%where we neglect for the moment the deviation (\ref{eq:58}) from the pure
%singlet case in I).
Disregarding sizable glueball admixtures to $f_0$ the decay amplitude
to two photons becomes proportional to the characteristic factor
well known from the corresponding decays of $\pi^{0}, \ \eta ,\ \eta^{\ '}$
involving the quark charges in units of the elementary one and proportional
to the number of colors; for a state with quark composition 
$ (c_u,\ c_d,\ c_s)$ one obtains
\begin{equation}
  \label{eq:40}
%  \begin{array}{l} 
  S_{\ \gamma \gamma} \ = 
  \ 3 \ \sum_{\ q = u,d,s} \ c_q \ Q_{\ q}^{\ 2}.
%\vspace*{0.3cm} 
\end{equation} 
% \begin{array}{c|cccc} 
%   &  I) & II) & T) & {\bf a^{\ 0}}
%   \\ \hline \vspace*{-0.3cm} \\
%   2 \ S_{\ \gamma \gamma}^{\ 2} &
%   \frac{24}{9} \ = \ 2.67 & \frac{25}{9} \ = \ 2.78 & \frac{2}{9} \ = \ 0.22 & 1  
%  \end{array}
%  \end{array}
%\end{equation}
%
\noindent
Then we obtain for the ratio of the two photon decay widths of $f_0$ and $a_0$
 with (practically) the same phase space 
\begin{equation}
  \label{eq:41}
 % \begin{array}{l} 
  R_{\gamma\gamma} \ = \ 
  \frac{ \Gamma \ ( f_0\ (980) \ \rightarrow \ \gamma \gamma \ )}
  { \Gamma \ ( a_{ 0} \ (980) \ \rightarrow \ \gamma \gamma \ )}\ ;
   \qquad 
    R_{\gamma\gamma} \ =  \ 2 \ S_{\ \gamma \gamma}^{\ 2}.
\end{equation}
The predictions for the various mixing schemes are given in Table
\ref{eq:39}.

%\vspace*{0.3cm} \\
%  \begin{array}{c|ccc} 
%   &  I) & II) & T) 
%   \\ \hline \vspace*{-0.3cm} \\
%   R  &
%   \frac{24}{9} \ = \ 2.67 & \frac{25}{9} \ = \ 2.78 & \frac{2}{9} \ = \ 0.22  
%  \end{array}
%  \end{array}
%\end{equation}
%
%
\begin{table}[ht]
\[
  \begin{array}{lllccccl}
\hline \vspace*{0.1cm}
 \mbox{case}&&   & c_{\ u}   &      c_{\ d} & c_{\ s} & 
  R_{\gamma\gamma}  & =  \ 2 \ S_{\ \gamma \gamma}^{\ 2} 
   \\ \hline  \vspace*{0.1cm}
 f_0(980)& (\mbox{Ia})  & \mbox{no mixing}
          & \frac{1}{\sqrt{3}} & \frac{1}{\sqrt{3}} & \frac{1}{\sqrt{3}}
                     & \frac{24}{9} & = \ 2.67 \\
         & (\mbox{Ib})  & \eta-\eta'\ \mbox{mixing}
           & \frac{1}{\sqrt{6}} & \frac{1}{\sqrt{6}} & \frac{2}{\sqrt{6}} 
                     & \frac{49}{27} &= \ 1.815 \\
         &  (\mbox{II})  & \mbox{OZI-mixing}
         & \frac{1}{\sqrt{2}} & \frac{1}{\sqrt{2}} &        0        
                     & \frac{25}{9} & = \ 2.78   \\
         &  (\mbox{T})  & \mbox{pure}\ s\overline s  
            &         0          &         0          &        1    
                    &  \frac{2}{9} &= \ 0.22     \\          
   a_0(980) & & 
           & \frac{1}{\sqrt{2}} & - \frac{1}{\sqrt{2}} &        0 
             & 1 & \vspace*{0.1cm} \\
\hline 
 \end{array}
 \]
\caption{Two photon branching ratio $R_{\gamma\gamma}$ defined in 
Eq. \protect(\ref{eq:41}) for different 
mixing schemes according to the quark composition  $c_q$.}
\label{eq:39}
\end{table}
\noindent
The Particle Data Group gives for the $a_0 (980)$ and the $f_0 (980)$
\begin{equation}
%  \begin{array}{l} 
\label{gamma2g}
  \begin{array}{lll} 
  \Gamma \ ( \ a_0 \ \rightarrow \ \gamma \gamma \ )
  & = & \left ( \ 0.24^{\ + 0.08}_{\ - 0.07} \ \right ) \ \mbox{keV}
  \ / \ B ( \  a_0 \ \rightarrow \ \eta \pi \ )
\vspace*{0.3cm} \\
  \Gamma \ ( \ f_0 \  \rightarrow \ \gamma \gamma \ )
  & = & ( \ 0.56 \ \pm \ 0.11 \ ) \ \mbox{keV}
  \end{array}
\end{equation}
and therefore
\begin{equation}
R_{\gamma\gamma} \ = \ ( \ 2.33 \ \pm \ 0.9 \ ) 
  \ B( a_0 \ \rightarrow \ \eta \pi \ ).
  \label{eq:42}
\end{equation}
\noindent
The branching fraction $B( \ a_0 \ \rightarrow \ \eta \pi \ )$
is not determined satisfactorily because of conflicting analyses
by Bugg et al. \cite{Bug} , Corden et al. \cite{Cord} 
and Defoix et al. \cite{Defoix}, but the PDG classifies the $\eta\pi$
mode as ``dominant''.

%These results get slightly 
%modified if we take into account the small singlet - octet
%mixing (\ref{eq:58}) as in the pseudoscalar sector suggested above.
%The ratio of two photon branching fractions, defined in
%Eq. (\ref{eq:41}) is then obtained by taking $c_q$ in (\ref{eq:40})
%from (\ref{eq:56}) as follows
%%
%\begin{equation}
%  \label{eq:59}
%  \begin{array}{l} 
%  R \ = 
%  \ \Gamma \ ( \ {\bf f}_{0} \ (980) \ \rightarrow \ \gamma \gamma \ ) \ /
%  \ \Gamma \ ( \ {\bf a^{\ 0}} \ \rightarrow \ \gamma \gamma \ )
%   \ \rightarrow \ \frac{49}{27} \ = \ 1.815.
%%
%%\vspace*{0.3cm} \\
%%\leftrightarrow \hspace*{0.3cm}
%R \ = \ ( \ 2.33 \ \pm \ 0.9 \ ) 
%  \ Br \ ( \ {\bf a^{\ 0}} \ \rightarrow \ \eta \pi \ )
%  \end{array}
%\end{equation}
We conclude from the measurement in Eq. (\ref{eq:42}) 
that case (T) with pure  $\overline{s} s$
composition of $f_0(980)$ is untenable. On the other hand,   
it becomes obvious that a distinction 
between alternatives (Ia), (Ib) and (II) by these measurements would need a
considerable increase in experimental precision.

Finally, we derive the corresponding prediction for the $f_{0}(1500)$
assuming its 
 $q \ \overline{q}$ composition according to Eq. (\ref{eq:60}), 
identical to its $\eta$ pseudoscalar counterpart.
Concerning the deviations from the Gell-Mann-Okubo mass square formula
in Eq. (\ref{eq:21b}) we refer to the well known stability of the
corresponding relation for the pseudoscalar nonet, with similar singlet
octet mixing angle. 
\vspace*{0.1cm}

%Neglecting any dynamical differences induced by the large mass
%difference between $f_{0} (980)$ and $f_{0} (1500)$ except
%for phase space 
We obtain for the
ratio of decay widths
into $2 \ \gamma$ 
\begin{equation}
  \label{eq:61}
  R_{\gamma\gamma}^{\ '} \ = 
  \frac{ \Gamma \ ( \ f_0 \ (1500) \ \rightarrow \ \gamma \gamma) }
  { \Gamma \ ( \ f_0 \ (980) \ \rightarrow \ \gamma \gamma \ )}
\end{equation}
 and the individual decay width of the $f_0\ (1500)$
 the following predictions
%\footnote{We assume here again that the 
%kinematic factor $p^3$ for this E1 transition is 
%reduced to $p^1$ at this high energy.}
\begin{equation}
  \label{rprpr}
\begin{array}{l}
  R_{\gamma\gamma}^{\ '} \ = 
 {\displaystyle \ \frac{32}{49} \  
\left (   \frac{ m \ ( \ f_{0} \ (1500) \ )}{ m \ ( \ f_{0} \ (980) \ )}
   \right )^p
%      \ \left ( \ \begin{array}{c} 
%      m \ ( \ f_{0} \ (1500) \ ) 
%        \vspace*{0.2cm} \\
%        \hline \vspace*{-0.2cm} \\
%      m \ ( \ f_{0} \ (980) \ ) 
%      \end{array}
%      \ \right )^{\ 3}
%   \ \frac{18}{49}
% \ \sim \ 2.3
}
\vspace*{0.3cm} \\
%\Rightarrow \hspace*{0.3cm}
  \Gamma \ ( f_0 \ (1500) \ \rightarrow \ \gamma \gamma \ )
   \ \sim \ 0.3 \ (\ 0.1 \ldots 0.6\ ) \ \mbox{keV}
\end{array}
\end{equation}
In Born approximation the power in (\ref{rprpr}) would be $p=3$ and this
power seems appropriate for the light pseudoscalars. At the higher energies,
formfactor effects (typically\footnote{The
transition amplitude should contain the nonperturbative constant
$\langle 0|\overline q q | f_0\rangle=m_0^2$, then $p=-3$ follows for
dimensional reasons.
Experimental data on the $\gamma\gamma$ decays of
the tensor mesons are consistent with $p\sim -3$
assuming ideal mixing.} 
$p=-3$)
become important. In (\ref{rprpr}) we give our best estimate, 
the lower limit  corresponds to  $p=-3$, 
the upper one corresponds to simple phase space with $p=1$.

The branching ratios into two photons have also been considered in the model
by Klempt et al. \cite{klempt}. Their $f_0(1500)$ with mixing angle
$\sin\phi=-0.88$ is 
 very close to the octet state ($\sin\phi=-0.82$), yet closer than in our
phenomenological analysis with  $\sin\phi=-0.58$.
Then they obtain for the above ratio 
$ R_{\gamma\gamma}^{\ '} \sim  0.086$. The results depend strongly
on the mixing angle as $ R_{\gamma\gamma}^{\ '}$  has a
nearby zero at  $\sin\phi = -0.96$,
corresponding to the mixture $(1,\ 1,\ -5)/\sqrt{27}$.
For a pure octet assignment we would obtain 
$\Gamma \ \sim \ 0.08 \ (0.03 \ldots 0.17)$ keV
instead of  (\ref{rprpr}).

It appears possible, that the $2 \ \gamma$ mode of $f_{0} \ (1500)$
can be detected in the double Bremsstrahlung reaction
$e^{\ +} \ e^{\ -} \ \rightarrow \ e^{\ +} \ e^{\ -} \ f_{0} \ (1500)$.
%
%\begin{equation}
%  \label{eq:62}
%  \begin{array}{l} 
%e^{\ +} \ e^{\ -} \ \rightarrow \ e^{\ +} \ e^{\ -} \ 2 \ \gamma_{\ V} 
%\ \rightarrow 
%\ e^{\ +} \ e^{\ -} \ f_{0} \ (1500).
%\ \rightarrow \ e^{\ +} \ e^{\ -} \ \eta \ \eta^{\ '} 
%  \end{array}
%\end{equation}
%
A first search by the ALEPH collaboration at LEP \cite{ALEPHgg} did not show
any $f_0(1500)$ signal. However, no clear signal of $f_0(980)$ has been
observed either although this process is well established. It appears that
the statistics is still too low. Also other
 decay modes such as $\eta\eta$ and $K\overline K$ 
of the  $f_{0} \ (1500)$ look promising to be studied.

\subsection{Relative signs of decay amplitudes}

Besides the branching ratios of the states
into various channels the relative signs of their
couplings is of decisive importance. They can also be deduced from Table
\ref{flavor}. The S-wave phases discussed in Sect. \ref{spectr} in the
mass range above 1 GeV 
are determined with respect to the phase of the
leading $f_2(1270)$ resonance which
is a nearly nonstrange $q\overline q$ state. In this case 
($\phi\approx 0$)
the coupling to all decay channels in Table \ref{flavor} is positive.

For the states discussed here we obtain the signs in Table \ref{signs}.
The predictions for the $f_0(1500)$ are in striking agreement with the data
on inelastic $\pi\pi$ scattering discussed in the previous section,
as can be seen from Fig. \ref{gbfig4}. The resonance loop
in the $K\overline K$ channel is oriented ``downwards'' in
opposite direction to the one in $\eta\eta$ and also opposite to the
$f_2(1270)$ resonance defined as ``upward''. 
This is consistent with our assignment
$(1,1,-1)/\sqrt{3}$ in (\ref{eq:60}) for the $f_0(1500)$. It is not
consistent in particular with the expectations for a glueball which would have
positive couplings to all decay channels. 
\begin{table}[ht]
\[
  \begin{array}{lccc}
\hline
  \mbox{decay} & \ f_0(980) & f_0(1500) & \mbox{glueball}
  \\ \hline
  K\overline{K} & + & - & + \\
  \eta \eta  & + & + & + \\
\hline
    \end{array} \]
\caption{Signs of  amplitudes for the decays of scalar states
into $ K\overline{K}$ and $\eta \eta$ relative to the  $f_2(1270)$.}            
\label{signs}         
\end{table}

As for the other two states in Table \ref{signs} we have only the small window
above the $K\overline K$ threshold.
The amplitude in this region  is composed of the tail of the $f_0(980)$ and the
``background'', i.e. the supposed glueball  state
according to our hypothesis. We note
that in both channels the amplitude has a qualitatively similar behaviour  
in accord with the expected positive signs of all components. 
At present we have no quantitative model for the absolute
phase around 230$\dgr$ for the superposition of these two states.

%-----------------------------------------------------------------

%\input{gb6v2.tex}

\section{The lightest glueball}

Adopting the {\em phenomenological hypotheses}  a) - c) 
in Sect. 1
we have exhausted in the previous analysis 
all positive parity mesons
in the PDG tables below 1600 MeV with the {\em notable} exception of the scalar
$f_{ 0}  (400-1200)$ and also 
the $f_0(1370)$ which we did not accept as standard
Breit-Wigner resonance.
We consider the spectrum in Fig. \ref{gbfig3} (the ``red dragon'') with the
peaks around 700 and 1300 MeV and possibly
with another one above 1500 MeV
as a reflection of a single very broad object (``background'') which interferes with
the $f_0$ resonances. In elastic $\pi\pi$ scattering this ``background''
is characterized by 
a slowly moving phase which passes through $90\dgr$ near 1000 MeV 
if the $f_0(980)$ is subtracted (see, for example \cite{mp}).
 This ``background'' with a slowly moving phase is also observed in the 1300
MeV region in the inelastic channels $\pi\pi\to \eta\eta,\ K\overline K$
as discussed above.
It is our hypothesis that this very broad object which couples to the 
$\pi\pi$, $\eta\eta$ and $K\overline K$ channels is
the lightest glueball 
\begin{eqnarray}
f_0(400-1200), \ f_0(1370)& \to & gb_0 (1000) \label{Gdef} \\
    \Gamma[gb_0(1000)]& \sim &  500-1000\ \mbox{MeV}.   \nonumber
\end{eqnarray}
The large width is suggested by the energy dependent fits in Table
\ref{tabres}. From the speed of the phase shift $\delta_0^0$ near 1000 MeV
-- $\frac{d\delta^0_0}{dm} \simeq 1.8$ GeV$^{-1}$ after the $f_0(980)$ effect 
has been subtracted out as in \cite{mp} -- one finds
using Eq. (\ref{gamma}) the larger value $\Gamma(gb_0)\sim 1000$ MeV.
The glueball mass (\ref{Gdef}) corresponds to alternative 1)  
%-- $gb_{ 0 , \infty} \ (0^{ +  +})$ -- with mass
($m_{gb_0, \infty} \ \lesssim \ m_{ a_0} \ \sim$ 1 GeV)
 as described at the beginning of Sect.
\ref{sec2}.

 We do not exclude some mixing with the scalar nonet
states but it should be sufficiently small such as to preserve the main 
characteristics outlined in the previous section.
In the following we will investigate whether our glueball assignment
for the states in (\ref{Gdef})
is consistent with general expectations.

\subsection{Reactions favorable for glueball production}
We first examine the  processes in which particles are expected to be
produced from gluonic intermediate states.
%\begin{description}

\noindent
 {\it (a)} $pp\to pp \ +\ X_{\rm central}$\\
In this reaction the double pomeron exchange mechanism should be favorable 
for the production of gluonic states.
A prominent enhancement at low $\pi\pi$ energies is observed
\cite{akesson,gamspp}
and can be interpreted in terms of the elastic $\pi\pi$ phase shifts 
\cite{amp,mp}. 

\noindent
 {\it
(b)  Radiative decays of $J/\psi$}\\
For our study of scalars the most suitable final states
 are those with the odd waves
forbidden. The simplest case is  $J/\psi\to \gamma \pi^0\pi^0$ which has been
studied by the Crystal Ball Collaboration \cite{crballpi}. The mass
spectrum shows a prominent  $f_2(1270)$ signal
 but is sharply cut down for masses below 1000 MeV
and the presentation ends at $m_{\pi\pi}\sim 700$ MeV. This cutoff in the mass
spectrum is not due to the limited detector efficiency 
 which is flat over the full mass region 
down to $\sim$ 600 MeV and drops sharply only below this mass value \cite{ral}. 

An incoherent background is fitted in \cite{crballpi} under the $f_2$ peak
which reaches the fraction 1/7.5 at the maximum of the peak. This is not
much smaller than 1/(5+1) expected for the S-wave from the counting of spin
states. No data have been presented on the azimuthal angle distribution
which would allow to estimate the amount of S-wave. 
No further hint can be obtained from
 the $\pi^+\pi^-$ channel analysed by
the Mark III collaboration \cite{Mk3a} because of the larger background.

It appears that -- contrary to our expectation -- there is no low mass
enhancement around 700 MeV in this channel related to the glueball; 
its production with higher mass of around 1300 MeV is not inconsistent with
data. For the moment we have no good explanation for the low mass
suppression.

\noindent
 {\it (c) Transition between radially excited states}.\\
\noindent
The radially excited states $\psi'$, $Y'$ and $Y''$ can decay by two gluon
emissions into the heavy quark ground state and give therefore rise to  
the production of gluonic states. The observed $\pi\pi$ mass spectra 
can be described consistently using
the elastic $\pi\pi$ $S$-wave phase shifts \cite{amp} although the
calculation is not very sensitive to their detailed behaviour. 
Another example of this kind  is the decay of the $\pi(1300)$, presumably a
radial excitation of the stable $\pi$; its decay mode into
$(\pi\pi)_{S-wave}\pi$ is seen \cite{Aaron}.
%\end{description}

\noindent
Finally, we comment on the different production phases of the glueball
amplitude with respect to the $f_0(980)$ discussed in Sect. \ref{spectr2}.
In most inelastic reactions the $f_0(980)$ appears as peak above the background (case b)
which is consistent with the phases of the decay amplitudes
for this state and the glueball to be
the same as expected from
 Table \ref{flavor}. The dip occurs in elastic scattering 
(case a) where  a peak is not allowed as the background is already
near the unitarity limit. In two reactions (case c) the large asymmetry
in the mass spectra  suggests a background out of phase by $90\dgr$ 
with respect to the $f_0(980)$ Breit-Wigner amplitude which may be a hint to
different production phases.

\subsection{Flavor properties}
Here we list a few observations which may give a hint towards the flavor
composition along the lines discussed for the $q\overline q$ nonet.\\
 
\noindent {\it Glueball production in $p\overline p$ annihilation}\\
The Crystal Barrel Collaboration has observed the $f_0(1370)$ 
in the processes
\begin{equation}
p\overline p \ \to\ f_0(1370)\ \pi^0;\quad f_0(1370)\ \to \ K_LK_L,\
\eta\eta \label{f1300dec}
\end{equation}
where clear peaks in the respective mass spectra have been seen.
The theoretical expectation for the ratio of reduced branching ratios
assuming $f_0(1370)$ to 
decay like a glueball according to Table \ref{flavor} is obtained as
\begin{equation}
R_g\ = \ \displaystyle \frac{\Gamma_{red}(f_0(1370)\to\eta\eta)}
   {\Gamma_{red}(f_0(1370)\to K\overline K)}
\ = \frac{(2+S)^2}{9 (1+S)^2}.
 \label{f1300r}
\end{equation} 
From the results  summarized by
Amsler \cite{amsams} we derive the quantity (\ref{f1300r}) after correction
for phase space and unseen $K\overline K$ decay modes 
%(see Table \ref{rgtab}).
%\begin{table}[ht]
%\[
%  \begin{array}{cccc}
%\hline
%     \mbox{exp. results}  &S  =  0 & S  =  0.5 & S =  1.0
%  \\ \hline %\vspace*{-0.6cm}  %-0.4
% % \\ & & &  &\\
%  R_g  \sim  0.44
%& \ .44 & \ .31 & \ .25\\
%\hline
%%  \\& & &  &\vspace*{-0.4cm} 
%    \end{array}  \]
%\caption{
%The ratio $R_g$ in (\ref{f1300r}) as determined from experiment 
%\protect\cite{amsams} and predicted for different strange quark amplitudes
%$S$.}
%\label{rgtab}
%\end{table}
\begin{equation}
\mbox{exp. result}: \qquad\qquad R_g \mbox{          } \sim  0.44.
\mbox{          } \qquad\qquad
\label{rgexp}
\end{equation}
This number is to be compared with the theoretical expectations for different 
 strange quark amplitudes $S$
\begin{equation}
\mbox{theor. result}:\qquad S=(0,\ 0.5,\ 1.0):\qquad R_g=(0.44,\ 0.31,\
0.25).
\label{rgth}
\end{equation}
The value extracted from the measurements
 is somewhat larger than expected but looking at the
difficulty to extract such numbers experimentally we consider the result
as encouraging.
Similar results for the $f_0(400-1200)$ cannot be extracted from the data in
Ref. \cite{amsams} because of the overlap with nearby other states.\\

\noindent{\it $J/psi$ decay into glueball +  vector mesons}\\
In analogy with the flavor  analysis  of the $f_0$ states above we 
now proceed
with our glueball candidate. In the final state $\phi \pi\pi$ DM2 observes
indeed a broad background under the $f_0(980)$ which  extends towards small
masses in the $\pi\pi$ invariant mass 
\cite{falvard}. On the other hand the mass spectrum in the 
final state $\omega \pi\pi$ looks very different with a peak at low masses 
around 500 MeV \cite{Augustin5pi}. Similar results are also seen by Mark-III
\cite{Lockman}.

If the low mass bump in the $\omega \pi\pi$ 
final state is a real effect and not due to background%
\footnote{In Ref. \cite{Lockman} the important background from
$\phi\eta,\ \eta\to 2\pi+\pi^0$ has been emphasized; it could also appear in
the $\omega$ channel}
it requires
quite a different dynamics in the two vector meson channels. One possibility
is the suppression of low mass $\pi\pi$ pairs from the decay of an
$s\overline s$ pair because of the heavier $s$-quark mass.
This problem could be avoided by a restriction of the
comparison to the mass region above 1 GeV.

\subsection{Suppressed production in $\gamma\gamma$ collisions}
If the mixing of the glueball with $q \overline q$ 
states is small then the same is true for  the two photon coupling.
We consider here the processes
\begin{equation}
\text{(a)}\quad \gamma\gamma\to \pi^+\pi^-\qquad 
\text{(b)}\quad \gamma\gamma\to \pi^0\pi^0. \label{ggpipi}
\end{equation} 
and distinguish two regions for the  mass $W\equiv m_{\pi\pi}$.\\

\noindent {\it Low energies $W\lesssim 700$ MeV}\\ 
The process (a) is dominated by the Born term with pointlike 
pion exchange. This
contribution is avoided in process (b) and the remaining cross section
is smaller by one order of magnitude in the same mass range; furthermore, it
is also very small compared to the dominant cross section at the $f_2(1270)$
resonance position. The reaction (b) has been studied by the Crystal Ball
\cite{Cball} and JADE \cite{JADE} Collaborations. 

We compare  the cross section in
$\gamma\gamma\to \pi^0\pi^0$ 
%(b) % from \cite{Cball}  
and in the isoscalar elastic $\pi\pi$ scattering
near the peak at $W\sim 600$ MeV. Only in the
second reaction the glueball should have a sizable coupling. We normalize
both cross sections to the $f_2(1270)$ meson peak representing a well
established $q\overline q$ state and obtain
\begin{eqnarray}
\displaystyle
R_\gamma & = & \frac{\sigma_{\gamma\gamma}(W_1=600\ \text{MeV})}
    {\sigma_{\gamma\gamma}(W_2=1270\ \text{MeV})} \nonumber \\
         &\simeq & 0.067 \label{Rgamma}\\
R_\pi & = & \frac{\sigma_{\pi\pi}^S(W_1=600\ \text{MeV})}
    {\sigma_{\pi\pi}^D(W_2=1270\ \text{MeV})}\ = 
    \ \frac{1}{5x_f}\frac{W_2^2}{W_1^2} \nonumber \\
        &\simeq & 1.05 \label{Rpi}.
\end{eqnarray}
Here we used for $R_\gamma$ the data from \cite{Cball} 
and for the $\pi\pi$  S-wave the cross section at the unitarity limit and 
the same for the $f_2$ meson in the D-wave but with elasticity
$x_f=0.85$. The ratios in (\ref{Rpi}) demonstrate that 
the low mass S-wave production in
 $\gamma\gamma$ collisions is suppressed by more than an order of
magnitude in comparison to  $\pi\pi$ collisions. 
 
The size of the cross sections in both charge states in 
(\ref{ggpipi}) can be understood \cite{Cball,mpgamgam}
by including the Born term in (a) only and a
rescattering contribution in both processes. So one can interpret the
reaction (b) as a two step process, first the two photons couple to charged
pions as in (a) then rescattering by charge exchange  $\pi^+\pi^-\to
\pi^0\pi^0$:
in this picture the photons do not
couple ``directly'' to the 
 ``quark or gluon constituents'' of the broad structure at 600 MeV
but only to the initial 
charged pointlike pions. This is at the same time a minimal
model for the production of the  
bare glueball according to our hypothesis without direct
coupling of the photons to the glueball state.\\

\noindent{\it Mass region around the $f_2(1270)$}\\
 One may look again for the presence of an S-wave state.   
The measurement of the angular distribution does not allow in general a
unique separation of the $S$-wave from the $D_\lambda$-waves 
in helicity states $\lambda=0$ and 
$\lambda=2$. It turns out, however, that the data 
are best fitted 
in the mass region $1.1 \leq W \leq 1.4$ GeV by the $D_2$ wave alone without
any $S$ and $D_0$ wave included \cite{Cball,Markii,JADE}. A restriction 
on the spin 0 contribution has been derived at the 90\% confidence limit
in \cite{Cball} as  
\begin{equation}
\displaystyle
\frac{\sigma_{\gamma\gamma}(\text{spin 0})}
{\sigma_{\gamma\gamma}(\text{total})}\ <\ 0.19
 \quad \text{for} \quad 1.1\ \leq \ W\ \leq \ 1.4\ \text{GeV} \label{spin0lim}
\end{equation}
which turns out not yet very restrictive. Taking all three experiments 
together a suppression of S wave under the $f_2(1270)$ is suggested.

In summary, the production of the broad S-wave enhancement
 is suppressed in $\gamma\gamma$
in comparison to $\pi\pi$ collisions, and this is very clearly seen
at the low energies. 
This we consider as a strong hint in favor of
our hypothesis of the mainly gluonic nature of this phenomena
both at low and high energies.
Clearly, the study of scalar states in $\gamma\gamma$ collisions will be of
crucial importance for the determination of their flavor content 
and classification into multiplets.

\subsection{Quark-antiquark and gluonic components in $\pi\pi$ scattering}

In our picture, the elastic  $\pi\pi$ scattering amplitude in the energy
region below $\sim 2$ GeV is not saturated by $q\overline q$ resonances
in the $s$- and $t$-channel alone.\footnote{
This would follow with "one-component-duality" between direct channel
resonances and $t$-channel Regge-poles as, for example, realized in the
Veneziano model \cite{veneziano} or, alternatively, in resonance pole
expansions in both channels simultaneously, as in  
 \cite{iiz}, or, more recently, in \cite{zoubugg}.} % end footnote
There is a second
component which corresponds to Pomeron exchange in the $t$-channel -- dual to
the so-called ``background'' in the $s$-channel. This dual picture with two
components, suggested by Freund and Harari 
\cite{fh}, has been very successful in the interpretation of the
$\pi N$ scattering data. 

In case of
the $\pi\pi$-interaction a similar situation was found by Quigg~\cite{quigg}: 
whereas the $I_t=1$ $t$-channel exchange amplitude 
can be saturated
by  $q\overline q$ resonances, 
the $I_t=0$ amplitude obtains a contribution of about
equal magnitude from the ``background'' as well. This background 
is present already in the low energy region around 1 GeV and is seen 
clearly in the S-wave amplitude 
corresponding to  $I_t=0$ \cite{quigg}; it also governs the exotic
$\pi^+\pi^+$ channel. 

The Pomeron 
exchange is naturally related to glueball exchange. 
Then, we consider a third component, obtained by crossing, with glueball
intermediate states in the s-channel and exotic four quark states 
in the t-channel. Indeed, the $\pi\pi$  
$I_t=2$ exchange amplitude in \cite{quigg}
shows resonance circles with little background and therefore could
correspond to
a glueball amplitude after appropriate averaging. This third component
with exotic exchange 
is expected to drop yet faster with energy than the $q\overline q$ resonance 
exchange amplitude.
We consider the phenomenological results on the low energy ``background''
 \cite{quigg} 
 as a further independent hint towards a gluonic component in the low energy
$\pi\pi$ scattering. 
  
%------------------------------------------------------------------
%\input{gb7v4.tex}
%gb7v4.tex
\section{Completing the basic triplet of gauge boson binaries}
\label{basic_triplet}
After we found the candidate for  $gb\ (0^{++})$ at $\sim 1$ GeV
we expect, as discussed in Sect. \ref{sec2}, 
the two remaining members of the basic triplet with $J^{PC}$
quantum numbers $0^{-+}$ and $2^{++}$ to be heavier than $gb\ (0^{++})$ and 
to exhibit a much smaller width because of the reduced strength of the
interaction (coupling $\alpha_s$) at higher mass
\begin{equation}
  \label{eq:66}
  \begin{array}{ll} 
  g b \ (0^{-+}): \hspace*{0.3cm}
  m_{ 2} \ > \ 1 \ \mbox{GeV}
  \hspace*{0.3cm} , \hspace*{0.3cm} &
  \Gamma_{ 2} \ \ll \ 1 \ \mbox{GeV};
  \vspace*{0.3cm} \\
  g   b \ (2^{++}): \hspace*{0.3cm}
  m_{ 3} \ \gsim \ m_{ 2} %1 \ \mbox{GeV}
  \hspace*{0.3cm} , \hspace*{0.3cm}&
  \Gamma_{ 3} \ \ll \ 1 \ \mbox{GeV}.
  \end{array}
\end{equation}

\noindent
Thus we are looking for two resonances, the width of which make them appear
much more similar to their prominent, relatively narrow $q \ \overline{q}$ 
counterparts. The mass range is tentatively set to $1 \ - \ 2$ GeV. 

We search for possible candidates in radiative $J/\psi$ 
decay, on which we focus next.
To this end we list in Table \ref{psirad} 
the most prominent radiative decay modes of 
$J/\psi \ \rightarrow \ \gamma \ X$
into a single resonance $X$ without charm content.
\begin{table}[ht]
\[  \begin{array}{l}
\hline 
  \begin{array}{l@{\hspace*{0.4cm}}l@{\hspace*{0.4cm}}r@{\hspace*{0.4cm}} 
  c@{\hspace*{0.4cm}}c} 
  & \mbox{name}\ (X)  & B(J/\psi\rightarrow \gamma X)\times 10^{3} 
        & \mbox{partial }B & \mbox{mode}
  \\ \hline 
  1 & \eta^{'}\ (958)      & 4.31 \ \pm \ 0.30 & &
  \vspace*{0.1cm} \\
  2 & \eta \ (1440) & > \ 3.01 \ \pm \ 0.44 &  &  
  \vspace*{0.1cm} \\
    &               &      &  1.7 \ \pm \ 0.4 & \varrho^{0} \varrho^{0}          
  \vspace*{0.1cm} \\
    &               &  & 0.91 \ \pm \ 0.18 & K \overline{K} \pi    
  \vspace*{0.1cm} \\
    &               &      &  0.34 \ \pm \ 0.07 & \eta\pi^+\pi^-
  \vspace*{0.1cm} \\
    &               &  & 0.064 \ \pm \ 0.014 & \gamma \varrho^{0}             
  \vspace*{0.1cm} \\
  3 & f_{4} \ (2050) & 2.7 \ \pm \ 1.0 & & \pi \pi
  \vspace*{0.1cm} \\
  4 & f_{2} \ (1270) & 1.38 \ \pm \ 0.14 & &  \pi \pi  
  \vspace*{0.1cm} \\
  5 & f_{J} \ (1710) & 0.85^{+1.2}_{-0.9} & & K \overline{K} \\
\hline
\end{array}
\end{array}
\]
\caption{Radiative decay modes of $J/\psi$ into single 
non-$c\overline c$ resonances 
with branching ratios $B \gsim \ 10^{-3}$
according to the PDG \protect\cite{PDG}.}
  \label{psirad}
\end{table}
\noindent
Among the 5 resonances % appearing in $J/\psi$ radiative decays with
%branching fraction $>10^{-3}$ 
we recognize
$\eta(1440)$ as a candidate for $gb\ (0^{-+})$ and $f_J(1710)$ with
spin $J$ either 0 or 2, as a candidate for $gb\ (2^{++})$.  

\subsection{The glueball with $J^{PC}=0^{-+}$}

A state with these quantum numbers is expected to decay into 3
pseudoscalars ($ps$). 
The first experiments on the radiative decays
$J/\psi \ \rightarrow \ \gamma \ 3 \ ps$ were performed by the
MarkII \cite{Mk2a} and Crystal Ball \cite{crball3} collaborations in
the channels $3 \ ps \ = \ K_{s} K^{\pm} \pi^{\mp}$ and
$3 \ ps \ = \ K^{+} K^{-} \pi^{0}$, respectively.

A spin analysis was performed by Crystal Ball \cite{crball3};
it revealed a major intermediary decay mode
\begin{equation}
  \label{eq:79}
  \begin{array}{l} 
\eta  (1440) \ \rightarrow \ a_{0}  (980) \pi \ \rightarrow \ K \overline{K} \pi
\end{array}
\end{equation}
\noindent
and $J^{PC} [\eta(1440)] \ = \ 0^{-+}$. While the branching fraction product
$B \ ( \ J/\psi \ \rightarrow \ \gamma \eta \ (1440) \ )
\times B \ (  \eta \ (1440) \ \rightarrow \ K \overline{K} \pi  )$
was overestimated in Refs. \cite{Mk2a,crball3},
the spin parity assignment was confirmed by Mark-III
\cite{Mk3eta} in the decay mode 
\begin{equation}
  \label{eq:80}
  \begin{array}{l} 
\eta  (1440) \ \rightarrow \ a_{0}  (980) \pi \ \rightarrow \ \eta \pi^{+} \pi^{-}
\end{array}
\end{equation}
\noindent
and by DM2 \cite{Augusteta} in both channels of Eqs. (\ref{eq:79}) and (\ref{eq:80}).
It is therefore natural to associate this state with its large radiative
$J/\psi$ decay mode with the $0^{-+}$ glueball.

On the other hand, in $pp$ and $\pi p$ collisions the central production of
this state is weak in comparison to the leading $q\overline q$ resonances
\cite{omega2} or not resolved at all \cite{Barb1}.

The glueball interpretation has a long 
history of debate \cite{Chan,Close}. Doubts have
been brought up, in particular, in view of 
the results from lattice QCD calculations referred to in Sect. 2
which suggest a heavier mass above 2 GeV. As we discussed there, we feel that
for a justification of such doubts, 
the more complete calculations should be awaited. 
However, because of the near absence in central production, the glueball
interpretation is at a more speculative level at present. 
 
\subsection{The glueball with $J^{PC}=2^{++}$}
\label{fJ1710}

This state is expected to decay into two pseudoscalars. $f_J(1710)$ has
long been a prime candidate. The problem for the classification of
this state was and still is \cite{PDG} the ambiguity 
in the spin assignment $J=0$ or $J=2$. 
In the following, we discuss the results of  spin analyses 
in various
experiments on $J/\psi$ decays and central hadronic collisions
 which will lead us to a definite conclusion concerning the existence of a
$J=2$ state.

\subsubsection{Radiative $J/\psi$ decays}
%Searches in the $\pi\pi$, $K\overline K$ and $\eta\eta$
%channels with overlapping resonances}

\noindent {\it Crystal Ball experiment}\\
The first observation of this state was obtained by 
the Crystal Ball collaboration at the SPEAR storage ring
in '81 \cite{crballeta}
in the decay channel
\begin{equation}
  \label{eq:67}
  \begin{array}{l} 
  J/\psi \ \rightarrow \ \gamma \ \eta \eta.
  \end{array}
\end{equation}
\noindent
The useful sample contained $~50$ events in
the $\eta \eta$ invariant mass range from 1200 - 2000 MeV. 
The resonance parameters were \cite{crballeta}:
\begin{equation}
  \label{eq:68}
  \begin{array}{l} 
  m \ = \ 1640 \ \pm \ 50 \ \mbox{MeV}
  \hspace*{0.3cm} , \hspace*{0.3cm}
  \Gamma \ = \ 220^{\ +100}_{\ -70} \ \mbox{MeV}.
  \end{array}
\end{equation}
\noindent
A spin analysis with respect to the two hypotheses $J = 2$ and $J=0$ was performed
with at least statistical preference of $J^{PC} \ = \ 2^{++}$.
\vspace*{0.1cm}
The same resonance could not be resolved in a significant way by the
same collaboration in the channel 
$J/\psi \ \rightarrow \ \gamma \ \pi^{0} \pi^{0}$ \cite{crballpi}.
The scarcity of events is matched by the scarcity
of precise description of the analysis.

\noindent {\it Mark-III and DM2 experiments}\\
A significant improvement in statistics is next reported by
the Mark-III collaboration \cite{Mk3a} in the channels
\begin{equation}
  J/\psi \ \rightarrow \ \gamma 
 \  \pi^{+} \ \pi^{-}, \quad 
 \gamma \   K^{+} \ K^{-}. \label{eq:69}
\end{equation}
\noindent
We first discuss the results in the $\pi^{+} \ \pi^{-}$ subchannel.
%Extending the marginal results, due to limited statistics, 
%of the crystal ball collaboration \cite{crballpi}, 
The two resonances
$f_2(1270)$ and $ f_J(1710)$ 
are clearly resolved % in Ref. \cite{Mk3a} 
and a 
small indication of $f^{\ '}_{2}  (1525)$ is visible in the
projected $\pi^{+} \ \pi^{-}$ invariant mass distribution.
 A full exposition is given of the
relevant angular acceptances and efficiencies.
Now a fit of four interfering resonances is performed:
\begin{displaymath}
  \begin{array}{l} 
  f_{2} \ (1270) \ , \ f^{'}_{2} \ (1525) \ ,
  \ f_{J} \ (1710) \ , \ f \ (2100).
  \end{array}
\end{displaymath}
The same reaction was investigated by the DM2 collaboration at the
DCI storage ring in Orsay %, in the $\pi^{+} \ \pi^{-}$ channel in 
 \cite{Augustpi} %, in the $K^{+} \ K^{-}$ channel in Ref. \cite{AugustK} 
with rather similar results. The product 
of branching ratios in both experiments is given as
\begin{equation}
 \label{eq:70}  
\begin{array}{l}
\begin{array}{l} 
  B  (  J/\psi \rightarrow \gamma f^{'}_{2}  (1525)  )\ \times 
  \ B  (  f^{'}_{2}  (1525) \rightarrow \pi^{+} \pi^{-}  ): 
\vspace*{0.3cm}\\
  \mbox{Mark-III}  :\ \sim   \ 3 \times 10^{-5};\qquad 
   \mbox{DM2}  :\ (2.5 \pm 1.0 \pm 0.4 ) \ 10^{-5} \mbox{     }
\end{array}
\end{array}
\end{equation}
\begin{equation}
 \label{eq:71}
 \begin{array}{l}
  B  (  J/\psi \rightarrow \gamma f_{J}  (1710)  )\ \times
  \ B  (  f_{J}  (1710) \rightarrow \pi^{+} \pi^{-}  ):  
\vspace*{0.3cm}\\                 
 \mbox{Mark-III}  :\   (  1.6  \pm  0.4  \pm  0.3  ) \ 10^{-4};\quad
 \mbox{DM2}  :\ (  1.03  \pm  0.16  \pm  0.15  ) \ 10^{-4}
  \end{array}
\end{equation}
We remark that both experiments reveal a background of $O ( 20\% )$ in the
$\pi^{+}  \pi^{-}$ channel. This we expect to be -- at least in part -- not
incoherent background, but the S-wave part, including the contribution from
$gb \ (0^{++})$ discussed in the last section.

Next we turn to the $K^{+} K^{-}$ channel with results again from Mark-III
\cite{Mk3a} and DM2 \cite{AugustK}. A full spin analysis
is performed by the Mark-III collaboration
for both invariant mass domains corresponding
to $f^{'}_{2}  (1525)$ and  $f_{J} (1710)$.
The likelyhood functions used to distinguish the two hypotheses
$J = 0$ and $J = 2$ strongly favor the $J = 2$ hypothesis for both resonances.
For the spin 0 assignment to  $f_{J}(1710)$ 
the purely statistical probability 
is estimated to be $2 \times 10^{-3}$ only. % by Mark-III \cite{Mk3a}.
Especially the non-uniform 
polar angle distribution in the resonance decay requires the
higher spin $J=2$.
This confirms the low statistics spin analysis of 
Crystal Ball \cite{crballeta}. 

No spin analysis is performed in this channel by DM2 in Ref. \cite{AugustK}.
However, one can see from the Dalitz plot that the density of points along
the $f_{J}(1710)$-band is peaked towards the edges, again favoring 
the presence of higher
spin. Furthermore,
in the projected $K^{+}  K^{-}$ invariant mass distribution
an interference effect between the two resonances is visible,
without any mention in Ref. \cite{AugustK}. Both phenomena, if
analyzed and eventually confirmed, would yield an independent
indication for the $J = 2$ quantum number of $f_{J}  (1710)$.

%The detection efficiencies for the two main resonances observed
%in the $K^{+} \ K^{-}$ channel by Mark-III turn out to be
%quite different. 
The branching fraction products corresponding to
Eqs. (\ref{eq:70}) and (\ref{eq:71}) are determined as
\begin{equation}
  \label{eq:72}
  \begin{array}{l} 
  B  (  J/\psi \rightarrow \gamma f^{'}_{2} \ (1525)  )\ \times 
  \ B  (  f^{'}_{2}  (1525) \rightarrow K^{+} K^{-}  ): 
\vspace*{0.3cm}\\
  \mbox{Mark-III}  :  \ ( 3.0 \pm 0.7 \pm 0.6 ) \ 10^{-4}; \quad
   \mbox{DM2}  :   \ ( 2.5 \pm 0.6 \pm 0.4 ) \ 10^{-4}.
 \end{array}
\end{equation}
\begin{equation}
  \label{eq:73}
  \begin{array}{l}
  B  (  J/\psi \rightarrow \gamma f_{J}  (1710)  )\ \times 
  \ B  (  f_{J}  (1710) \rightarrow K^{+} K^{-}  ):
\vspace*{0.3cm}\\
  \mbox{Mark-III}  : \  (  4.8  \pm  0.6  \pm  0.9  ) \ 10^{-4}; \quad
  \mbox{DM2}  :  \ (  4.6  \pm  0.7  \pm  0.7  ) \ 10^{-4}
  \end{array}
\end{equation}
From the branching fractions in Eqs. (\ref{eq:70}) - (\ref{eq:73})
we obtain % the -- first for  $f^{'}_{2} \ (1525)$ -- 
the following ratio in comparison with the PDG result: 
\begin{equation}
  \label{eq:74}
%  \begin{array}{l} 
  \begin{array}{llll} 
  \mbox{Mark-III and DM2}: \ & 
{ \displaystyle
    \frac{  B(f^{'}_{2}  (1525) \rightarrow \pi\pi) }
    { B(f^{'}_{2}  (1525) \rightarrow K\overline K )}
} %end displaystyle
 \ & = & 
 \ 0.075 \ \pm \ 0.030 % \\
  \vspace*{0.3cm} \\
  \mbox{PDG}:& \  & = & \ 0.0092 \pm \ 0.0018.
  \end{array}
%  \end{array}
\end{equation}
\noindent
%There is indeed a discrepancy as noted by PDG! 
The obvious discrepancy between both numbers may point 
towards larger
systematic errors in the relative efficiency of the two channels
in  (\ref{eq:69}) and eventually also to errors in the determinations
of $f^{'}_{2}  (1525)$ branching fractions in earlier experiments.
However, we tend to believe that the 
 discrepancy of deduced
branching fractions in Eq. (\ref{eq:74}) is too significant to be
``explained'' by some unknown source of large errors; rather we conclude that 
{\em the peaks `` $f^{\ '}_{2}  (1525)$ '' as seen in radiative decays 
$J/\psi \ \rightarrow \ \gamma \ \pi^{+} \pi^{-} \ , \ K^{+} K^{-}$ 
are not just $ f^{\ '}_{2}  (1525)$.
}
%
%The interpretation is quite clear: the assumption, that the spectral
%enhancement in the neighbourhood of 1500 MeV in the invariant mass distribution
%of two pions or kaons corresponds uniquely to the tensor meson
%
%\begin{displaymath}
%  \begin{array}{l} 
%f^{\ '}_{2}  (1525) \ \mbox{with} \ J^{PC} \ = \ 2^{++}
%  \end{array}
%\end{displaymath}
%
%$f^{\ '}_{2}  (1525)$
%is incorrect. Rather 
There are further states, in particular in the S-wave, which are not
resolved in the analysis. One candidate is $f_0(1500)$, not yet
established at the time of the Mark-III and DM2 experiments under discussion.
Because of the small branching fraction of
$f_{0} (1500)$ into $K\overline K$ deduced by Amsler \cite{amsams}, 
the effect is expected to be especially important in the $\pi\pi$ channel.
Furthermore, there could be contributions from the high-mass tail of the
$0^{++}$ glueball or other states in this partial wave.
Such contributions may also affect the spin determinations of  
$f_J(1710)$.\\
\noindent
{\it Reanalysis of the  $f_{J}  (1710)$ spin by Mark-III}\\
The spin analysis in the $K  \overline{K}$ channel was subsequently
extended by the Mark-III collaboration with higher statistics and including
the $K_sK_s$ final states 
\cite{thesisbo,thesische}. In a mass-independent analysis, both the $J=0$ and
the $J=2$ components have been studied, preliminary results became available
as conference reports \cite{chenrep,chenrep1}. 
In these analyses the earlier Mark-III results  \cite{Mk3a} are contradicted
in favor of a large $J = 0$ component of $f_{J}  (1710)$, although a
contribution of up to 25\% from spin 2 was not excluded.

Looking into these results in more detail, we observe a considerable
qualitative difference between the $K^+K^-$ and the $K_sK_s$ results.
Whereas in the former channel the $J=0$ component dominates over $J=2$ by a
factor 4.5 in the mass range 1600-1800 MeV, the opposite is true for the
neutral kaon mode: in this case, the $J=2$ component dominates by a factor
2.8 over $J=0$. It is interesting to note that the efficiency in the
azimuthal angle $\phi$ is much better in the neutral mode: for $K^+K^-$
pairs the acceptance drops towards its minimum
at $\phi=0,\pi$ to $\sim$15\% of its maximal value, but
for  $K_sK_s$ pairs only to 57\%. Therefore, the 
results from the neutral mode are very important 
despite the somewhat lower statistics.

Breit-Wigner resonance fits to the combined $K\overline K$ 
data sample are
presented in Fig.2a of \cite{chenrep1}. 
In this data compilation, a significant spin 2 component of
$f_J(1710)$ is clearly visible and is comparable in its overall size 
with the $f_2'(1525)$ signal. The fitted curve does not describe 
the data well near $f_J(1710)$ and
underestimates the observed rates by roughly a factor of two.
% 
%We feel that the above form of presenting new
%and relevant results
%cannot be considered as an equivalent of a fully published paper;
%we are unable to assess these sources except for the very short
%account by L.P. Chen \cite{chenrep} with the six small scale figures composing
%his Fig. 7.
In view of the preliminary character of these studies, one might conclude
%that these results constitute
%partial evidence indicating
%, {\em together} with the ones
%discussed in this subsection, 
that both hypotheses $J = 2$ and $J = 0$ 
should be considered.

\noindent {\it BES experiment}\\
The situation became considerably clarified by the recent results of
the BES collaboration \cite{besK}. At the BEBC facility in Beijing,
the decay $J/\psi \ \rightarrow \ \gamma \ K^{+} K^{-}$ was analyzed
with specific determination of all helicity amplitudes for $J = 0,\  2$.
The region around 1700 MeV for the $K^{+} K^{-}$ invariant mass spectrum
-- beyond $f^{\ '}_{2}  (1525)$ -- 
reveals a dominant resonant structure with spin 2. Furthermore, the ana\-lysis
provides evidence indeed for a $0^{++}$ resonance, although  
 weaker  and less significant and at a slightly larger mass value.
 The parameters of the resonance fit
are given in Table \ref{eq:75a}.
\begin{table}[ht]
\[
  \begin{array}{l} 
  \begin{array}{lc@{\hspace*{0.5cm}}cc}
\hline
  J^{\ PC}(X) & \mbox{mass (MeV)} & \mbox{width (MeV)} & 
  \begin{array}[t]{l}
  B \ ( J/\psi \ \rightarrow \ \gamma X \ ) 
  \vspace*{0.1cm} \\
  \times   B \ ( \ X \ \rightarrow \ K^{+} K^{-} \ ) \times 10^{\ 4}
  \end{array}
  \\ \hline \vspace*{0.1cm}
%  & & & \\
  2^{++} & 1696 \ \pm \ 5^{\ +9}_{\ -34} & 103 \ \pm \ 18^{\ +30}_{\ -11} &
  2.5 \ \pm \ 0.4^{\ +0.9}_{\ -0.4} \vspace*{0.1cm} \\
%  & & & \\
  2^{++} & 1516 \ \pm \ 5^{\ +9}_{\ -15} & 60 \ \pm \ 23^{\ +13}_{\ -20} &
  1.6 \ \pm \ 0.2^{\ +0.6}_{\ -0.2} \vspace*{0.1cm} \\
  0^{++} & 1781 \ \pm \ 8^{\ +10}_{\ -31} & 85 \ \pm \ 24^{\ +22}_{\ -19} &
  0.8 \ \pm \ 0.1^{\ +0.3}_{\ -0.1} \vspace*{0.1cm} \\
\hline
  \end{array}
  \end{array}
\]
\caption{Resonance parameters from fit to 
mass regions near $f^{\ '}_{2}(1525)$ and
$f_J(1710)$ as  obtained by the BES collaboration \protect\cite{besK}.}
 \label{eq:75a}
\end{table}
The results on $f^{\ '}_{2}(1525)$ are now in good agreement with the PDG
results. 
In comparison
with the earlier results in (\ref{eq:73})
both the smaller branching ratios 
of $f_J(1710)$ into spin $J=2$ alone and the reduced statistical errors
 are to be noted. 

In comparison with the preliminary Mark-III results \cite{chenrep1},
we note the good agreement with their branching ratio into $f_2(1525)$
of (1.7 $\pm\ 0.3)\times 10^{-4}$ (for $K^+K^-$ mode
as defined in our Table \ref{eq:75a}). The corresponding
fraction for $f_J(1710)$ with $J=2$ reads (1.0 $\pm \ 0.4)\times 10^{-4}$
\cite{chenrep1} -- which we would increase by a factor 2
to $\sim 2.0 \times 10^{-4}$ as explained above --
to be compared with 2.5$\times 10^{-4}$ in our Table \ref{eq:75a}.
So there are no gross differences in the identification of the
$J=2$ objects in these experiments.

\subsubsection{Hadronic decays 
$J/\psi \ \rightarrow \ \omega X \ ; \ \varphi X $}

A new interesting chapter in studying
hadronic $J/\psi$ decays has been opened up by the Mark-III collaboration in the channels 
\begin{equation}
  \label{eq:76}
  \begin{array}{l} 
  J/\psi \ \rightarrow \ \gamma \ K \overline K; \hspace*{0.3cm} \ \omega \ K \overline{K}
  \hspace*{0.3cm} ; \hspace*{0.3cm}  
  \ \varphi \ K \overline{K}
  \end{array}
\end{equation}
\noindent
discussed by L. K\"{o}pke \cite{kop}. The $K\overline K$ 
invariant mass  distributions in the charge state $K^+K^-$ in the three
channels (\ref{eq:76}) 
are compared in Fig. 2 of Ref. \cite{kop}. 
%
%\begin{equation}
%  \label{eq:77}
%  \begin{array}{l} 
%  J/\psi \ \rightarrow 
%  \ \left \lbrace
%  \begin{array}{l} 
%  \gamma \ K^{+} K^{-} 
%\vspace*{0.3cm} \\
%   \omega \ K^{+} K^{-} 
%\vspace*{0.3cm} \\
%  %\hspace*{0.3cm} ; \hspace*{0.3cm}  
%   \varphi \ K^{+} K^{-}
%  \end{array}
%  \right.
%  \end{array}
%\end{equation}
%

In the $\omega \ K^{+} K^{-}$ channel the (mainly) $f^{\ '}_{2}  (1525)$
signal -- clearly visible in the other two decay modes in Eq. (\ref{eq:76})
-- is absent. 
The most interesting channel is $\varphi \ K^{+} K^{-}$, where $f_{J}  (1710)$
is visible only as a broadening shoulder of the dominant
$f^{\ '}_{2}$ resonance. K\"{o}pke presents two fits to the
acceptance/efficiency corrected invariant mass distributions, one
admitting interference between $f^{\ '}_{2}  (1525)$ and $f_{J}  (1710)$
and one with incoherent addition of the two resonances.
He shows, that only the coherent superposition admits to assign 
mass and width to $f_{J}  (1710)$ compatible with the same parameters as
determined in other channels. 
%We are not certain, whether the
%{\em indispensible} acceptance/efficiency corrections were actually applied
%to the interference study in Ref. \cite{kop} and in the susequent
%experient by the DM2 collaboration, discussed below \cite{falvard}.
For angular integrated mass spectra
the crucial consequence of coherent superposition 
is that the two resonances
have to have the same spin. The quantitative distinction between
the two fits is however not disclosed in Ref. \cite{kop}.

Precisely this question is taken up by the DM2 collaboration \cite{falvard}.
Falvard et al. perform three fits, two with 
% (1-3): 1 and 2 
coherent superposition
and one with incoherent superposition. 
The respective $\chi^{2}$ (p.d.f.) clearly
favor the two fits with coherence.
%With all doubts as to what type of coherence is
%observed and analyzed in Refs. \cite{kop} and \cite{falvard}
We take these results together %with weight on 
as further indication of a large spin $J = 2$ component in 
$f_{J}  (1710)$. 

\subsubsection{Hadronic collisions}

\noindent{\it Central production}\\
If $f_{J}  (1710)$ is a glueball it should 
also be produced  centrally in hadronic collisions. Indeed, 
the WA76 collaboration working with the Omega spectrometer \cite{Omega}
has observed a clear signal in the $K^+K^-$ and $K_sK_s$ mass spectra
in 
\begin{equation}
pp \rightarrow p_{fast}(K\overline K) p_{slow}
\label{ppcent}
\end{equation}
 at 85 and 300 GeV. 
Similar to the case of radiative $J/\psi$ decays, two peaks  
appear above a smooth
background from  $f_2'(1525)$ and $f_J(1710)$. The 
polar angle decay distribution in both resonance regions 
is rather similar and largely non-uniform. It is concluded that the spin of 
 $f_J(1710)$ is $J=2$ and the assignments $J^P=1^-$ and $J^P=0^+$
are excluded.

Very recently, new results on reaction (\ref{ppcent}) at 800 GeV have been
presented by the E690 collaboration at Fermilab \cite{Reyes}. In the region
of interest, the mass spectrum again shows two peaks. Surprisingly, the
first peak is now dominated by $f_0(1500)$ besides a smaller contribution 
presumably from
$f_2(1525)$. Looking at the small branching ratio into $K\overline K$
(see  Table \ref{flavor}) the process (\ref{ppcent}) could serve as a real
 $f_0(1500)$ factory if confirmed. 

In the region of  $f_J(1710)$, there are two solutions with large and
small spin $J=2$ component, respectively. No attempt has been made to 
find the most appropriate decomposition into Breit-Wigner resonances
consistent with other knowledge. For the moment, the most accurate data leave
us with a large uncertainty.

\noindent{\it Peripheral production}\\
Finally, we quote the work by Etkin et al. \cite{bnl}  measuring 
$\pi^- p \rightarrow K_sK_s n$ collisions which we discussed already before in 
Sect. \ref{sectnonet} in connection with the $f_0(1500)$ state. In the
higher mass region, the same experiment gave evidence of another scalar state 
at 1771 MeV and $\Gamma \ \sim$ 200 MeV which is produced through the one
pion exchange mechanism. It is natural to identify this state with the one
observed in the $f_J(1710)$ region. The higher mass agrees well with the one
observed by BES (see Table \ref{eq:75a}).

\subsubsection{Summary on spin assignments to $f_J(1710)$}

We summarize the experimental indications for both
$J = 2$ and $J=0$ in Table \ref{sumspin}.
\begin{table}[ht]
\[  \begin{array}{l} 
  \begin{array}{l@{\hspace*{0.5cm}}cc ll}
\hline \vspace*{0.1cm}
  \mbox{Collaboration} & J \ = \ 2 & J \ = \ 0 & \mbox{channel} 
  & \mbox{method}
%\vspace*{0.1cm} 
  \\ \hline 
(1)\ J/\psi\ \mbox{decays}:&&&& \vspace*{0.1cm}  \\
  \mbox{Crystal Ball} \ \cite{crballeta} 
       & \mbox{yes} & \mbox{no} & \gamma \ \eta \eta &
  \mbox{spin analysis (sb)}
\vspace*{0.1cm} \\
\mbox{Mark-III} \ \cite{Mk3a}            
        & \mbox{yes} & \mbox{no} & \gamma \ K^{+} K^{-} &
  \mbox{spin analysis (sb)}
\vspace*{0.1cm} \\
\mbox{Mark-III prel.}  \  \cite{chenrep1}
        & \sim 25\% -40\% & \mbox{yes} & \gamma \ K \overline{K} &
  \mbox{spin analysis (mi)}
\vspace*{0.1cm} \\
\mbox{BES} \ \cite{besK}              
        & 75\% & 25\%  & \gamma \ K^{+} K^{-} &
  \mbox{spin analysis (mi)}
\vspace*{0.1cm} \\
\mbox{Mark-III} \ \cite{kop} \ , \ \mbox{DM2} \ \cite{falvard} &
\mbox{yes} & \mbox{no} & \varphi \ K^{+} K^{-} &
\mbox{interference} \vspace*{0.1cm}\\
(2)\  \mbox{central production}:&&&& \vspace*{0.1cm}\\
\mbox{WA76} \ \cite{Omega}
        & \mbox{yes} & \mbox{no}  & p \ K \overline K \ p &
  \mbox{spin analysis (sb)} \vspace*{0.1cm} \\
\mbox{E690} \ \cite{Reyes}
        & \mbox{yes} & \mbox{yes}  & p \ K_s K_s\ p &
  \mbox{spin analysis (mi)} \vspace*{0.1cm} \\
(3)\  \mbox{peripheral production}:&&&& \vspace*{0.1cm}\\
\mbox{BNL} \ \cite{bnl}
        & \mbox{no} & \mbox{yes}  & n \ K_s K_s  &
  \mbox{spin analysis (mi)} \vspace*{0.1cm} \\
\hline
%\hspace*{0.3cm} ; \hspace*{0.3cm}  
\end{array}
\end{array}\]
\caption{Summary of spin assignments to the $f_J(1710)$ 
in the various ana\-lyses for three reaction types. The spin determination
is carried out in a single mass bin (sb) or mass-independent analysis (mi).}
\label{sumspin}
\end{table}
\noindent
The experiments in group (1) and (2), 
analyzing a single mass interval around 1700 MeV, all prefer
$J=2$ clearly over  $J=0$. The more refined experiments with higher statistics
performing a mass-independent ana\-lysis find a spin zero component
in addition.
As a $J=0$ state is found in peripheral collisions (3) in this mass range,
it is most natural to associate it with a scalar quarkonium state
$f_0(1770)$ MeV, slightly higher in mass than $f_J(1710)$.
On the other hand, the prominent
peak in the $J=2$ wave only appears in the gluon-rich reactions (1) and (2),
and is therefore  
our primary
glueball candidate 
\begin{equation}
  \label{eq:75c}
  \begin{array}{l} 
J=2:\qquad  f_{J} \ (1710)% \rightarrow 1700) \ \leftrightarrow
 \rightarrow  \ g  b \ ( \ 2^{++} \ )
  \end{array}
\end{equation}
which completes the basic triplet of binary glueballs.
%
%-----------------------------------------------------------------
%
%\input{gb8v1.tex}
\section{Conclusions}

In this paper we have reanalysed the spectroscopic evidence for 
various hadronic states
with the aim to find the members of the 
$q\overline q$ nonet with  $J^{PC}=0^{++}$  of lowest mass 
and to identify the triplet of lightest binary glueballs. We draw the
following conclusions from our study:

\begin{description}
\item {\it 1. The $0^{++}$ $q\overline q$ nonet}\\
 As members of this multiplet we identify the 
isoscalar states $f_0(980)$ and $f_0(1500)$ together with $a_0(980)$ and
$K_0^*(1430)$. The mixing between the isoscalars is about the same as in 
the pseudoscalar nonet, i.e. little mixing between singlet and
octet states,  with the correspondence and approximate flavor decomposition
$(u\overline u,d\overline d,s\overline s)$
\[
\begin{array}{lll} 
\eta \ \leftrightarrow\  f_0(1500)&\quad \frac{1}{\sqrt{3}}\ (1,\ 1,\ -1) 
      &\quad\mbox{close to octet},\\
\eta'\ \leftrightarrow\  f_0(980)&\quad \frac{1}{\sqrt{6}}\ (1,\ 1,\ 2)
     &\quad \mbox{close to singlet},
\end{array}
\]
whereby the $(\eta',\  f_0(980))$ pair forms a parity doublet
approximately degenerate in mass. 

The support for this assignment comes
from the Gell-Mann-Okubo mass formula 
(after rejecting the $K\overline K$ bound state interpretation of 
$a_0(980)$),
the $J/\psi\to \phi/\omega + f_0(980)$
decays, the branching ratios for decays of the scalars 
into pairs of pseudoscalars 
as well as the amplitude signs we obtained. 
The most important information comes from phase shift analyses of
elastic and inelastic $\pi\pi$ scattering as well as from the recent
analyses of $p\overline p$ annihilation near threshold. 

\item {\it 2. The  $0^{++}$ glueball of lowest mass}\\
The broad object which extends in mass from 400 MeV up to about 1700 MeV
is taken as the lightest $0^{++}$ glueball. In this energy range, the 
$\pi\pi$ amplitude describes a full loop in the Argand diagram after the 
$f_0(980)$ and $f_0(1500)$ states are subtracted. In particular, we do not
consider the occasionally suggested  $\sigma(700)$ and the $f_0(1370)$, listed
by the Particle Data Group as genuine resonances, since
the related phase movements are too small.

This hypothesis is further supported by the occurrence of this state
in most reactions which provide the gluon-rich environment favorable for
glueball production, also by the decay properties in the 1300 MeV region and
especially by the strong suppression in $\gamma\gamma$ collisions. 
An exception is perhaps the decay $J$/$\psi\to
\pi\pi\gamma$, but no complete amplitude ana\-lysis is available yet in this
case.
\item  {\it 3. The $0^{-+}$ and $2^{++}$ glueballs}.\\
The triplet of binary glueballs is completed by the state  $f_J(1710)$,
for which by now  overwhelming evidence exists in favor of 
a dominant spin 2 component, 
and the $0^{-+}$ state $\eta(1440)$. They appear with large branching
ratios in the radiative decays of the $J$/$\psi$, in
agreement with the expectations for a glueball. 
Central production in  $pp$ collisions is observed for $f_2(1710)$,
but less significantly for $\eta(1440)$, so this
assignment is at a more tentative level. 
\end{description}
\begin{table}[ht]
\[
%\begin{equation}
%  \label{eq:82} 
%  \begin{array}{l}
%  \eta \ (1440) \ \rightarrow \ g \ b \ ( \ 0^{-} \ )
%  \hspace*{0.3cm} , \hspace*{0.3cm}
%  f_{J} \ (1710) \ \rightarrow \ g \ b \ ( \ 2^{+} \ )
%  \vspace*{0.5cm} \\
\begin{array}{c@{\hspace*{0.6cm}}c@{\hspace*{0.4cm}}
l@{\hspace*{0.4cm}}c@{\hspace*{0.4cm}}c}\\ \hline
  \mbox{name} & \mbox{PDG} & \mbox{mass (MeV)}& \mbox{mass}^2 \mbox{(GeV)}^2 
& \mbox{width (MeV)}
  \\ \hline 
   gb \ ( \ 0^{++} \ ) & f_0(400-1200)& \sim 1000 & \sim 1. &
      \hspace*{0.2cm}   500-1000
  \vspace*{0.1cm}   \\
  & f_0(1300) &&&
        \vspace*{0.1cm}\\
   g b \ ( \ 0^{-+} \ ) & \eta  (1440)& 1400\ -\ 1470 & 2.07 &
      \hspace*{0.2cm} 50\ -\ 80
  \vspace*{0.1cm} \\
   gb \ ( \ 2^{++} \ ) & f_{J}  (1710)& 1712 \ \pm \ 5 & 2.93
 & 133 \ \pm \ 14 \\
\hline
\end{array}
%\end{array}
%\end{equation}  
\]
\caption{Properties of the basic triplet of binary glueballs $gb$.}
\label{tabsum}
\end{table}
\noindent
The properties of the basic triplet of binary glueballs are
summarized in Table \ref{tabsum}.
Interestingly, the mass square of these states are separated by about 1
GeV$^2$ as in case of the $q\overline q$ Regge recurrences.

Whereas this overall picture of the low-mass $q\overline q$ and $gg$ states 
seems to accommodate the relevant experimental facts, there is certainly a
need for further improvements of the experimental evidence, 
for which we give a few examples:
\begin{description}
\item {\it 1. Elastic and inelastic $\pi\pi$ scattering}\\
The status of elastic $\pi\pi$ scattering above 1.2 GeV is still not
satisfactory. The phase shift analysis of available $\pi^0\pi^0$ data
could be of valuable help to establish the parameters of the $f_0(1500)$ in
this channel and to determine the behaviour of the ``background'' amplitude,
the same applies for the $\eta\eta$ channel. It will be interesting to
obtain a decomposition of the ``background'' from $f_0(980)$ and find the
relative signs of the components. 
\item {\it 2. Branching ratios of scalar mesons}\\
Of particular interest are the tests of predictions on the decays
$J/\psi\to \phi/\omega + f_0(1500)$ to further establish the quark
composition of this state. The same applies for the 2$\gamma$ widths
of both isoscalar states.
\item{\it 3. Production and decay of the lightest glueball}\\
 The radiative decays of the $J/\psi$ into $ \pi\pi$ and other pseudoscalars 
are naturally expected to show a signal
from the lightest glueball. So far, the experimental results have been
plagued by background problems and the dominance of 
higher spin states like $f_2(1270)$; a spin ana\-lysis is required to get
more clarity. The decays of this object into other pseudoscalars 
above 1 GeV is of interest.
\item{\it 4. Glueballs in $\gamma\gamma$ collisions}\\
If the mixing with $q\overline q$ is small, the production of the 
glueballs should be suppressed. For the lightest glueball this is observed
 in the mass region below 1 GeV. It is of crucial importance to demonstrate 
this suppression in the region above 1 GeV in the $0^{++}$ wave. 
Here, only
$f_0(980)$ and the $f_0(1500)$ should remain as dominant features.
\end{description}

Our hypotheses on the spectroscopy of low-lying glueballs and $q\overline q$
states are not in contradiction with theoretical expectations. 
The masses in Table \ref{tabsum} are in good agreement with the expectations
from the bag model. Also the QCD sum rules suggest a strong gluonic coupling
of $0^{++}$ states around 1 GeV.
It will be interesting
to see whether the more complete lattice calculations now on their way
yield a ``light''
gluonic $0^{++}$ state around 1 GeV 
 as well as
``light'' scalar $q\overline q$ mesons. It is expected that a light glueball
is much broader than the heavier brothers, and this is consistent 
with our scheme in
Table \ref{tabsum}.

We found the 
most general effective potential for the scalar nonet sigma variables 
to be compatible with the $a_0 - f_0$ mass degeneracy, 
independently of the strange
quark mass $m_s$. The mass splitting $O(m_s)$ shows a continuum of breaking
patterns not necessarily along the OZI rule, as often assumed from the
beginning. It remains an open question in this approach,
 though, what  the physical origin
of the $a_0 - f_0$ mass degeneracy is and 
the same holds for the mirror symmetry of the mixing patterns
in the scalar and pseudoscalar nonets.
A possible explanation for the latter
structure is suggested by a renormalizable model with an instanton induced
$U_A(1)$-breaking interaction.

%  Dragon picture
\vfill
\begin{center}
\resizebox*{4cm}{4cm}{\includegraphics{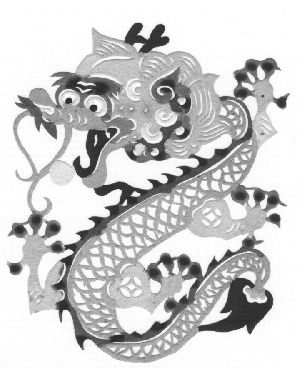}}
\end{center}
\vfill
%
%-------------------------------------------------------------------
%\input{gbref3.tex}

%****************************************************************

% References

%****************************************************************

%****************************************************************

% End of References

%****************************************************************

%-----------------------------------------------------------------

%\input{gbfigv2}
%
%gbfigv2
\newpage
\section*{Figure Captions}

\begin{figure}[t]
\begin{center}
\mbox{\epsfig{file=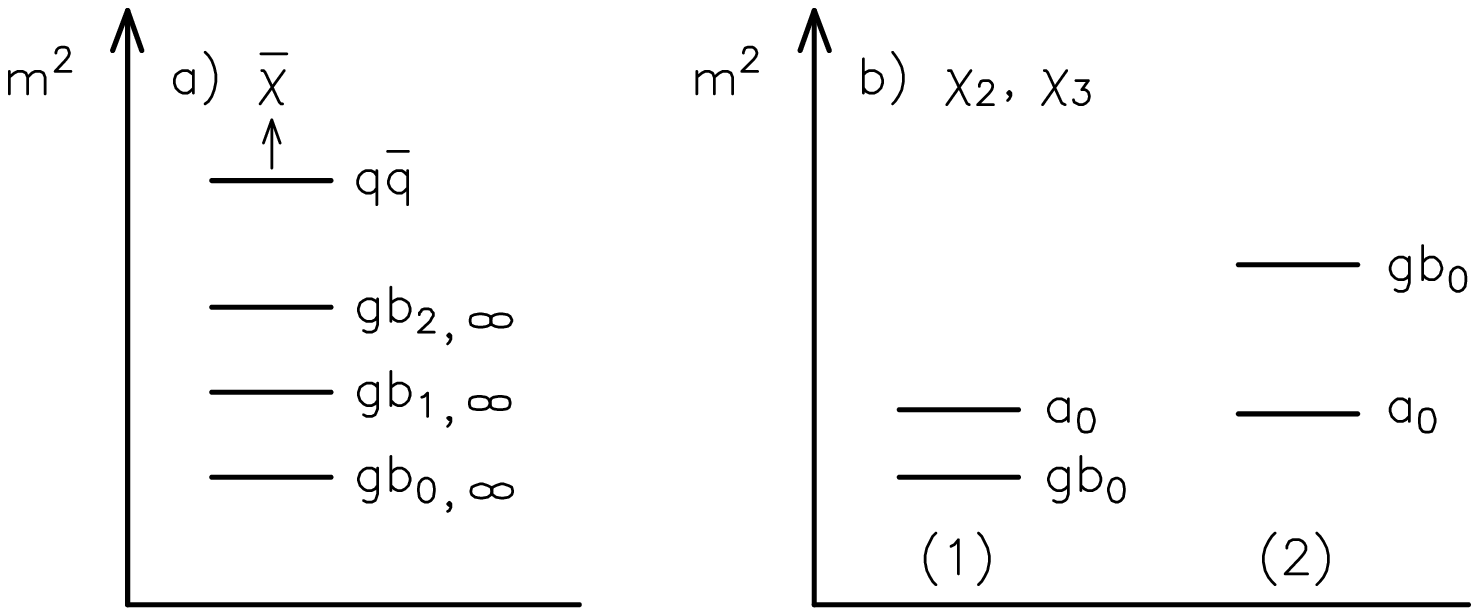,width=130mm}}
\end{center}
\caption{}
\label{gbfig1}   
\end{figure}

\noindent  
{\bf Fig.~\ref{gbfig1}}.
Schematic representation of the mass spectra of glueballs and
$q\overline q$ states in the antichiral limit $\overline{\chi}
\ ( m_q \rightarrow  \infty)$ and in the
chiral limits $\chi_2,\ \chi_3$ of Eq. (\ref{eq:1}).\\

\begin{figure}[ht]
\begin{center}
\mbox{\epsfig{file=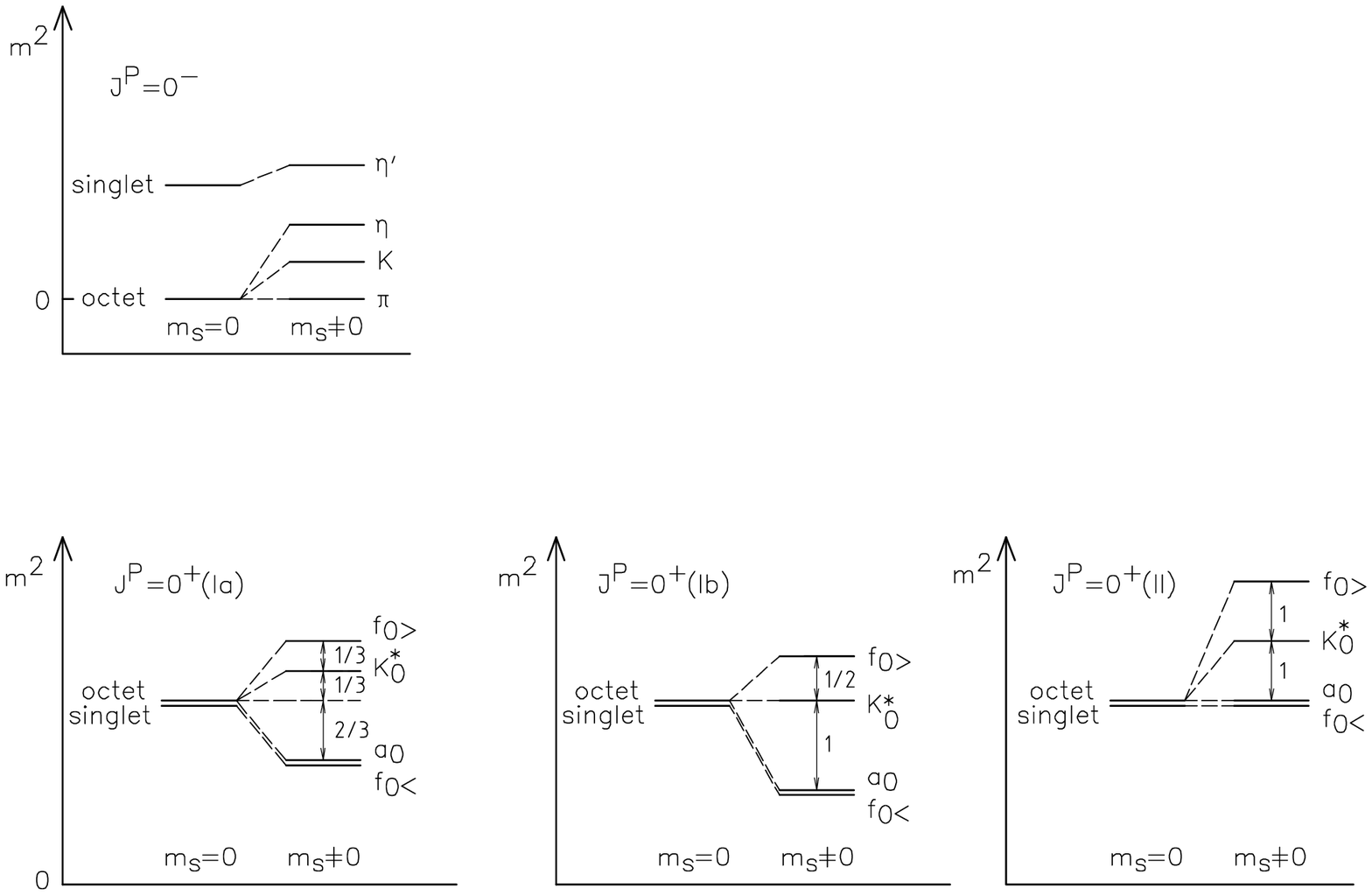,width=130mm}}
\end{center}
\caption{}
\label{gbfig2}
\end{figure}

\noindent  
{\bf Fig.~\ref{gbfig2}}.
Mass splitting of the pseudoscalar states
in the chiral limit $\chi_3$ with all $m_q=0$ and after the strange
quark mass $m_s$ is set nonzero, 
and the same for the scalar states for three different choices 
of the singlet octet mixing angle: 
Ia) mixing angle $\Theta \ = \ 0$ (no mixing),
Ib) small mixing angle $\Theta \ = \ \arcsin \ \frac{1}{3}$ (as for pseudoscalars),
and II) mixing angle $\Theta \ = \ - \arcsin \ \frac{1}{\sqrt{3}}$ 
(according to OZI-rule).

\begin{figure}[t]
\begin{center}
\mbox{\epsfig{file=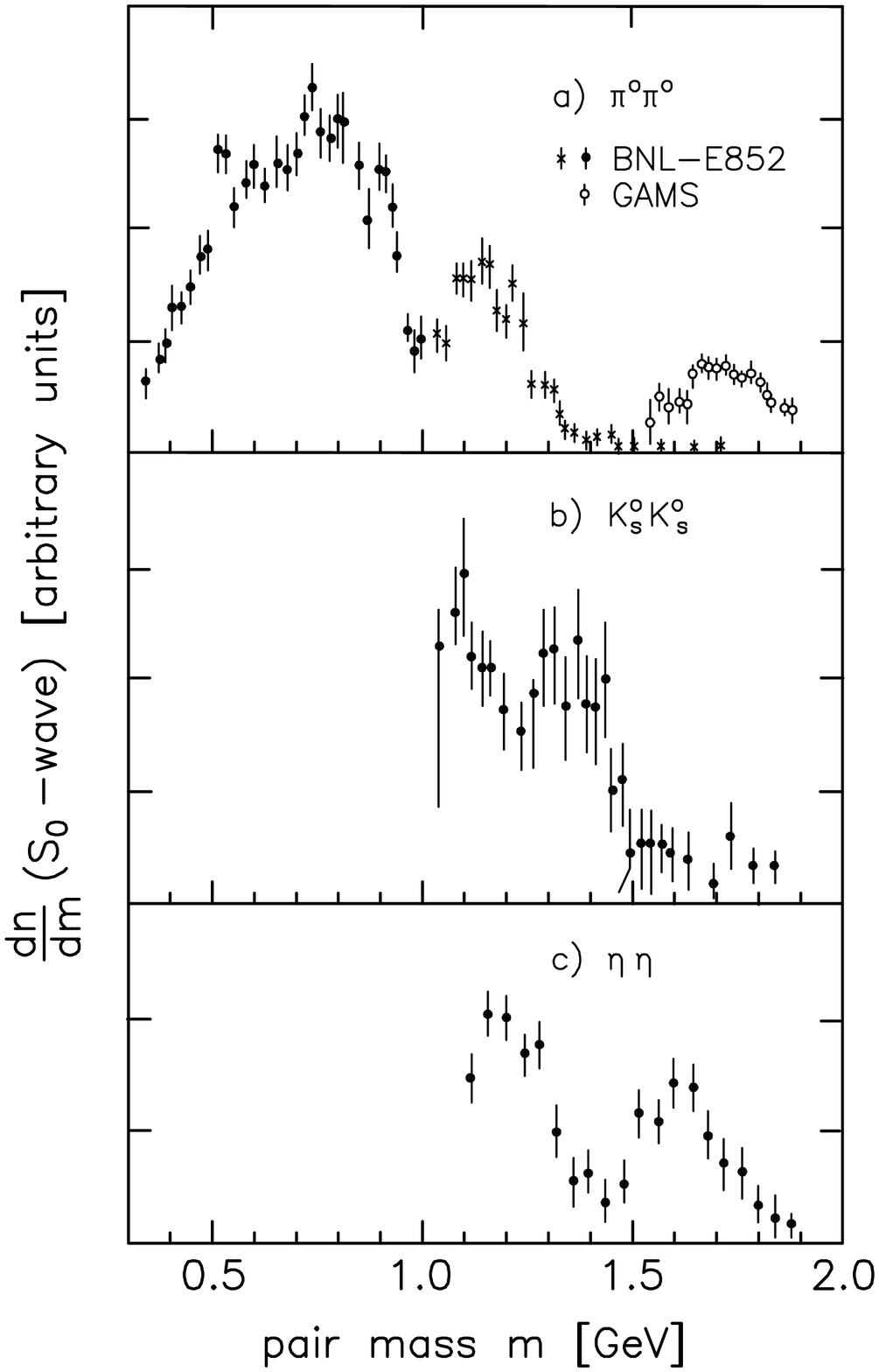,width=120mm}}
\end{center}
\caption{}
\label{gbfig3}
\end{figure}

\noindent
{\bf Fig.~\ref{gbfig3}}.
Isoscalar $S$-wave components of the mass spectra of pseudoscalar pairs
produced in $\pi p$-collisions at small momentum transfers $t$,
(a) $\pi^0\pi^0$ spectrum, the prefered solution for $m<1.5$ GeV
by the BNL-E852 experiment \protect\cite{gunter} 
(preliminary results) and an alternative solution
for higher masses by GAMS \protect\cite{aps}; (b) $K^0_s K^0_s$ spectrum 
by Etkin et al. \protect\cite{bnl} which is similar to the results by Cohen
et al. \protect\cite{argonne} below 1600 MeV and (c) $\eta\eta$ spectrum 
by Binon et al. \protect\cite{IIL1}.\\

\begin{figure}[t]                                                          
\begin{center}
\mbox{\epsfig{file=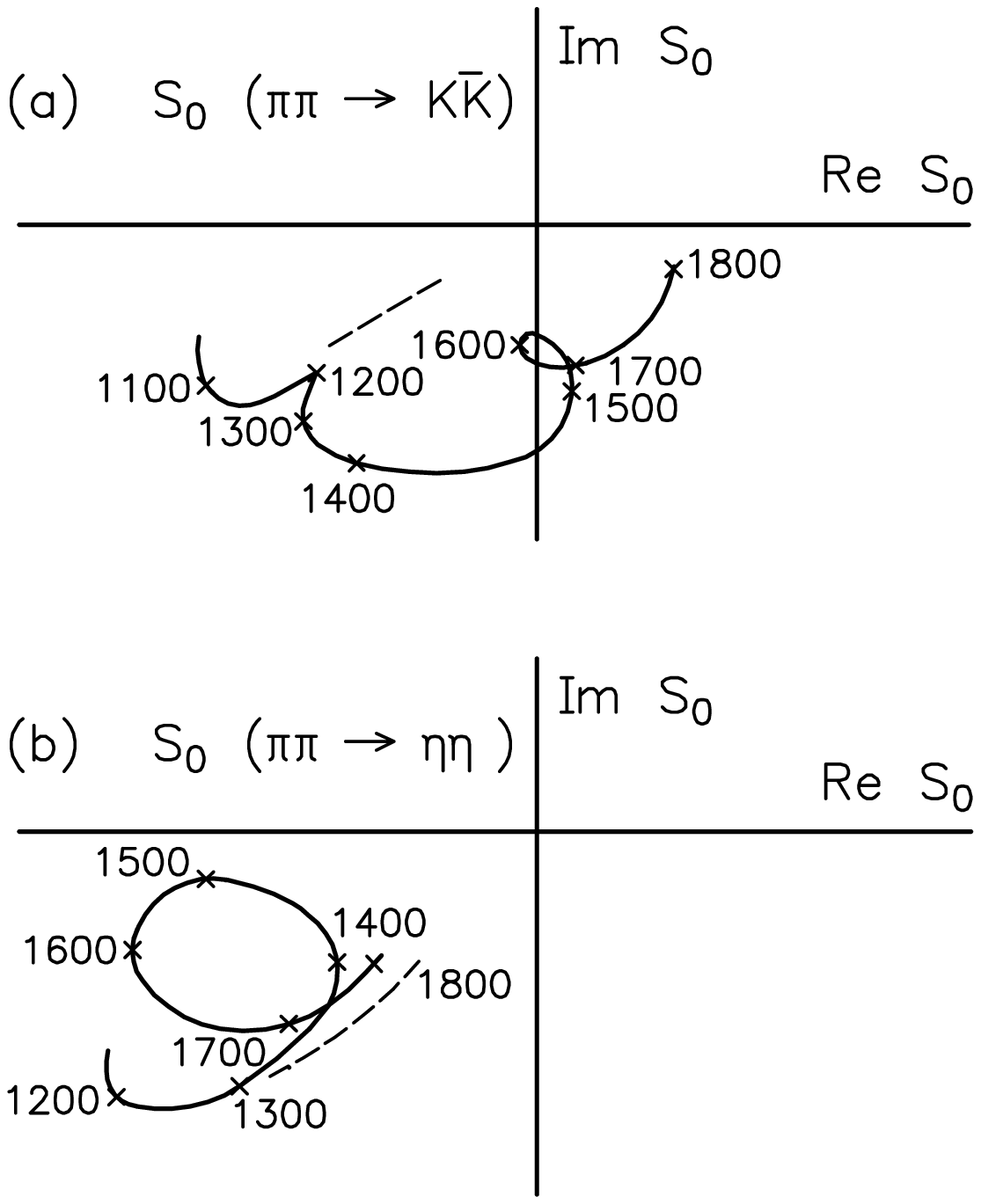,width=100mm}}
\end{center} 
\caption{}
\label{gbfig4}
\end{figure}

\noindent
{\bf Fig.~\ref{gbfig4}}.
Argand diagrams of the isoscalar $S$-wave amplitudes representing
the data on the mass spectra shown in Fig. \ref{gbfig3}
and the relative phases between $S$ and $D$-waves assuming a Breit-Wigner
form for the latter (data from  \protect\cite{bnl} and  \protect\cite{IIL1}).
The numbers indicate the pair masses in MeV, the dashed curve represents an
estimate of the background.\\

\newpage
\hoffset -0.5cm

\newpage

\newpage

\end{document}